\documentclass[]{aa}
\usepackage{natbib}
\bibpunct{(}{)}{;}{a}{}{,}
\usepackage{amssymb}
\usepackage{amsmath}
\usepackage{graphicx}
\usepackage{times}
\usepackage{lscape}
\usepackage{supertabular}
\usepackage{longtable}

\begin{document}

\title{Keplerian discs around post-AGB stars: \\ a common phenomenon?
\thanks{Based on observations with the P7 photometer at the $0.7\,$m
Swiss Telescope, La Silla (Chile) and at the Flemish Mercator
Telescope, La Palma (Spain) and on observations with the Submillimetre
Common-User Bolometer Array (SCUBA) at the James Clerk Maxwell
Telescope (JCMT), Mauna Kea, Hawaii.}}

\author{S. De Ruyter\inst{1}
\and H. Van Winckel\inst{2}
\and T. Maas\inst{2,3}
\and T. Lloyd Evans\inst{4}
\and L.~B.~F.~M Waters\inst{2,5}
\and H. Dejonghe\inst{1}}

\institute{Sterrenkundig Observatorium, Universiteit Gent, Krijgslaan
281 S9, 9000 Gent, Belgium \and Instituut voor Sterrenkunde,
K.U. Leuven, Celestijnenlaan 200B, 3001 Leuven, Belgium \and
Department of Astronomy, University of Texas, Austin, TX 78712 \and
School of Physics and Astronomy, University of St. Andrews, North
Haugh, St. Andrews, Fife, Scotland KY16 9SS \and Sterrenkundig
Instituut `Anton Pannekoek', Universiteit Amsterdam, Kruislaan 403,
1098 Amsterdam, The Netherlands}

\offprints{S. De Ruyter, \email{stephanie.deruyter@UGent.be}}

\date{Received <date> / Accepted <date>}

\abstract
{}
{We aim at showing that the broad-band SED characteristics of our
sample of post-AGB stars are best interpreted, assuming the
circumstellar dust is stored in Keplerian rotating passive discs.}
{We present a homogeneous and systematic study of the Spectral Energy
Distributions (SEDs) of a sample of 51 post-AGB objects. The selection
criteria to define the whole sample were tuned to cover the broad-band
characteristics of known binary post-AGB stars. The whole sample
includes 20 dusty RV\,Tauri stars from the General Catalogue of
Variable Stars (GCVS). We supplemented our own Geneva optical
photometry with literature data to cover a broad range of fluxes from
the UV to the far-IR.}
{All the SEDs display very similar characteristics: a large IR\,excess
with a dust excess starting near the sublimation temperature,
irrespective of the effective temperature of the central
star. Moreover, when available, the long wavelength fluxes show a
black-body slope indicative of the presence of a component of large mm
sized grains.}
{We argue that in all systems, gravitationally bound dusty discs are
present. The discs must be puffed-up to cover a large opening angle
for the central star and we argue that the discs have some similarity
with the passive discs detected around young stellar objects. We
interpret the presence of a disc to be a signature for binarity of the
central object, but this will need confirmation by long-term
monitoring of the radial velocities. We argue that dusty RV\,Tauri
stars are those binaries which happen to be in the Population\,{\rm
II}\,instability strip.}


\maketitle

\section{Introduction}
Post-AGB stars are low and intermediate initial mass
($\leq\,8$-$9$\,M$_{\sun}$) stars which have suffered a large and
dusty mass-loss phase at the end of the Asymptotic Giant Branch (AGB)
during which almost the whole stellar envelope was expelled. They are
evolving at constant luminosity on a fast evolutionary track in which
the central star crosses the HR-diagram from a cool AGB photosphere to
the ionizing temperature of the central star of a Planetary Nebula
(PN). Given the small evolutionary timescale of about $10^4$\,years,
post-AGB stars are rare and not many are known
\citep[e.g][]{01Szczerba,03VanWinckel}.

Discussions on the morphology of PNe usually start with a display of
the aesthetic pictures of the Hubble Space Telescope (HST) showing the
complex geometry and structure of the nebulae, immediately followed by
the puzzling and contradictory finding that, on the AGB, the mass-loss
is found to be spherically symmetric. During the transition time, the
star and circumstellar envelope must undergo fundamental and rapid
changes in structure, mass-loss mode and geometry which are still
badly understood. The debate on which physical mechanisms are driving
the morphology changes gained even more impetus from the finding that
also resolved cooler post-AGB stars or Proto-Planetary Nebulae (PPNe)
display a surprisingly wide variety in shapes and structures, very
early in their post-AGB evolution \citep[][and references
therein]{02Balick}.

In a survey of 27 PPNe, 21 were found to be resolved
\citep{00Ueta}. Moreover, the degree of asymmetry could be linked to
the pole-to-equator density contrast, as determined on the basis of
high spatial resolution mid-infrared images
\citep{99Meixner}. Recently also \citet{05Gledhill} found evidence for
axi-symmetry in the dust density in his polarimetric imaging survey of
candidate post-AGB stars. The detached shells correspond to stars with
an optically thin expanding circumstellar envelope whereas the bipolar
and unresolved targets have optically thick dust structures, probably
in the form of discs. It is suggested once again that this bifurcation
in morphology is rooted in the presence or absence of a binary
companion, which determines whether or not a disc forms.

Stunning kinematic information resulted from the extensive CO survey
of \citet{01Bujarrabal}: it appears to be a fundamental property of
the omnipresent fast molecular outflow in PPNe, that it carries a huge
amount of linear momentum, up to $1000$ times the momentum available
for a radiation driven wind. Clearly, other momentum sources have to
be explored. Some molecular jets are resolved by high spatial
resolution imaging \citep{04Sahai}. The formation process of the
strongly collimated jets is, however, still badly understood. An
intriguing suggestion is that the processes similar to the jet
formation in low-mass young stellar objects operate and that the jets
are born in accretion discs. This mechanism requires a significant
amount of mass orbiting the post-AGB star. Such a disc could be
present, but likely only in binary stars. Testing of such an
hypothesis is severely hampered by the lack of observational
information on binarity in PNe and PPNe but also on our poor
theoretical understanding of AGB evolution in binary systems. Note
that most objects which were detected in CO are strongly embedded and
the sample is probably biased towards more massive PPNe.

Probing binarity in PNe and embedded PPNe directly with radial
velocity monitoring is not easy. For optically bright post-AGB stars,
this is different and some famous examples exist in which it is clear
that the binary nature of the central object must have played a key
role in the evolution of the system.

The most famous example is certainly HD\,44179, the central star of
the Red Rectangle nebula. It displays a huge and broad IR\,excess. The
IR luminosity is $33$ times stronger than the optical luminosity
\citep{89Leinert,96Waelkens}. \citet{95VanWinckel} have shown that
HD\,44179 is a spectroscopic binary with an orbital period of $298$
days. The eccentricity is a remarkably high $e=0.37$. Since its
discovery by \citet{75Cohen}, HD\,44179 has often been used as an
archetypical example of a C-rich post-AGB object, but it is now
generally accepted that many of the remarkable phenomena and the
peculiar morphology of the nebula \citep[for an overview
see][]{04Cohen} are closely related to the presence of a stable
circumbinary disc around the binary central star. The longevity of the
disc was dramatically confirmed by the detection of cool O-rich
crystalline silicate dust grains in the disc \citep{98Waters}. The
mixed chemistry is best explained assuming the formation of the O-rich
disc predated the more recent C-rich transition of the central
object. Recent Spitzer data show, however, that also far from the
central star, the nebula appears to have a dust component which is
O-rich \citep{05Markwick-Kemper}. The (chemical) history of the
binary, disc and nebula is therefore far from understood. The disc is
resolved in ground-based high spatial-resolution imaging at optical
and near-IR wavelengths \citep[][and references
therein]{02Menshchikov} as well as in HST optical images
\citep{04Cohen}. The disc was also resolved in interferometric CO(2-1)
maps, and the Keplerian kinematics of the disc were directly detected
\citep{03Bujarrabal,05Bujarrabal}. HD\,44179 shows a considerable
amount of dust processing in the disc with indications of the presence
of very large grains \citep{97Jura} and possibly even macro-structures
\citep{98Jura}. 

Another remarkable evolved object with a long-lived disc is
HR\,4049. It is a binary with an orbital period of $430$ days with a
remarkably high eccentricity of $e=0.30$. Also in this object, the
circumstellar material shows a mixed chemistry with both carbon rich
and oxygen rich features. The SED of the dust is also very peculiar as
it can be fitted with a single black-body of about 1150\,K, from
$1\,\mu$m down to $850\,\mu$m. These SED characteristics are very
constraining and the best model for the circumstellar material is,
that the dust is trapped in a very opaque dust torus at Keplerian
rotation \citep{03Dominik}. It is clear that also in this object the
dusty disc plays a lead role in the (future) evolution. Other
examples exist and there is substantial observational evidence that
the systems are all likely surrounded by a circumstellar orbiting disc
\citep{03VanWinckel}. Note that so far only for the Red Rectangle the
Keplerian disc is spatio-kinematically resolved by CO interferometric
maps \citep{05Bujarrabal}.

To gain insight in the evolution of binary systems and their
circumstellar material, we report in this contribution on a
homogeneous and systematic study of a sample of objects with similar
IR\,characteristics as the known binary post-AGB stars. The main aim
is to study the broad-band SEDs in a systematic way, in order to gain
insight in the possible evolutionary link between the different
objects.

Our sample and the selection criteria are presented in
Sect.~\ref{section:stars}. In Sect.~\ref{section:data} we present our
photometric data complemented with the literature values and the
results of the large all sky surveys. In Sect.~\ref{section:SED} we
present the detailed Spectral Energy Distribution (SED) construction
of all individual objects. Distances to the objects are estimated in
Sect.~\ref{section:D}. We end this extensive study with a detailed
discussion in Sect.~\ref{section:discussion}. Conclusions are
summarized in Sect.~\ref{section:conclusions}.

\section{Programme stars}
\label{section:stars}
Our total sample of 51 stars was defined on several observational
criteria inspired by the characteristics of known post-AGB stars in
binary systems. The whole sample is given in
Table~\ref{tab:coordinates}. Besides these proven binaries, we discuss
also RV\,Tauri stars with IRAS detections of good quality and a sample
of newly identified objects, originally found by one of us (T.~Lloyd
Evans) in search of new candidate RV\,Tauri stars.
\begin{landscape}
\begin{table}
\centering
\caption{The IRAS\,number, the HD\,number or the name from the GCVS,
the spectral type, the equatorial coordinates $\alpha$ and
$\delta$\,(J2000), the effective temperature $T_{\mathrm{eff}}$, the
surface gravity $\log g$, the metallicity [\element{Fe}/\element{H}],
the reference for the model parameters, the type of object (post-AGB,
RV\,Tauri or New Sample) and a reference where the orbital motion of
the object can be found, are given. Note that IRAS\,11472$-$0800 is a
strongly depleted object added to our sample
stars.\label{tab:coordinates}}
\begin{scriptsize}
\begin{tabular}{llllllllrlll}
\hline
\hline 
1	&2			&3			&4	&5			&6
&7			&8		&\multicolumn{1}{l}{9}		&10			&11		&12\\
No	&IRAS\,number		&HD\,number or		&Spectral	&$\alpha$ (J2000)	&$\delta$ (J2000)	
&$T_{\mathrm{eff}}$	&$\log g$	&\multicolumn{1}{l}{[Fe/H]}	&Reference Model Parameters		&Type		&Reference Binarity\\
	&			&GCVS name		&type	&(h m s)		&($\degr\,\arcmin\,\arcsec$) 	
&(K)			&(cgs)		&				&			&		&\\
\hline
1	&IRAS\,04166$+$5719	&TW\,Cam		&G3I	&04 20 48.1		&$+$57 26 26			
&4800			&0.0		&$-0.5$				&\citet{00Giridhar}			&RV\,Tauri	&\\
2	&IRAS\,04440$+$2605	&RV\,Tau		&G2I	&04 47 06.8		&$+$26 10 44			
&4500			&0.0		&$-0.4$				&\citet{00Giridhar}			&RV\,Tauri	&\\
3	&IRAS\,05208$-$2035	&\,			&M0e+F	&05 22 59.423		&$-$20 32 53.03			
&4000			&0.5		&$0.0$				&from spectral type			&New Sample	&\\
4	&IRAS\,06034$+$1354	&DY\,Ori		&G0I	&06 06 12.3		&$+$13 53 09			
&6000			&1.5		&$-2.0$				&\citet{97aGonzalez}			&RV\,Tauri	&\\
5	&IRAS\,06072$+$0953	&CT\,Ori		&F9I	&06 09 57.4		&$+$09 52 35			
&5500			&1.0		&$-2.0$				&\citet{97bGonzalez}			&RV\,Tauri	&\\
6	&IRAS\,06108$+$2743	&SU\,Gem		&G5I	&06 14 00.8		&$+$27 42 12			
&5750			&1.125		&$-0.7$				&\citet{92Wahlgren}			&RV\,Tauri	&\\
7	&IRAS\,06160$-$1701	&UY\,CMa		&G0V	&06 18 16.367		&$-$17 02 34.72			
&5500			&1.0		&$0.0$				&from spectral type			&RV\,Tauri	&\\
8	&IRAS\,06176$-$1036	&HD\,44179		&F1I	&06 19 58.2160		&$-$10 38 14.691		
&7500			&0.8		&$-3.3$				&\citet{92Waelkens}			&post-AGB	&\citet{95VanWinckel}\\
9	&IRAS\,06338$+$5333	&HD\,46703		&F3I	&06 37 52.4253		&$+$53 31 01.957		
&6250			&1.0		&$-1.5$				&from spectrum				&post-AGB	&Hrivnak, private communication\\
10	&IRAS\,06472$-$3713	&ST\,Pup		&F7I	&06 48 56.4131		&$-$37 16 33.332		
&5750			&0.5		&$-1.5$				&\citet{96Gonzalez}			&RV\,Tauri	&\citet{96Gonzalez}\\
11	&IRAS\,07008$+$1050	&HD\,52961		&F6I	&07 03 39.6314		&$+$10 46 13.067		
&6000			&0.5		&$-4.8$				&\citet{91bWaelkens}			&post-AGB	&\citet{92Waelkens}\\
12	&IRAS\,07140$-$2321	&SAO\,173329		&F5I	&07 16 08.271		&$-$23 27 01.61			
&7000			&1.5		&$-0.8$				&\citet{97VanWinckel}			&post-AGB	&\citet{95VanWinckel}\\
13	&IRAS\,07284$-$0940	&U\,Mon			&G0I	&07 30 47.5		&$-$09 46 37			
&5000			&0.0		&$-0.8$				&\citet{00Giridhar}			&RV\,Tauri	&\citet{95Pollard}\\
14	&IRAS\,08011$-$3627	&AR\,Pup		&F0I	&08 03 01.1		&$-$36 35 47			
&6000			&1.5		&$-1.0$				&\citet{97aGonzalez}			&RV\,Tauri	&\\			
15	&IRAS\,08544$-$4431	&\,			&F3	&08 56 14.182		&$-$44 43 10.73			
&7250			&1.5		&$-0.5$				&\citet{03Maas}				&New Sample	&\citet{03Maas}\\
16	&IRAS\,09060$-$2807	&\,			&F5	&09 08 10.1		&$-$28 19 10			
&6500			&1.5		&$-0.5$				&\citet{05Maas}				&New Sample	&\\
17	&IRAS\,09144$-$4933	&\,			&G0	&09 16 09.1		&$-$49 46 06			
&5750			&0.5		&$-0.5$				&\citet{05Maas}				&New Sample	&\\
18	&IRAS\,09256$-$6324	&IW\,Car		&F7I	&09 26 53.4		&$-$63 37 48			
&6700			&2.0		&$-1.0$				&\citet{94Giridhar}			&RV\,Tauri	&\\
19	&IRAS\,09400$-$4733	&\,			&M0	&09 41 52.9		&$-$47 47 03			
&			&		&				&					&New Sample	&\\
20	&IRAS\,09538$-$7622	&\,			&G0	&09 53 58.5		&$-$76 36 53			
&5500			&1.0		&$-0.5$				&\citet{05Maas}				&New Sample	&\\
21	&IRAS\,10158$-$2844	&HR\,4049		&A6I	&10 18 07.5903		&$-$28 59 31.201		
&7500			&1.0		&$-4.5$				&\citet{03Dominik}			&post-AGB	&\citet{91bWaelkens}\\
22	&IRAS\,10174$-$5704	&\,			&K:rr	&10 19 18.1		&$-$57 19 36			
&			&		&				&					&New Sample	&\\
23	&IRAS\,10456$-$5712	&HD\,93662		&K5	&10 47 38.3965		&$-$57 28 02.679		
&4250			&0.5		&$0.0$				&from spectral type			&New Sample	&\\
24	&IRAS\,11000$-$6153	&HD\,95767		&F0I	&11 02 04.314		&$-$62 09 42.84			
&7600			&2.0		&$0.1$				&\citet{97VanWinckel}			&post-AGB	&\citet{95VanWinckel}\\
25	&IRAS\,11118$-$5726	&GK\,Car		&M0I	&11 14 01.3		&$-$57 43 09			
&			&		&				&					&RV\,Tauri	&\\
26	&IRAS\,11472$-$0800	&\,			&F5I	&11 49 48.5		&$-$08 17 21			
&5750			&1.0		&$-2.5$				&from spectrum				&		&\\
27	&IRAS\,12067$-$4508	&RU\,Cen		&F6I	&12 09 23.7		&$-$45 25 35			
&6000			&1.5		&$-2.0$				&\citet{02Maas}				&RV\,Tauri	&\\
28	&IRAS\,12185$-$4856	&SX\,Cen		&G3V	&12 21 12.6		&$-$49 12 41			
&6000			&1.0		&$-1.0$				&\citet{02Maas}				&RV\,Tauri	&\citet{02Maas}\\
29	&IRAS\,12222$-$4652	&HD\,108015		&F3Ib	&12 24 53.501		&$-$47 09 07.51			
&7000			&1.5		&$-0.1$				&\citet{97VanWinckel}			&post-AGB	&\\
30	&IRAS\,13258$-$8103	&\,			&G5 De	&13 31 07.1		&$-$81 18 30			
&			&		&				&					&New Sample	&\\
31	&\,			&EN\,TrA		&F2Ib	&14 57 00.6847		&$-$68 50 22.879		
&6000			&1.0		&$-0.5$				&\citet{97VanWinckel}			&RV\,Tauri	&\citet{95VanWinckel}\\
32	&IRAS\,15469$-$5311	&\,			&F3	&15 50 44.0		&$-$53 20 44			
&7500			&1.5		&$0.0$				&\citet{05Maas}				&New Sample	&\\
33	&IRAS\,15556$-$5444	&\,			&F8	&15 59 32.1		&$-$54 53 18			
&			&		&				&					&New Sample	&\\
34	&IRAS\,16230$-$3410	&\,			&F8	&16 26 20.29		&$-$34 17 12.3			
&6250			&1.0		&$-0.5$				&\citet{05Maas}				&New Sample	&\\
35	&IRAS\,17038$-$4815	&\,			&G2p(R)e&17 07 36.3		&$-$48 19 08			
&4750			&0.5		&$-1.5$				&\citet{05Maas}				&New Sample	&\\
36	&IRAS\,17233$-$4330	&\,			&G0p(R)	&17 26 57.5		&$-$43 33 13			
&6250			&1.5		&$-1.0$				&\citet{05Maas}				&New Sample	&\\
37	&IRAS\,17243$-$4348	&LR\,Sco		&G2	&17 27 56.1		&$-$43 50 48			
&6250			&0.5		&$0.0$				&\citet{05Maas}				&New Sample	&\\
38	&IRAS\,17534$+$2603	&89\,Her		&F3I	&17 55 25.1889		&$+$26 02 59.966		
&6500			&1.0		&$0.0$				&\citet{93Waters}			&post-AGB	&\citet{93Waters}\\
39	&IRAS\,17530$-$3348	&AI\,Sco		&G4I	&17 56 18.5		&$-$33 48 47			
&5000			&0.0		&$0.0$				&from spectrum				&RV\,Tauri	&\\
40	&IRAS\,18123$+$0511	&\,			&G5	&18 14 49.4		&$+$05 12 55			
&5000			&0.5		&$0.0$				&from spectral type			&New Sample	&\\
41	&IRAS\,18158$-$3445	&\,			&F6	&18 19 13.6		&$-$34 44 32			
&6500			&1.5		&$0.0$				&from spectral type			&New Sample	&\\
42	&IRAS\,18281$+$2149	&AC\,Her		&F4I	&18 30 16.2		&$+$21 52 00			
&5500			&0.5		&$-1.5$				&\citet{98VanWinckel}			&RV\,Tauri	&\citet{98VanWinckel}\\
43	&IRAS\,18564$-$0814	&AD\,Aql		&G8I	&18 59 08.1		&$-$08 10 14			
&6300			&1.25		&$-2.1$				&\citet{98Giridhar}			&RV\,Tauri	&\\
44	&IRAS\,19125$+$0343	&\,			&F2	&19 15 00.8		&$+$03 48 41			
&7750			&1.0		&$-0.5$				&\citet{05Maas}				&New Sample	&\\
45	&IRAS\,19163$+$2745	&EP\,Lyr		&A4I	&19 18 17.5		&$+$27 50 38			
&7000			&2.0		&$-1.5$				&\citet{97aGonzalez}			&RV\,Tauri	&\\
46	&IRAS\,19157$-$0247	&\,			&F3	&19 18 22.5		&$-$02 42 09			
&7750			&1.0		&$0.0$				&\citet{05Maas}				&New Sample	&\\
47	&IRAS\,20056$+$1834	&QY\,Sge		&G0 De	&20 07 54.8		&$+$18 42 57			
&5850			&0.7		&$-0.4$				&\citet{02KameswaraRao}			&New Sample	&\\
48	&IRAS\,20117$+$1634	&R\,Sge			&G0I	&20 14 03.8		&$+$16 43 35			
&5750			&0.0		&$-0.5$				&\citet{97aGonzalez}			&RV\,Tauri	&\\
49	&IRAS\,20343$+$2625	&V\,Vul			&G8I	&20 36 31.8		&$+$26 36 17			
&5250			&1.0		&$0.0$				&from spectral type			&RV\,Tauri	&\\
50	&IRAS\,22327$-$1731	&HD\,213985		&A2I	&22 35 27.5259		&$-$17 15 26.889		
&8250			&1.5		&$-1.0$				&from spectrum				&post-AGB	&\citet{95VanWinckel}\\
51	&\,			&BD$+$39$^{\circ}$4926	&A9I	&22 46 11.2273		&$+$40 06 26.294		
&7500			&1.2		&$-2.9$			       &\citet{95VanWinckel}			&post-AGB	&\citet{70Kodaira}\\
\hline
\end{tabular}
\end{scriptsize}
\end{table}
\end{landscape}

\subsection{Confirmed binary post-AGB stars}
\label{subsection:binaries}
The group of binary post-AGB stars in our sample was assembled in a
rather coincidental way on the basis of independent detailed
star-to-star analyses. We included the proven binaries in our
programme star sample and in Col.~$12$ of Table~\ref{tab:coordinates}
we give a reference where the orbital motion of the objects was
discussed. The orbital elements themselves are listed in
Table~\ref{tab:Porbit}.

The photospheres of several binary post-AGB stars (e.g. HR\,4049
\citep{88Lambert, 91aWaelkens}, HD\,44179 \citep{92Waelkens},
HD\,52961 \citep{92VanWinckel} and BD$+$39$^{\circ}$4926
\citep{70Kodaira}) are strongly affected by a poorly understood
selective depletion process. The basic scenario of this process is
that circumstellar gas is separated from the dust and subsequently
re-accreted onto the star \citep{92Mathis, 92Waters}. Since that gas
is devoid of refractories, the photosphere will be altered chemically,
and this may result in very Fe-poor stars which are rich in
non-refractories like Zn and S. This process is very efficient in the
four named binaries where there is observational evidence for the
presence of a dusty stable reservoir \citep{95VanWinckel}. Efficient
separation of circumstellar gas and dust is not evident and
\citet{92Waters} argued that the most favourable circumstances may
occur if the circumstellar dust is indeed trapped in a stable
disc. Note that for one strongly depleted object,
BD$+$39$^{\circ}$4926, there was no IR\,excess detected by IRAS.

In the few cases where ISO spectra are available, the dust processing
is strong, which results in a large crystallinity fraction of the
grains \citep[e.g.][]{02aMolster}. Also the dust grain size
distribution is different in discs than in outflows and where long
wavelength fluxes are available, they indicate the presence of large,
$\mu$m sized and even cm sized dust grains \citep[e.g.][]{95Shenton}.

Obviously such a disc must play an important role in these systems
and, although the indirect observational evidence for the presence of
a stable disc is well established in the binary stars, the actual
structure let alone the formation, stability and evolution are not
well understood. Note that only for HD\,44179 is this disc spatially
resolved. For all other objects, the presence of the disc was
postulated.

\subsection{Classical RV\,Tauri stars from the General Catalogue of Variable Stars with strong IR\,excess}
\label{subsection:rvtauries}
RV\,Tauri stars form a class of classical pulsating variables. They
are luminous (I-II) supergiant pulsating variables, the light curve of
which shows alternating deep and shallow minima with a formal period
(measured between one deep minimum and the next) of 30 to 150 days and
a brightness range of up to four magnitudes. The spectral type is
typically F to G at minimum and G to K at maximum. There are two main
photometric varieties of RV\,Tauri stars: the RVa types maintain a
roughly constant mean brightness; RVb types show on top of their
pulsational period a longer term ($600$ to $1500$ days) periodicity.

With their high luminosity and often large IR\,excesses due to thermal
radiation from circumstellar dust, there is general agreement that
RV\,Tauri stars are low-mass objects in transition from the Asymptotic
Giant Branch (AGB) to white dwarfs \citep{86Jura}. It was noted
already in the seventies that many RV\,Tauri stars show a considerable
near-IR\,excess caused by a hot dust component \citep{72aGehrz,
72bGehrz, 85LloydEvans}, which was attributed to the possible presence
of a dusty disc. \citet{87Morris} suggested that circumstellar dust
could indeed exist in a disc structure in binary systems, either in a
circumbinary disc or in a disc around one of the components. The weak
circumstellar CO emission \citep[e.g.][]{88Bujarrabal} together with
the black-body spectral index at long wavelengths observed in some
RV\,Tauri stars \citep{05DeRuyter} corroborate the conclusion that the
circumstellar dust is not freely expanding but confined.

In recent years it has become clear that also many RV\,Tauri
photospheres show chemical anomalies pointing to an efficient
depletion of refractory elements \citep{94Giridhar, 95Giridhar,
98Giridhar, 00Giridhar, 97aGonzalez, 97bGonzalez, 98VanWinckel,
02Maas}. The many affected stars show that the depletion process is a
very common phenomenon in evolved stars. \citet{05Giridhar} show that
depletion in RV\,Tauri stars is less strong when the central star is
cooler, which is interpreted as pointing to a stronger dilution due to
a deeper convective envelope in the cooler stars. The depletion
patterns are not seen in objects for which the intrinsic metallicity
is smaller than about one tenth solar
([\element{Fe}/\element{H}]$\,\leq\,-1.0$).

It is clear that the observational restriction that depletion is only
active in binary stars, which was formulated when only four extremely
depleted objects were known, has become much less evident with the
many new detections. Direct evidence for binarity, from radial
velocity measurements, is difficult to obtain since RV\,Tauri stars
have large pulsational amplitudes. Moreover, they often show the
presence of shocks in the line-forming region of the photosphere
making the very determination of the radial velocity difficult
\citep[e.g.][]{90Gillet}. Nonetheless orbital elements have been
determined for quite a few classical RV\,Tauri stars: U\,Mon
\citep{95Pollard}, AC\,Her \citep{98VanWinckel}, EN\,TrA
\citep{99VanWinckel}, SX\,Cen \citep{02Maas} and orbital motion is
also found for IW\,Car \citep{97Pollard}, EP\,Lyr \citep{97aGonzalez}
and RU\,Cen \citep{02Maas}.

In this contribution we analyse the broad-band SEDs of the RV\,Tauri
stars of the GCVS with a reliable IRAS measurement at $60\,\mu$m
\citep[e.g.][]{86Jura} and compare them with the SEDs of similar
objects, which are not in the Pop\,{\rm II}\,Cepheid instability
strip.

\subsection{New Sample of candidate RV\,Tauri stars}
\label{subsection:TLEvans}
\citet{85LloydEvans} and \citet{89Raveendran} showed that the
RV\,Tauri stars are located in a well-defined and relatively
thinly-populated part of the IRAS colour-colour diagram. The defining
rectangle is:
\begin{equation}
\label{eq:colour-colour}
\left\{
\begin{array}{l}
\left[12\right] - \left[25\right] = 1.56 + 2.5
\log\left[F\left(25\right)/F\left(12\right)\right] = 1.0 -1.55\\
\left[25\right] - \left[60\right] = 1.88 + 2.5
\log\left[F\left(60\right)/F\left(25\right)\right] = 0.20 -1.0
\end{array}
\right.
\end{equation}

\citet{97LloydEvans} searched for new examples of RV\,Tauri stars, as
stars with dusty discs, in the IRAS Catalogue, and found the stars of
the `new sample' discussed in this paper. The presence in this sample
of stars with the infrared properties of the GCVS sample of RV\,Tauri
stars, but which fell outside the instability strip as judged from
their spectral types and which lacked the large-amplitude RV\,Tauri
variability, suggested that RV\,Tauri stars are those stars with dusty
discs which are currently located within the instability strip
\citep{99LloydEvans}.

\citet{05Maas} presented a chemical abundance analysis on the basis of
high signal-to-noise and high resolution spectra for 12 stars of the
newly defined sample. They found that 9 stars are affected by the
depletion process. In a detailed study of one object, IRAS\,08544
\citep{03Maas}; orbital elements were found which show that this star
must have undergone severe binary interaction when it was an
(asymptotic) giant.

In this contribution we analyse the broad-band SEDs of all 20 newly
characterized stars of this sample.

\subsection{Total Sample}
\label{subsection:totalsample}
The 51 objects of our sample are presented in
Table~\ref{tab:coordinates}. In Col.~11 we differentiate between the
several subtypes. Statistical studies on the occurrence and
characteristics of post-AGB stars with respect to AGB stars and/or
Planetary Nebulae (PNe) do not exist yet. In the most extensive
catalogue description found in the literature, \citet{01Szczerba}
consider a Galactic sample of about 220 post-AGB stars known at that
time. Although the overlap between the specific sample discussed here,
and the total sample of \citet{01Szczerba} is far from complete, the
sample of 51 objects discussed here, forms a significant population of
post-AGB stars known to date. Note that our sample was defined on the
basis of a specific set of criteria explained above and our sample is
not meant to cover all known evolved stars with similar SED
characteristics.

\section{Broad-band photometric data}
\label{section:data}
In order to reconstruct the complete SEDs of our programme stars, we
combined different sets of broad-band photometric data: our own Geneva
optical photometry, supplemented with UV, optical, near-IR, far-IR and
submillimetre photometry found in the literature.

The main difficulty in constructing the SEDs of pulsating stars with
large amplitudes, like the RV\,Tauri objects, is the acquisition of
equally phased data over a wide spectral domain. Since these data are
not available, we limited our study of the broad-band energetics to
the phases of minimal and maximal covered brightness.

\subsubsection*{Geneva optical photometry}
We acquired Geneva optical photometry at random epochs with the
$0.7\,$m Swiss Telescope at La Silla and with the Flemish Mercator
Telescope at La Palma, using the refurbished Geneva photometer P7
\citep{04Raskin}. Our total data set was scanned for the maximum and
minimum magnitudes (see Table~\ref{DATA:Geneva}). Observation dates,
the number of measurements and the total timebases of these maxima and
minima are given as well. Additional Geneva optical photometry was
found in the Geneva database: the General Catalogue of Photometric
Data (GCPD,\texttt{http://obswww.unige.ch/gcpd/gcpd.html}). Note these
data are only given for completeness, but were not used in our
analysis; except for IRAS\,10456 for which the GCPD data are our only
Geneva fluxes.

\subsubsection*{$UBVRI$ photometry}
$UBVRI$ photometry for the newly-identified objects was obtained from
the South African Astronomical Observatory (SAAO) service observing
programme (Table~\ref{DATA:optical}). This uses the $0.5\,$m Telescope
and the $0.75\,$m Automatic Photometric Telescope (APT) at
Sutherland. If the $UBVRI$ data of the star show large variations
mainly due to a large variability of the star, we plot both minima and
maxima. If the number of data points available is too low or if there
are no big variations, we use a mean $UBVRI$.

We also searched for Johnson and Cousins broad-band photometry in the
literature and these data are given in Table~\ref{DATA:optical} as
well.

\subsubsection*{Near-IR data}
$JHKL$ photometry of the newly-identified objects was obtained by
T.~Lloyd Evans with the $0.75\,$m reflector of the South African
Astronomical Observatory.

These data points were complemented with near-IR data from the 2\,MASS
and DENIS projects and with other $JHKLM$ data points from the
literature (Table~\ref{DATA:NIR}).

\subsubsection*{Far-IR data}
For the characterization of the longer wavelength part of the SEDs we
use far-IR data from the IRAS Point Source Catalogue
\citep[][Table~\ref{DATA:IRAS}]{88Beichman}; for some stars data from
the MSX Infrared Point Source Catalogue \citep{03Egan} are also
available (Table~\ref{DATA:MSX}). The IRAS satellite had four
passbands, at $12$, $25$, $60$ and $100\,\mu$m, respectively. The
\mbox{SPIRIT\,III} instrument on board the MSX satellite had 6
passbands, at $4.29$, $4.35$, $8.3$, $12.1$, $14.7$ and $21.3\,\mu$m.

\subsubsection*{Submillimetre data}
For six stars (TW\,Cam, RV\,Tau, SU\,Gem, UY\,CMa, U\,Mon and AC\,Her)
submillimetre data (see Table~\ref{DATA:SCUBA}) are available from
\citet{05DeRuyter}, for HR\,4049 and IRAS\,20056 we use data from the
other literature. For HR\,4049, HD\,52961 and 89\,Her, we have some
newly determined SCUBA 850 micron continuum measurements. The
observations were carried out with the 15\,m James Clerk Maxwell
Telescope (JCMT) at Mauna Kea, Hawaii, during March-April 1999
(programme M99AN08). SCUBA \citep{99Holland} was operated in
photometry mode to simultaneously obtain data at $850\,\mu$m and
$450\,\mu$m. For the data reduction, the standard software SURF was
used with Mars as a flux calibrator. Table~\ref{DATA:SCUBA:NEW}
details the new $850\,\mu$m continuum measurements, while for
$450\,\mu$m, no significant signal was detected. Note that the new
$850\,\mu$m SCUBA data point for HR\,4049 is, within the errors, the
same as the one in \citet{03Dominik}.
\begin{table} [!h]
\centering
\caption{New observational data at $850\,\mu$m from
SCUBA.\label{DATA:SCUBA:NEW}}
\begin{tabular}{llr@{\,$\pm$\,}l}
\hline
\hline
No	&Name	&\multicolumn{2}{c}{$F_{850}$ (mJy)}\\
\hline
11	&HD\,52961	&2.8	&1.9\\

21	&HR\,4049	&8.7	&2.8\\

38	&89\,Her	&40.9	&2.4\\
\hline
\end{tabular}
\end{table}

For AC\,Her we have a flux point at $1.1$\,mm from \citet{95Shenton}.

\subsubsection*{Other}
At the short wavelength side, we use the IUE data (International
Ultraviolet Explorer, $0.115\,\mu$m - $0.320\,\mu$m) from the newly
extracted spectral data release (INES Archive Data Server). If there
were multiple spectra available, we only considered the maximal flux.

\section{Spectral Energy Distributions}
\label{section:SED}
We analyse the SEDs of our sample stars in a homogeneous and
systematic way.

\subsection{The stars photosphere: model parameters}
\label{subsection:parameters_model}
The determination of the model atmospherical parameters
$T_{\mathrm{eff}}$ and $\log g$, and the overall metallicity
[\element{Fe}/\element{H}] is based on the analysis of high resolution
spectra used in the chemical analyses of the stars. Those spectra are
preferentially taken at maximum light, because of the minimal
molecular veiling during the hotter phase in the light curve
\citep{00Giridhar}.

For most stars we use model parameters deduced from our own spectra
\citep[e.g][]{97VanWinckel, 03Maas, 05Maas}, for others we use values
found in the literature. For some of our sample stars (IRAS\,05208,
UY\,CMa, IRAS\,10456, IRAS\,18123, IRAS\,18158, V\,Vul), however, we
don't have a spectrum nor do we find any estimates for the
photospheric parameters in the literature. Here we deduce the
parameters on the basis of the spectral type of the star. And for some
others (IRAS\,09400, IRAS\,10174, GK\,Car, IRAS\,13258, IRAS\,15556)
we lack optical photometry. The model parameters are given in
Table~\ref{tab:coordinates}, Cols.~7 to 9. The references where we
found the parameters are shown in Col.~10.

An appropriate photospheric Kurucz model -based on $T_{\mathrm{eff}}$,
$\log g$ and [\element{Fe}/\element{H}]- is used for the unattenuated
stellar photospheres \citep{79Kurucz}.

\subsection{Colour excess $E(B-V)$ determination}
\label{subsection:colour_excess}
Light coming from a star is attenuated and reddened by material in the
line-of-sight. Note that the total line-of-sight extinction is likely
to contain both a circumstellar and an interstellar component.

We estimate the total colour excess $E(B-V)$ by using the average
interstellar extinction law given by \citet{79Savage} to deredden the
observed maximal UV-optical fluxes. $E(B-V)$ is found by minimizing
the difference between the dereddened observed fluxes in the
UV-optical, and the appropriate Kurucz model \citep{79Kurucz}. We
scale to the $J$ filter which is the reddest filter where no dust
excess can be expected. We assume the circumstellar reddening law to
be similar to the ISM law. Determining the $E(B-V)$ in this way
implies that we don't correct for the contribution of the grey
extinction.

\begin{figure}[!h]
\centering
\includegraphics[width=0.50\textwidth]{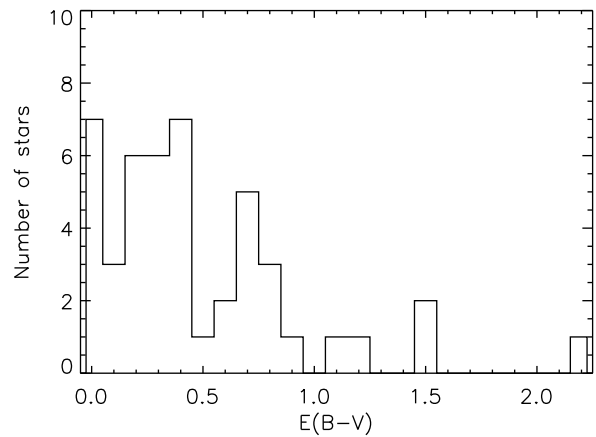}
\caption{The distribution of the total $E(B-V)$ found for our
sample. Remark that most colour excesses are less than $1.0$.}
\label{fig:ebminv}
\end{figure}

Results are in Table~\ref{tab:dustmodel}. The error on $E(B-V)$ is
typically $0.1$. But changing the stellar models by $\pm\,250$\,K in
effective temperature causes a change in $E(B-V)$ of about
$0.2$. Thus, together with the error of $0.2$ induced by the
uncertainty of the temperature of the underlying photosphere, we have
an uncertainty of $0.3$ on the total extinction during maximal
light. A distribution of the total colour excesses found for our
sample stars is shown in Fig.~\ref{fig:ebminv}. Note that the total
reddening is small for most of the stars. We may thus assume that a
situation with only grey extinction would be rather exceptional.

\subsection{A first-order approximation: an optically thin dust fit}
\label{subsection:fit}
A broad-band SED has limited diagnostic value for constraining the
chemistry and spatial distribution of the circumstellar dust. We apply
an optically thin dust model as a simple first-order approximation to
fit the SEDs with the prime goal to quantify the differences in SED
characteristics between the different objects. We use the optically
thin dust model described by \citet{85Sopka}. In this model the dust
is assumed to be distributed spherically symmetric around the star
from inner radius $r_{\mathrm{in}}$ to outer radius
$r_{\mathrm{out}}$. Since we apply an optically thin model, the
geometry of the dust is not constrained as we observe the whole volume
and all dust particles contribute to the observed flux.

For more details of the model we refer to \citet{05DeRuyter}. The flux
$F_{\nu}$ is determined by five parameters: the normalization
temperature $T_0$, the inner radius of the dust shell
$r_{\mathrm{in}}$, the outer radius of the dust shell
$r_{\mathrm{out}}$, the spectral index $p$ and the density parameter
$m$. Applying a least square minimization the free parameters
$r_{\mathrm{in}}$, $r_{\mathrm{out}}$, $p$ and $m$ are
determined. Results are given in Table~\ref{tab:dustmodel}.

\subsection{SEDs}
\label{subsection:SEDs}
\begin{figure*}[!t]
\centering
\includegraphics[width=0.49\textwidth]{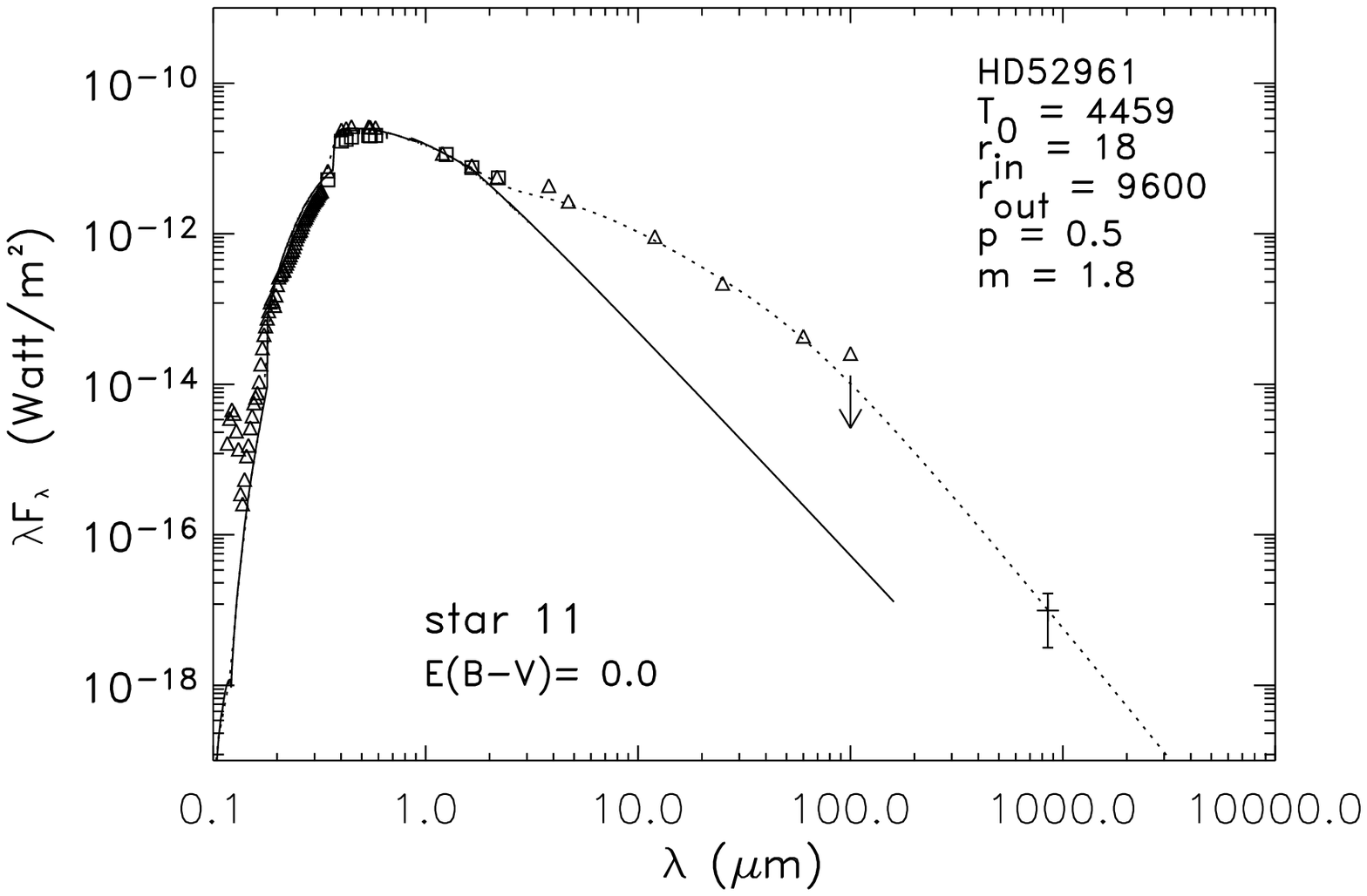}
\includegraphics[width=0.49\textwidth]{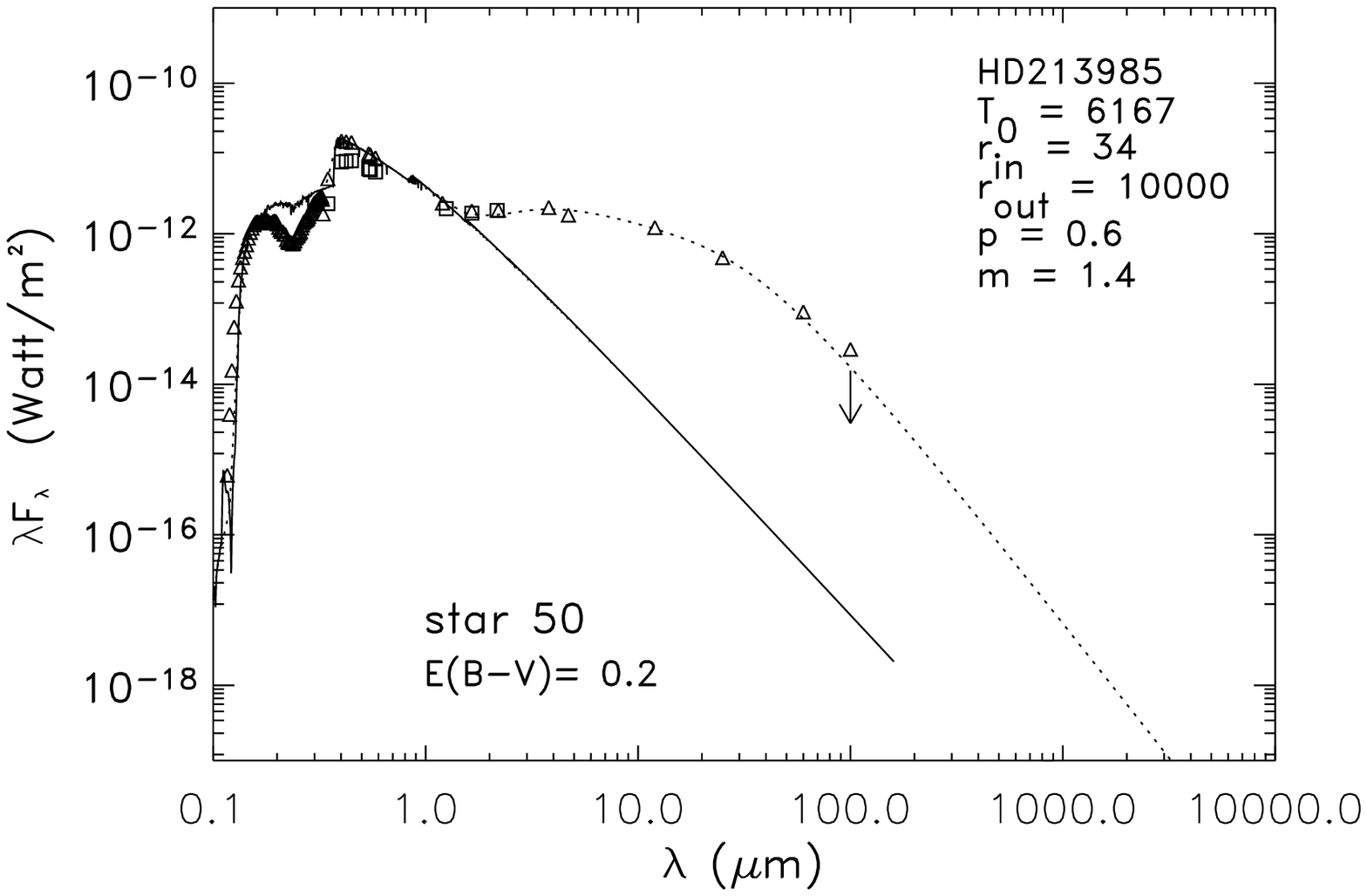}\\
\includegraphics[width=0.49\textwidth]{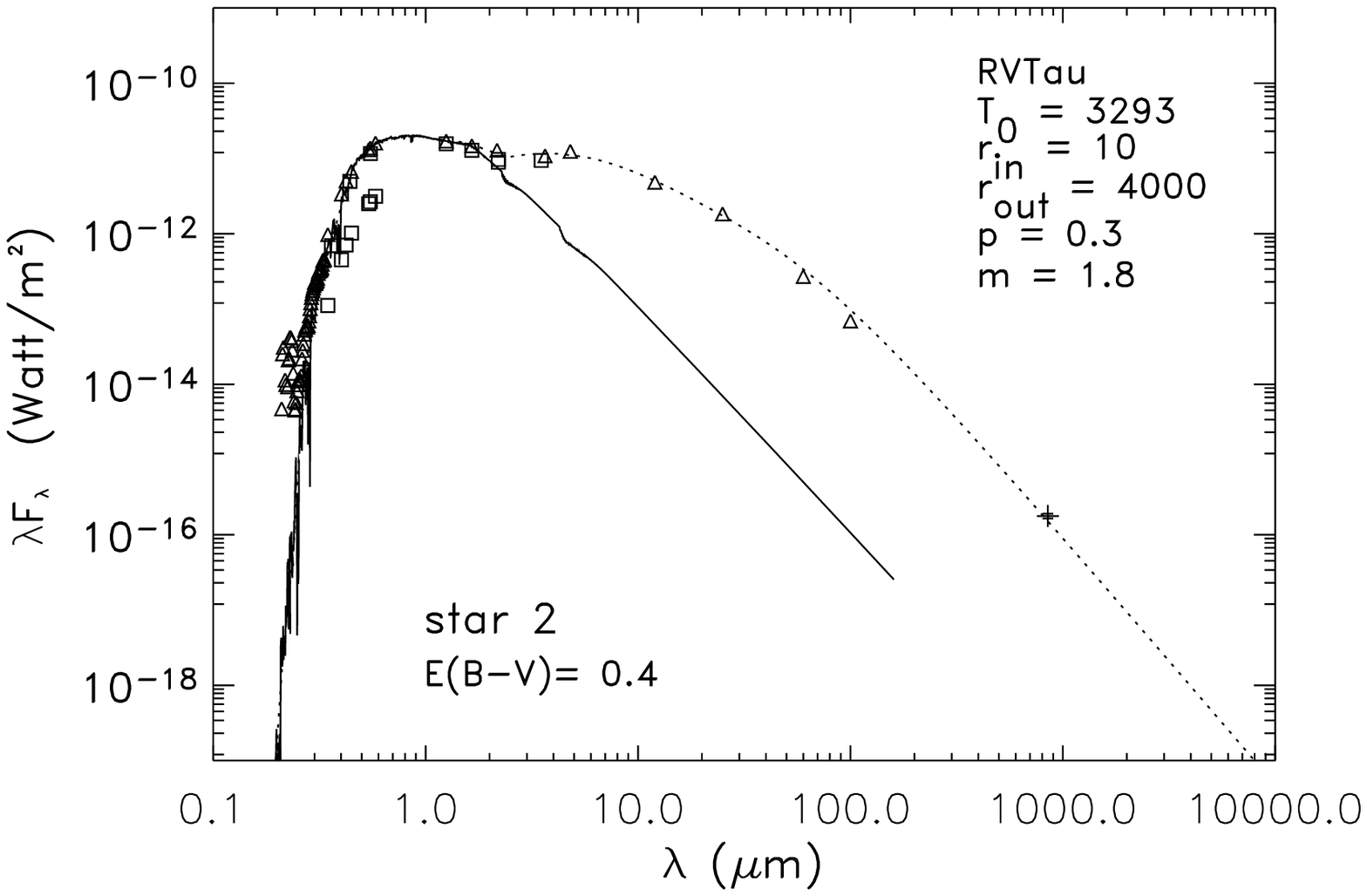}
\includegraphics[width=0.49\textwidth]{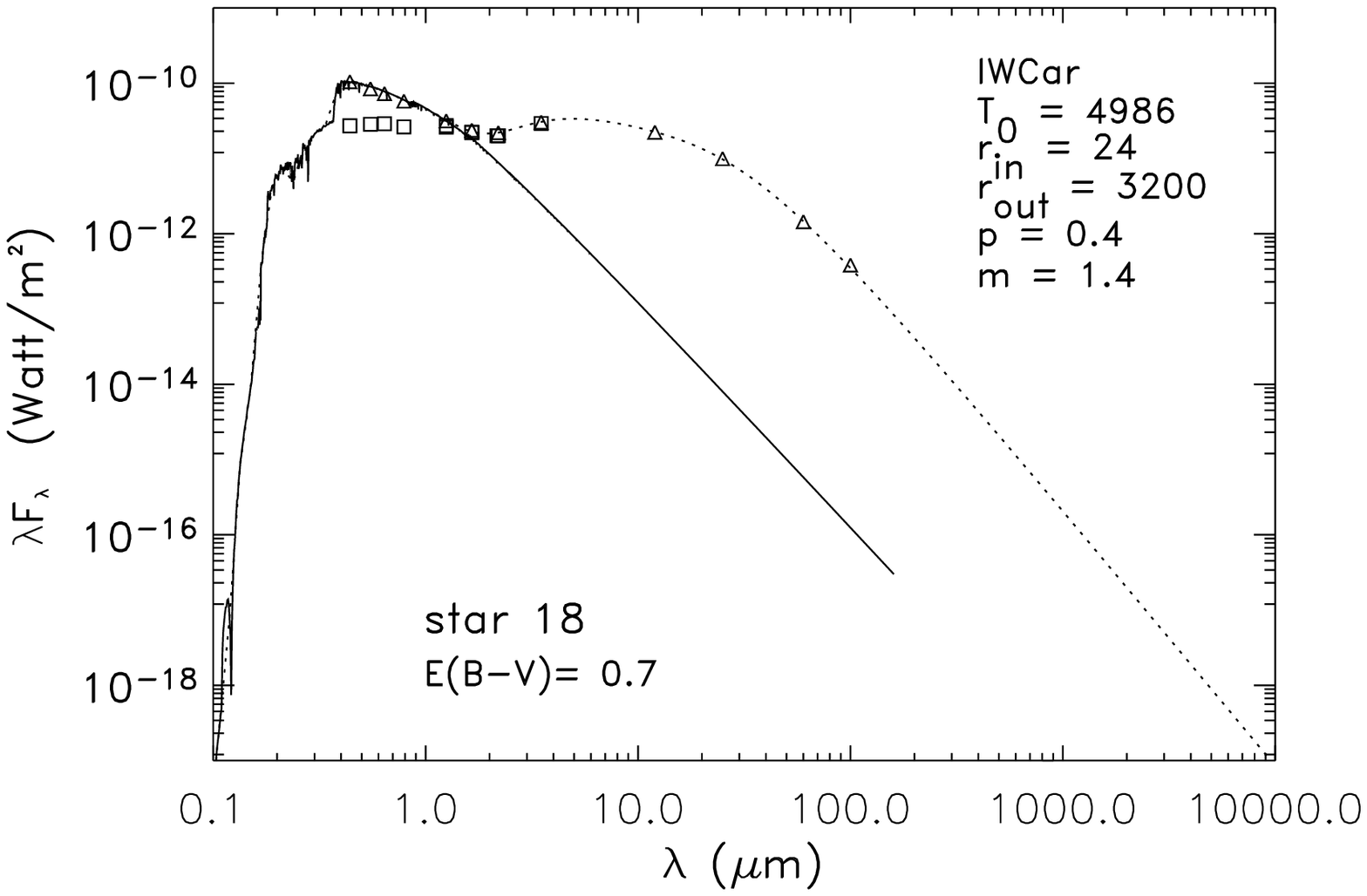}\\
\includegraphics[width=0.49\textwidth]{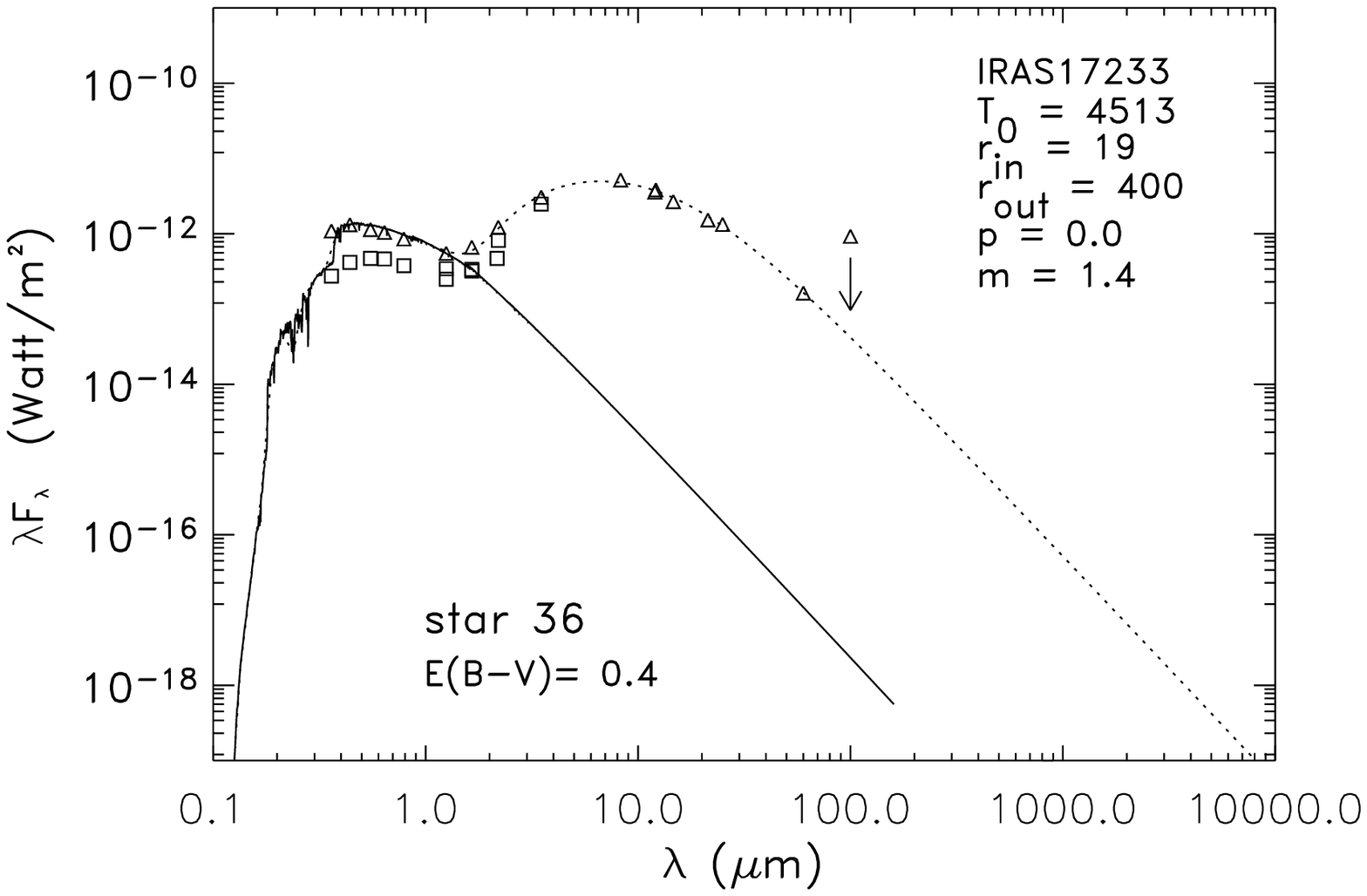}
\includegraphics[width=0.49\textwidth]{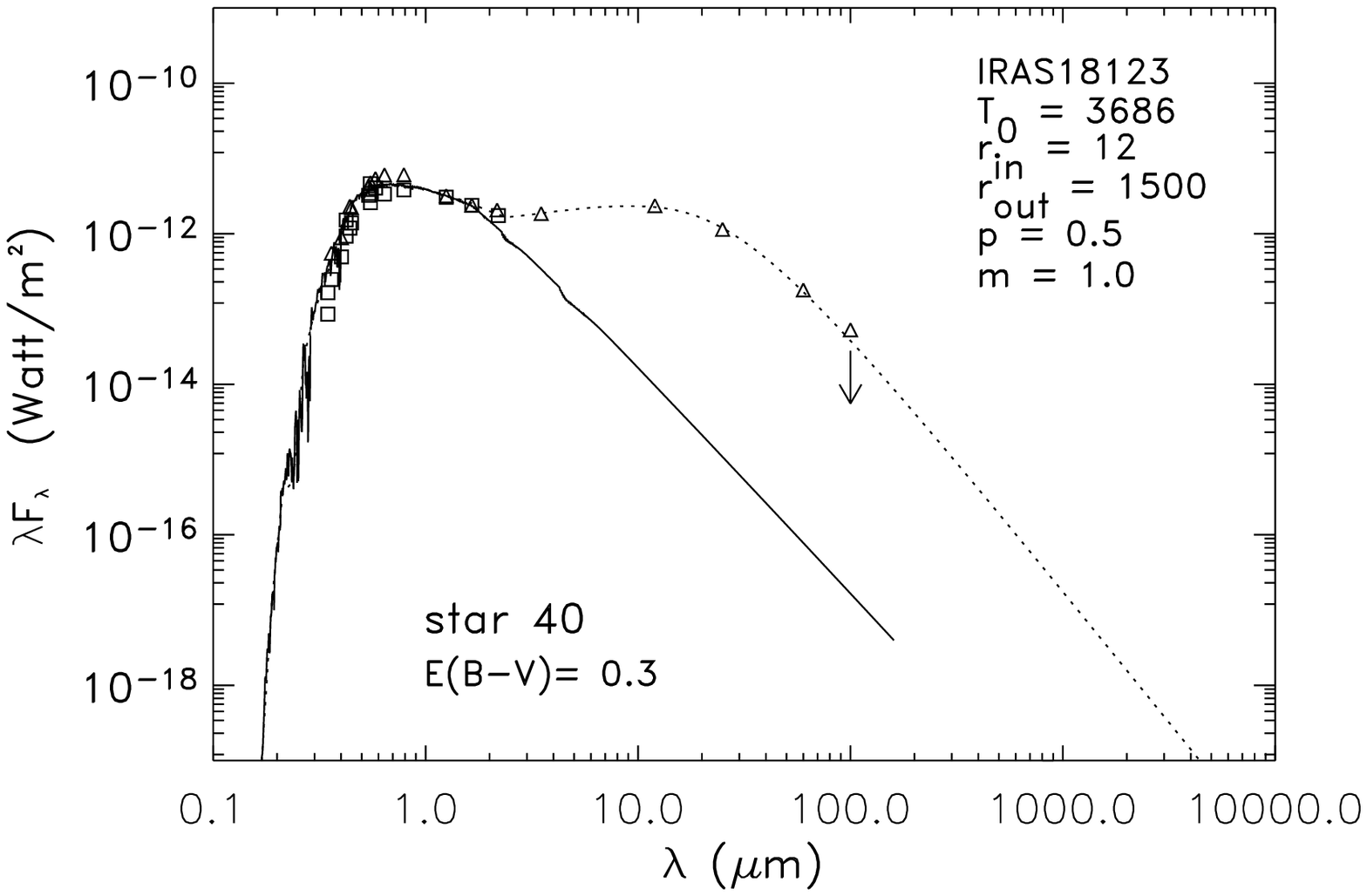}
\caption{The SEDs of a selection of our sample of post-AGB stars:
HD\,52961 and HD\,213985 are examples of confirmed binary post-AGB
stars, RV\,Tau and IW\,Car are genuine RV\,Tauri stars and IRAS\,17233
and IRAS\,18123 are RV\,Tauri like objects from the new list of
T.~Lloyd Evans. The dereddened fluxes are given together with the
scaled photospheric Kurucz model representing the unattenuated stellar
photosphere (solid line). An optically thin dust model was used to fit
the IR\,excess (dotted line). Data found in the literature together
with our 7 band Geneva photometry (only the maxima) are plotted as
triangles. The minimal data points (squares) were not used for the
determination of $E(B-V)$. They are shown to give an indication of the
amplitude of the pulsations. Where available, crosses represent our
$850\,\mu$m SCUBA data point. Error bars on the $850\,\mu$m SCUBA data
point are plotted as well, for some objects though smaller than the
symbols. The arrow at the $100\,\mu$m flux point signifies an upper
limit. \label{plots}}
\end{figure*}

In Fig~\ref{plots} the SEDs of a selection of post-AGB stars are
given. The other SEDs are given in Fig.~\ref{app:plots}. The
dereddened fluxes are plotted together with the scaled photospheric
model representing the unattenuated stellar photosphere. We plotted
both minimal (open triangles) and maximal (open rectangles)
photometry.

HD\,44179 and AR\,Pup are two rather exceptional stars, for which we
adopted another strategy to determine the SEDs. The dust excess for
HD\,44179 and AR\,Pup starts even at shorter wavelengths making it
impossible to scale the model photosphere to the $J$ filter. We
therefore scale to the Geneva $G$ band data point in our least squares
procedure.

The IR\,excess due to the presence of circumstellar dust, is in all
cases very significant with BD$+$39$^{\circ}$4926 a noticeable
exception for which no IR\,excess was found. A general characteristic
is that the dust energy distribution peaks at very high temperatures
and that there is no evidence for large amounts of cool
($T_{\mathrm{d}} \leq 100$\,K) dust: the peak of the dust SEDs lies
around $10\,\mu$m and in some cases even bluer. In most cases, the
dust excess starts near the dust sublimation temperature.

\begin{figure} [!h]
\centering
\includegraphics[width=0.50\textwidth]{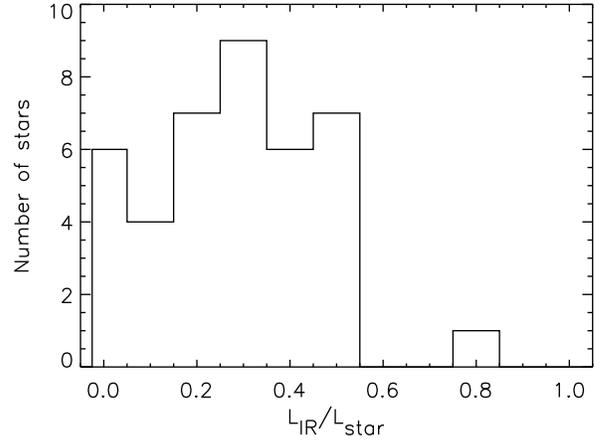}
\caption{The distribution of the energy conversion ratio
$L_{\mathrm{IR}}/L_\ast$ found for our sample.}
\label{fig:Lratio}
\end{figure}

To determine the amount of energy reprocessed by the dust grains in
the circumstellar environment, we compute the energy ratio
$L_{\mathrm{IR}}/L_\ast$. We first calculate the stellar flux $L_\ast$
by numerically integrating the scaled Kurucz model between $145\,$nm
and $850\,\mu$m. This gives a good estimate of the unattenuated
stellar flux. The energy radiated at even shorter and longer
wavelengths is only a negligible fraction ($\lesssim\,10^{-4}$) of the
total stellar flux so we evaluate our wavelength integral boundaries
as adequate. Integrating over the IR\,excess model described in
Section~\ref{subsection:fit} yields $L_{\mathrm{IR}}$. We note that
the ratio $L_{\mathrm{IR}}/L_\ast$ (Table~\ref{tab:dustmodel}) is high
for $78$\% of the objects (larger than $20$\%). Assuming the presence
of $20$\% grey or non-selective extinction and a mean $E(B-V)$ of
$0.5$, reduces $L_{\mathrm{IR}}/L_\ast$ with $25$\%, which remains,
nevertheless, still large. The absorption and thermal re-radiation of
the stellar radiation by the circumstellar dust is on average very
efficient. In Fig.~\ref{fig:Lratio} the distribution of the energy
conversion ratios is shown. Note that ratios larger than $1.0$ (for
HD\,44179, AR\,Pup, IRAS\,17233, IRAS\,18158 and IRAS\,20056) are
omitted in the figure. Like HD\,44179 \citep{04Cohen} and IRAS\,20056
\citep{88Menzies,02KameswaraRao} we suspect that we only see the
photosphere of those sources through scattered light. This means that
we see these systems nearly edge-on.

For those stars where we have a submillimetre data point (see
Table~\ref{DATA:SCUBA}) the SED follows a Rayleigh-Jeans slope from
the $60$-$100\,\mu$m flux point redwards. Assuming that the dust
emissivity at far-IR wavelengths follows a power law ($Q_{\nu} \sim
\nu^p$), the spectral index $p$ as determined from the slope between
the $60$-$100\,\mu$m emission and the $850\,\mu$m flux point is close
to zero. The small spectral indices in the range from $0.0$ to $0.5$
for most objects are consistent with the presence of a component of
large grains (radius $\gtrsim0.1$\,mm) in the circumstellar
environment of the stars. For those stars where we lack a
submillimetre data point, the spectral index of the long wavelength
slope is not well constrained.

Note that the $100\,\mu$m IRAS fluxes for some stars are affected by
Galactic cirrus, so they present upper limits. In the figures an arrow
is drawn. This makes it still more difficult to be sure about the
slope from the $60$-$100\,\mu$m flux point redwards.

From all SED characteristics mentioned, we infer that the dust excess
in the SEDs is clearly different from what we expect to observe in a
post-AGB star where the excess represents the expanding and cooling
relic of the strong AGB mass-loss episode(s).

\section{Distance estimates}
\label{section:D}
The typical luminosity of a lower mass post-AGB star is expected to be
between $1000$ and $10000\,L_\odot$. Accurate luminosities of
individual post-AGB objects are still largely unavailable since they
are generally too far to obtain reliable parallaxes.

In the few cases where reliable Hipparcos parallaxes are available we
have a direct probe of the distance and hence luminosity. Results are
given in Table~\ref{tab:distances}. Note the large errors on the
parallaxes translate in a large error box for the distance and the
luminosity determinations. The parallax of HD\,44179 can not be used
to infer a luminosity because we know that we see only scattered light
\citep{04Cohen}. The same is probably true for AR\,Pup (see
Sect.~\ref{subsection:SEDs}).

\citet{98Alcock} reported the discovery of RV\,Tauri stars in the
Large Magellanic Cloud (LMC). In light- and colour-curve behaviour,
the classical RV\,Tauri stars appeared to be a direct extension of the
type {\rm II} Cepheids to longer periods. We use the P-L relation of
\citet{98Alcock} -derived for the LMC RV\,Tauri stars- to derive the
luminosity for the pulsating objects in our sample. For variables with
\mbox{$P/2 > 12.6$} days we use
\begin{multline}
\label{eq: P-Lrelation}
M_V = 2.54 \left(\pm 0.48\right) - \left(3.91 \left(\pm
0.36\right)\right) \log\left(P/2\right),\\ \sigma = 0.35.
\end{multline}

Note that the intrinsic scatter is large, implying very significant
uncertainties in the estimated absolute magnitudes of the objects. In
Eq.~\ref{eq: P-Lrelation} we use the fundamental period $P/2$, defined
as the time between a deep and a shallow minimum. The formal period
$P$ is shown in Table~\ref{tab:distances}. We apply this extension of
the period-luminosity relation known for type {\rm II} Cepheids, for
the RV\,Tauri stars from the GCVS to derive an estimate for their
luminosities. Results, together with the bolometric corrections BC$_V$
\citep{98Bessell} used, are listed in Table~\ref{tab:distances}. The
error propagation starting from the errors in the P-L relation are
given in the table as well. The objects are indeed luminous, as can be
expected from post-AGB objects, but the large scatter in the P-L
relation prevents more accurate luminosity estimates.

For the other objects, where we don't have a pulsation period
determination, we take \mbox{$L = 5000\,\pm\,2000\,L_\odot$}. Note
that it is not known whether the P-L relation of the LMC stars is
directly applicable to Galactic stars. In general, the luminosities
found are smaller than, or equal to, about $5000\,L_\odot$ indicating
that the population is of rather low initial mass.

By comparing the integrated fluxes of the scaled Kurucz model with the
luminosity deduced from the P-L relation or the default assumed value
of \mbox{$L = 5000\,L_\odot$}, we calculate the distance $D$ to the
star. The determined distances are shown in
Table~\ref{tab:distances}. The propagated error of the poorly
calibrated P-L relation, yields very uncertain luminosities and
therefore also uncertain distances.

Remark that the luminosities determined by the P-L relation will not
be appropriate to stars seen with the dusty disc nearly
edge-on. Besides for the well-known example HD\,44179, this is likely
the case for AR\,Pup, IRAS\,17233, IRAS\,18158 and IRAS\,20056 for
which the distances listed in Table~\ref{tab:distances} are clearly
upper limits.

\section{Discussion}
\label{section:discussion}
We presented the SEDs of all stars in our sample based on literature
data supplemented with our own Geneva photometry and -where available-
continuum measurements at $850\,\mu$m (Fig.~\ref{plots} and
Fig.~\ref{app:plots}). Despite the different criteria for the
selection of the three subsamples, it is clear that the broad-band
SEDs display a high degree of uniformity. Between extremes like
HR\,4049 (most compact SED with a single dust temperature) and
HD\,44179 (largest $L_{\mathrm{IR}}/L_\ast$), the shape of the SEDs is
very similar for all stars.

\begin{figure}[!h]
\centering
\includegraphics[width=0.50\textwidth]{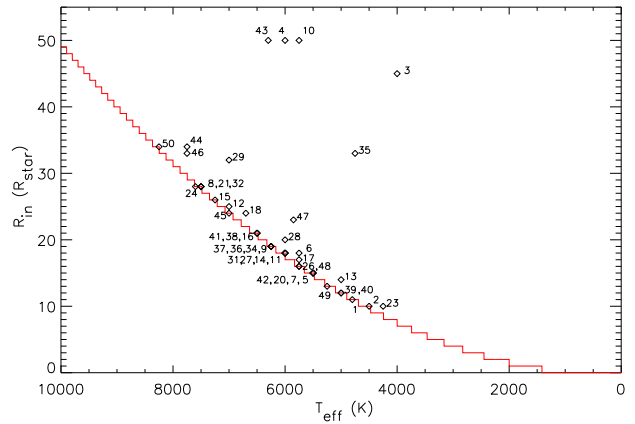}
\caption{The inner radii are very small. The dust must be located very
close to the star: the dust excess starts for nearly all objects near
the dust sublimation temperature. All SEDs show a clear near-IR
excess.}
\label{fig:R0-Teff}
\end{figure}

One of the most remarkable features is the start of the dust excess.
In Fig.~\ref{fig:R0-Teff} we plot the inner radii of the dust
components against the effective temperature for all stars and in all
cases, there is dust at or very near sublimation temperature. With
typical luminosity estimates, this sublimation temperature edge is at
a distance smaller than about \mbox{10\,AU} from the central
source. Moreover, the SEDs reveal that the presence of dust, very
close to the object, is irrespective of the effective temperature of
the central star. Note that none of the objects shows evidence for a
present-day dusty mass-loss. We therefore infer that part of the dust
must be gravitationally bound: any typical AGB outflow velocity would
bring the dust to cooler regions within years.

Within our sample of post-AGB objects considered - confirmed binary
post-AGB stars, classical RV\,Tauri stars and new RV\,Tauri-like
objects - there is a wide range in the strength of the total
IR\,excess, but the shape of the IR\,excess thus indicates that in all
systems, the circumstellar shell is not freely expanding but stored in
the system. We argue therefore that the same inner geometry as in the
resolved system HD\,44179 applies to the whole sample: the objects
seem to be surrounded by a Keplerian disc.

From the dust modelling fit it is also clear that the outer radii are
not very large either. Objects similar to the enigmatic extreme
HR\,4049 are UY\,CMa, U\,Mon and IRAS\,17233 for which the dust
temperature gradient observed is very small.

\noindent
Other indirect indications for the presence of gravitionally bound
dust are :
\begin{itemize}
\item[-] Despite the significant IR\,excess, the total $E(B-V)$ for
most stars is very small. Efficient redistribution of the stellar flux
is clearly in contradiction with the lack of line-of-sight
reddening. This implies a significant grey extinction and/or a
non-spherically symmetric dust distribution. Since the objects all
have a rather small total extinction, the latter is more plausible. A
good independent estimate of the ISM contribution to the total
extinction in the line-of-sight of all individual objects is still
lacking. It must, however, be a rewarding experiment to measure those
since it would help in quantifying the characteristics of the
broad-band energetics of the individual objects.

\item[-] At far-IR and millimetre wavelengths, we are sampling the
Rayleigh-Jeans tail of the flux distribution. The spectral index $p$
lies, for $60$\% of the sample stars, in the range $0.0$-$0.5$. The
IRAS-submillimetre flux distributions are thus indicative of emission
by large ($\gtrsim 0.1$\,mm) grains. The presence of such large grains
is another indication of discs, where grain growth is
facilitated. However, since we do not have submillimetre data points
for all stars at our disposal, this conclusion is uncertain for those
objects. Extension of the Spectral Energy Distributions (SEDs) to the
submillimetre region would be helpful in constraining the disc
characteristics and the grain size distribution.
\end{itemize}

Assuming the IR emission is indeed produced by an infinitely optically
thick disc, we can use the ratio $L_{\mathrm{IR}}/L_\ast$ to estimate
the opening angle $\theta$ of the disc as seen from the star. Suppose
the thick disc is at a distance $R$ from the star and at a height $H$
above the mid-plane (total thickness of the disc is $2H$). The total
inner surface of the disc is $2 \pi R \times 2H$. As
$L_{\mathrm{IR}}/L_\ast$ of the stellar energy should be absorbed by
the disc, its surface must cover a solid angle of
$L_{\mathrm{IR}}/L_\ast \times 4 \pi R^2$ \citep{03Dominik} or
\begin{equation}
\label{eq: height1}
\tan\frac{\theta}{2} = \frac{H}{R} = \frac{L_{\mathrm{IR}}}{L_\ast}.
\end{equation}
If the disc is not optically thick, the height $H$ would have to be
even larger. If grey extinction is important for the optical fluxes,
we underestimate the intrinsic stellar luminosity which has an
opposite effect on the scale height. For the objects with
$L_{\mathrm{IR}}/L_\ast$ approximately equal to $0.40$, this implies
$\theta \sim 44\,\degr$. Clearly, the scale height of such a disc must
be significant and this disc is likely to be gas-rich. A famous
example is HR\,4049 for which the energy emitted in the IR is due to
the presence of a circumbinary disc with a large height \citep[$H/R
\sim 1/3$, ][]{03Dominik}.

As the material around the evolved objects is not expanding, we prefer
to use the word `disc' or `gravitionally bound dust' instead of
torus. Dust tori are often resolved around Proto-Planetary Nebulae,
but these have very different SED characteristics since they are much
colder and are probably expanding \citep[e.g.][]{04Sanchez}. Moreover,
the physical sizes of the resolved tori are much larger than we expect
the circumstellar material to be in the evolved objects. Recently, the
very compact nature of the $N$ band flux of the dusty environments
around SX\,Cen and HD\,52961 was directly proven by \citet{05Deroo}
thanks to the Science Demonstration Time measurements of the $N$ band
interferometric instrument (MIDI) on the VLTI. Despite the baseline of
$45$\,m, SX\,Cen was not resolved which implies an upper limit of
$11$~mas (or $18$~AU at the estimated distance of the object) at
$10\,\mu$m. HD\,52961 was resolved but also here, the dust emission at
$10\,\mu$m comes from a very small angle of about $42$~mas ($\sim
60$~AU).

As argued in Sect.~\ref{section:SED} the shape of the SEDs clearly
differs from the SEDs of post-AGB stars for which dusty outflows are
detected. The large optical-to-infrared energy conversion ratios
indicate that the scale heights and opening angles of the discs are
large. The inner rims of the discs are probably puffed up by the
pressure of the hot gas in the disc.

\subsection*{Comparison with young stellar objects}
We note that the broad-band SED characteristics, are, at first glance,
very similar to the SEDs observed in young stellar objects (YSOs) in
which the circumstellar passive disc is a relic of the star formation
process. Despite this resemblance, differences in formation history of
the discs and in the luminosity of the central stars make it likely
that the characteristics will not be identical. Here we explain the
similarities and differences between the structure of the post-AGB
discs and the YSO discs.

Herbig Ae/Be (HAEBE) stars are the somewhat more massive analogues of
the T\,Tauri stars, which are low-mass young stellar objects. The
Spectral Energy Distributions (SEDs) of HAEBE stars are characterized
by the presence of a flux excess in the infrared due to circumstellar
dust and gas \citep{04aAcke} and the geometry of this circumstellar
matter is believed to be disc-like \citep[e.g.][]{97Mannings,
00Mannings, 03Fuente}. \citet{01Meeus} classified a sample of $14$
isolated HAEBE sample stars into two groups, based on the shape of the
SED. Group I contains sources in which a rising mid-IR
($20$-$200\,\mu$m) flux excess is observed; these sources have an SED
that can be fitted with a power law and a black-body continuum. Group
II sources have more modest mid-IR\,excesses; their IR SEDs can be
reconstructed by a power law only. \citet{01Meeus} suggested that the
difference between the two groups is related to the disc geometry. The
irradiated passive disc models developed by \citet{97Chiang} and later
extended by \citet{01Dullemond} show that the difference in SEDs of
both classes can be theoretically understood as indeed originating in
the disc geometry: the mid-IR\,excess of the group I sources forms an
indication of the flaring of the outer disc while in the group II
sources, the inner rim is such that it shadows the whole disc and no
flaring will occur.

\begin{figure} [!h]
\centering
\includegraphics[width=0.50\textwidth]{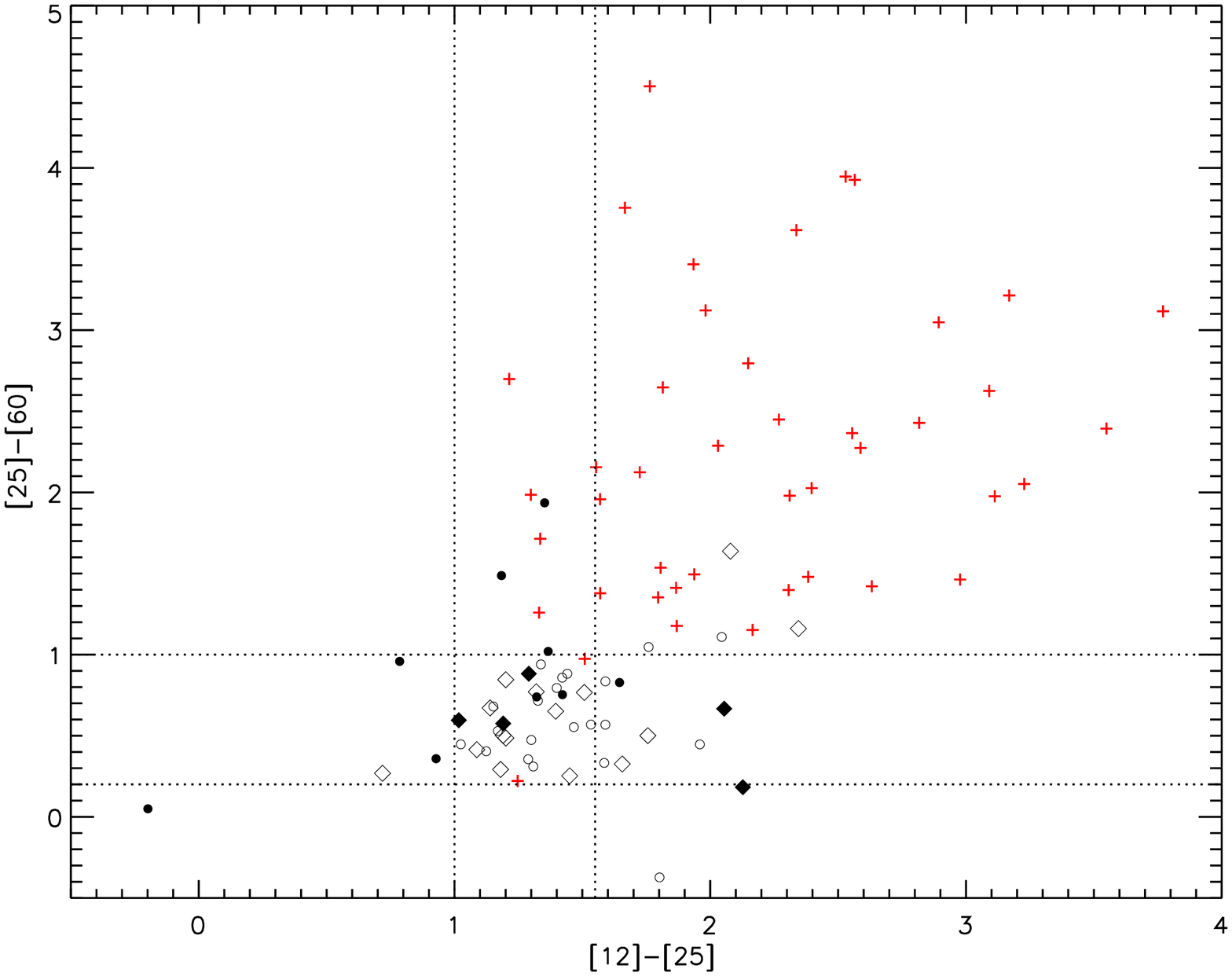}
\includegraphics[width=0.50\textwidth]{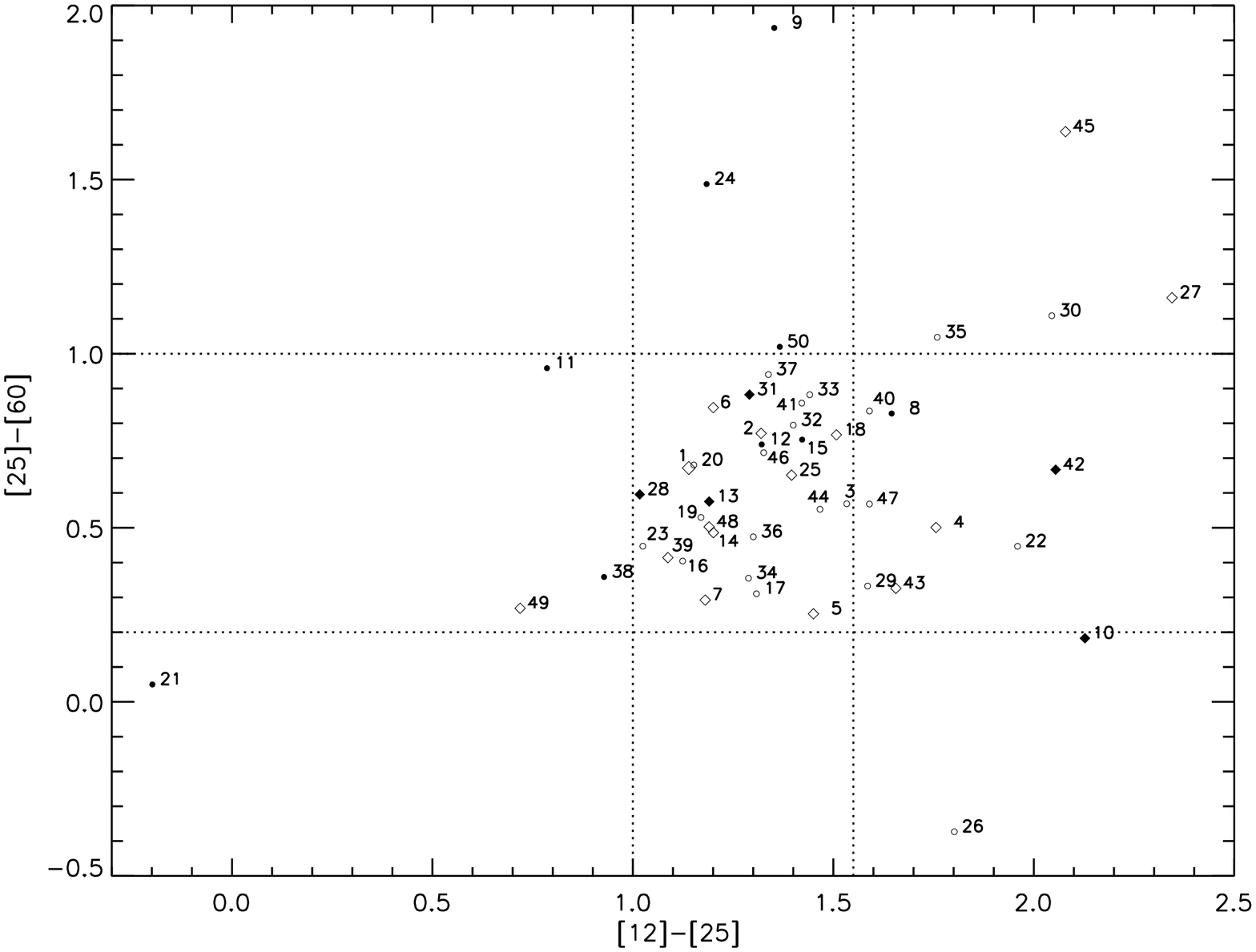}
\caption{\textbf{Upper-panel}: The IRAS colour-colour plot in which we
mark the `RV\,Tauri' box as defined in
Eq.~\ref{eq:colour-colour}. Note that the post-AGB objects we consider
(squares and circles, filled as well as not filled) are all situated
in or close to the `RV\,Tauri' box. The HAEBE objects (crosses) on the
other hand are redder and are not in the `RV\,Tauri'
box. \textbf{Lower-panel}: A closer view to the `RV\,Tauri' box. The
numbers of the post-AGB objects can be found in
Table~\ref{tab:coordinates}. The classical RV\,Tauri stars (squares)
as well as the post-AGB objects (circles) are mainly situated in the
box. Note that the filled symbols are objects for which we know the
orbital parameters.}
\label{fig:colour-colour}
\end{figure}

To gain more insight in the differences and similarities between the
stars of our sample and the young HAEBE stars, we plot the objects in
a colour-colour diagram (Fig.~\ref{fig:colour-colour}) with
\mbox{$\left[12\right] - \left[25\right]$} and \mbox{$\left[25\right]
- \left[60\right]$} as defined in Eq.~\ref{eq:colour-colour}. Nearly
all the post-AGB stars fall in the IR-box defined by T.~Lloyd Evans
(Sect.~\ref{subsection:TLEvans}), while the HAEBE stars are situated
on very different positions. The dust excesses of the post-AGB stars
are bluer with reduced mid- and far-IR colours compared to the HAEBE
stars. The post-AGB discs are therefore likely to be more compact.

\begin{figure} [!h]
\centering
\includegraphics[width=0.50\textwidth]{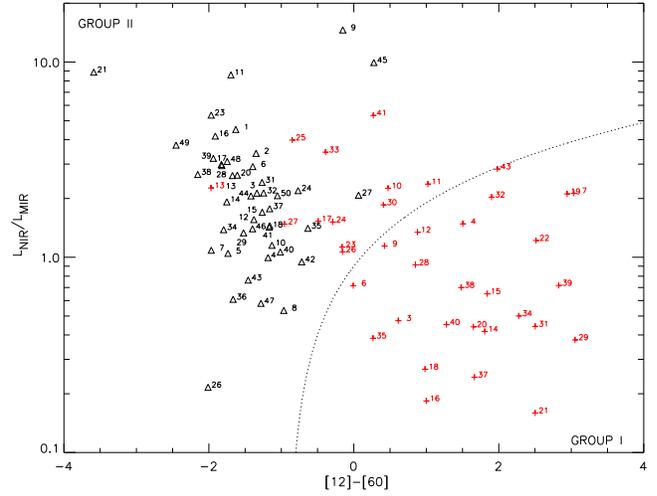}
\caption{Diagram based on \citet{04aAcke} and \citet{03VanBoekel}. The
ratio $L_{\mathrm{NIR}}/L_{\mathrm{MIR}}$ is plotted versus the
non-colour corrected IRAS $[12]-[60]$ colour. The dashed line
empirically separates the group II (left) and the group I sources
(right). Our sample stars are shown as triangles, the HAEBE stars as
crosses. The numbers of the post-AGB objects can be found in
Table~\ref{tab:coordinates}, those of the HAEBE objects in
\citet{04aAcke}.}
\label{fig:combi_verh}
\end{figure}

Another way of characterizing the differences between the young HAEBE
stars and the old post-AGB objects can be seen in
Fig.~\ref{fig:combi_verh}. The IR SED of the sample sources can be
characterized by two quantities: the ratio of the near-IR luminosity
$L_{\mathrm{NIR}}$ (derived from the broad-band $J$, $H$, $K$, $L$ and
$M$ photometry) and the mid-IR luminosity $L_{\mathrm{MIR}}$ (the
corresponding quantity based on IRAS $12$, $25$ and $60\,\mu$m
measurements), and the non-colour corrected IRAS $[12]-[60]$
colour. Both parameters compare the near-IR photometric data to the
mid-IR IRAS measurements \citep{03VanBoekel,03Dullemond}. The
luminosity ratio $L_{\mathrm{NIR}}/L_{\mathrm{MIR}}$ represents the
strength of the near-IR compared to the mid-IR\,excess, which is lower
for group I than group II sources.

Following \citet{03VanBoekel}, we use these quantities to distinguish
between group I and II in the classification of \citet{01Meeus}. We
plotted all our sample stars together with the HAEBE sample stars of
\citet{04aAcke} in Fig.~\ref{fig:combi_verh}. The dashed line in the
diagram represents $L_{\mathrm{NIR}}/L_{\mathrm{MIR}} =
([12]-[60])+0.9$, which empirically provides the best separation
between the two groups. The post-AGB stars are definitely similar to
group II sources and the HAEBE group I objects are significantly
redder than the post-AGB objects. Note that the IR\,excess of our
stars starts later (near $K$) than that of the HAEBE stars. Given the
fact that -in contrast with the HAEBE stars- for the post-AGB objects
the $JHKLM$ photometry is thus influenced by the stellar photosphere,
we dereddened the $JHKLM$ data points to make sure we don't take into
account the effects of the star. To make the comparison between the
HAEBE stars and the post-AGB stars more reliable, we dereddened
likewise also the $JHKLM$ data of the young objects. We conclude that
the dust excess around our sample stars do indeed resemble the group
II HAEBE sources and no indication of disc flaring is found.

The shape of the SEDs point for both the HAEBE and the post-AGB
objects to a disc. Evolution, disc formation and the central stars'
features, however, lead to different characteristics of the discs
around both types of objects. The discs around post-AGB stars are
smaller (Fig.~\ref{fig:colour-colour}) and there is little evidence
for flared discs (Fig.~\ref{fig:combi_verh}).

\section{Conclusions}
\label{section:conclusions}
A homogeneous and systematic study of the Spectral Energy
Distributions (SEDs) of a sample of 51 post-AGB objects was
presented. The selection criteria to define the sample were tuned to
cover the broad-band characteristics of known binary post-AGB
stars. Our whole sample included 20 dusty RV\,Tauri stars from the
General Catalogue of Variable Stars.

We conclude that the broad-band SED characteristics of the sample are
best interpreted, assuming the circumstellar dust is stored in
Keplerian rotating passive discs. First-order estimates indicate that
the discs are small in the radial direction and, given the large
distances of these stars, they are likely only resolvable with
interferometric observations.

The actual structure of those discs, let alone their formation,
stability and evolution are not well understood. In
Table~\ref{tab:Porbit}, we list the orbital elements of $15$ objects
of our total sample for which the orbital elements are covered in the
literature. These systems are now not in contact (assuming typical
post-AGB luminosities), but the orbits are all too small to accomodate
a full grown AGB star. It is clear that those systems did not evolve
on single star evolutionary tracks. They must have suffered a phase of
strong interaction while the primary was at giant dimension. It is
well known that those orbital elements are not very well understood
yet, since most objects display a significant eccentricity as well
\citep[][and references therein]{03VanWinckel}. One of the models
argues that it is the feedback of the disc on the orbital dynamics,
which will induce a large eccentricity \citep{91Artymowicz}.

We postulate that the same formation mechanism could apply to the
whole sample, which means that the objects must all be
binaries. Direct detection of orbital motion and, even better,
determination of the orbital elements are clearly needed but poses a
non-trivial observational challenge and certainly requires a long-term
project.

As our sample contains a significant fraction of all known post-AGB
stars, this would lead to the conclusion that binarity and binary
interaction are a widespread phenomenon among \mbox{(post-)} AGB
stars. Dusty RV\,Tauri stars are those evolved binaries which happen
to be in the Population\,{\rm II}\,Cepheid instability strip
\citep{99LloydEvans}.

Since the orbital periods detected till today, span already a wide
range, from $115$ to about $2600$ days, we conclude that the formation
of a stable orbiting disc is common and appears more and more to be a
key ingredient in understanding the final evolution of a very
significant fraction of binary post-AGB stars.

\begin{acknowledgements}
The staff and the service observers of the Mercator Observatory at La
Palma are acknowledged for the photometric Geneva data.

This publication makes use of data products from the Two Micron All
Sky Survey, which is a joint project of the University of
Massachusetts and the Infrared Processing and Analysis
Center/California Institute of Technology, funded by the National
Aeronautics and Space Administration and the National Science
Foundation.

We also made use of data from the DENIS project. This is partly funded
by the European Commission through SCIENCE and Human Capital and
Mobility grants. It is also supported in France by INSU, the Education
Ministry and CNRS, in Germany by the Land of Baden-W\"urtenberg, in
Spain by DGICYT, in Italy by CNR, in Austria by the Fonds zur
F\"orderung der Wissenschaftlichen Forschung and Bundesministerium
f\"ur Wissenschaft und Forschung.

This publication was also based on INES (IUE Newly Extracted Spectra)
data from the IUE (International Ultraviolet Explorer) satellite.

We also like to thank S.~Hony for his contribution to this paper.
\end{acknowledgements}

\bibliographystyle{aa}
\bibliography{4062ref.bib}

\clearpage

\onecolumn
\appendix
\section{Results}
\label{app:results}
\vspace{0.5cm}
\tablecaption{First the estimated total colour excess $E(B-V)$ is
given. Then the five parameters of the dust model are shown: the
normalization temperature $T_0$, the inner radius of the dust shell
$r_{\mathrm{in}}$, the outer radius of the dust shell
$r_{\mathrm{out}}$, the spectral index $p$ and the density parameter
$m$. The last column lists the energy ratio
$L_{\mathrm{IR}}/L_\ast$.\label{tab:dustmodel}}
\begin{center}
\tablehead{
\hline
\hline
No	&Name			&\multicolumn{2}{c|}{$E(B-V)$}	&$T_0$	&$r_{\mathrm{in}}$	&\multicolumn{2}{c}{$r_{\mathrm{in}}$} &$r_{\mathrm{out}}$
&\multicolumn{2}{c}{$r_{\mathrm{out}}$}	&$p$	&$m$	&\multicolumn{2}{c}{$L_{\mathrm{IR}}/L_\ast$}\\
        &			&\multicolumn{2}{c|}{}		&(K)	&($R_\ast$)		&\multicolumn{2}{c}{(AU)}		&($R_\ast$)
&\multicolumn{2}{c}{(AU)}		&	&	&\multicolumn{2}{c}{}\\
\hline}
\tabletail{\hline}
\begin{supertabular}{ll|r@{\,$\pm$\,}l|ccr@{\,$\pm$\,}lcr@{\,$\pm$\,}lcc|r@{\,$\pm$\,}l}
1	&TW\,Cam		&0.4	&0.3			&3489	&11			&4.5	&1.7				&5200
&2100	&800				&0.4	&1.8	&$0.38$		&$0.05$\\

2	&RV\,Tau		&0.4	&0.3			&3293	&10			&4.7	&1.7				&4000
&1900	&700				&0.3	&1.8	&$0.51$		&$0.07$\\

3	&IRAS\,05208		&0.0	&0.3			&2952	&45			&30.9	&7.3				&1200
&800	&200				&0.6	&1.4	&$0.06$		&$0.01$\\

4	&DY\,Ori		&1.1	&0.3			&4437	&50			&8.5	&2.9				&1800
&300	&100				&0.4	&1.2	&$0.12$		&$0.02$\\

5	&CT\,Ori		&0.2	&0.3			&4161	&15			&5.9	&2.3				&3600
&1400	&600				&0.9	&0.8	&$0.22$		&$0.03$\\

6	&SU\,Gem		&0.8	&0.3			&4256	&18			&2.9	&1.0				&10000
&1600	&500				&0.5	&1.8	&$0.81$		&$0.11$\\

7	&UY\,CMa		&0.0	&0.3			&4008	&15			&5.2	&2.0				&400
&100	&100				&0.2	&1.0	&$0.37$		&$0.05$\\

8	&HD\,44179		&0.0	&0.3			&5644	&28			&5.5	&1.2				&9100
&1800	&400				&0.8	&1.0	&$14.89$	&$2.11$\\

9	&HD\,46703		&0.1	&0.3			&4569	&19			&5.3	&1.1				&10000
&2800	&600				&0.2	&1.4	&$0.01$		&$0.00$\\

10	&ST\,Pup		&0.0	&0.3			&4307	&50			&3.8	&1.1				&3300
&200	&100				&0.7	&1.2	&$0.19$		&$0.03$\\

11	&HD\,52961		&0.0	&0.3			&4459	&18			&5.5	&1.2				&9600
&2900	&600				&0.5	&1.8	&$0.11$		&$0.02$\\

12	&SAO\,173329		&0.4	&0.3			&5215	&25			&5.6	&1.2				&4400
&1000	&200				&0.6	&1.2	&$0.24$		&$0.03$\\

13	&U\,Mon			&0.0	&0.3			&3625	&14			&5.3	&2.0				&400
&200	&100				&0.1	&1.2	&$0.12$		&$0.02$\\

14	&AR\,Pup		&0.6	&0.3			&4534	&18			&3.6	&1.3				&6300
&1300	&400				&0.9	&1.4	&$4.04$		&$0.57$\\

15	&IRAS\,08544		&1.5	&0.3			&5434	&26			&5.4	&1.1				&8300
&1700	&400				&0.8	&1.2	&$0.31$		&$0.04$\\

16	&IRAS\,09060		&0.8	&0.3			&4878	&21			&5.5	&1.2				&8900
&2300	&500				&0.8	&1.6	&$0.43$		&$0.06$\\

17	&IRAS\,09144		&2.2	&0.3			&4307	&17			&5.6	&1.2				&4600
&1500	&300				&0.8	&1.4	&$0.44$		&$0.06$\\

18	&IW\,Car		&0.7	&0.3			&4986	&24			&3.5	&1.2				&3200
&500	&200				&0.4	&1.4	&$0.41$		&$0.06$\\

19	&IRAS\,09400		&\multicolumn{2}{c|}{}		&	&			&\multicolumn{2}{c}{}	                &
&\multicolumn{2}{c}{}			&	&	&\multicolumn{2}{c}{}\\

20	&IRAS\,09538		&0.4	&0.3			&4032	&15			&5.4	&1.2				&2700
&1000	&200				&0.3	&1.8	&$0.53$		&$0.07$\\

21	&HR\,4049		&0.2	&0.3			&5541	&28			&5.5	&1.2				&100
&20	&4				&0.4	&1.8	&$0.31$		&$0.04$\\

22	&IRAS\,10174		&\multicolumn{2}{c|}{}		&	&			&\multicolumn{2}{c}{}			&
&\multicolumn{2}{c}{}			&	&	&\multicolumn{2}{c}{}\\

23	&IRAS\,10456		&0.3	&0.3			&3149	&10			&6.1	&1.4				&1800
&1100	&300				&0.7	&1.6	&$0.36$		&$0.05$\\

24	&HD\,95767		&0.7	&0.3			&5519	&28			&5.3	&1.1				&10000
&1900	&400				&0.4	&1.6	&$0.29$		&$0.04$\\

25	&GK\,Car		&\multicolumn{2}{c|}{}		&	&			&\multicolumn{2}{c}{}			&
&\multicolumn{2}{c}{}			&	&	&\multicolumn{2}{c}{}\\

26	&IRAS\,11472		&0.4	&0.3			&4357	&16			&5.3	&1.2				&1800
&600	&100				&0.9	&0.0	&$0.41$		&$0.06$\\

27	&RU\,Cen		&0.3	&0.3			&4529	&18			&3.2	&1.1				&7900
&1400	&500				&0.8	&0.6	&$0.03$		&$0.00$\\

28	&SX\,Cen		&0.3	&0.3			&4452	&20			&2.1	&0.7				&3000
&300	&100				&0.5	&1.6	&$0.29$		&$0.04$\\

29	&HD\,108015		&0.2	&0.3			&5248	&32			&7.2	&1.5				&7500
&1700	&400				&0.9	&1.2	&$0.57$		&$0.08$\\

30	&IRAS\,13258		&\multicolumn{2}{c|}{}		&	&			&\multicolumn{2}{c}{}			&
&\multicolumn{2}{c}{}			&	&	&\multicolumn{2}{c}{}\\

31	&EN\,TrA		&0.3	&0.3			&4424	&18			&2.1	&0.7				&4000
&500	&100				&0.4	&1.6	&$0.38$		&$0.05$\\

32	&IRAS\,15469		&1.5	&0.3			&5499	&28			&5.5	&1.2				&8500
&1700	&300				&0.3	&1.8	&$0.58$		&$0.08$\\

33	&IRAS\,15556		&\multicolumn{2}{c|}{}		&	&			&\multicolumn{2}{c}{}			&
&\multicolumn{2}{c}{}			&	&	&\multicolumn{2}{c}{}\\

34	&IRAS\,16230		&0.7	&0.3			&4651	&19			&5.3	&1.1				&1800
&500	&100				&0.6	&1.0	&$0.26$		&$0.04$\\

35	&IRAS\,17038		&0.2	&0.3			&3492	&33			&16.0	&3.6				&10000
&4900	&1100				&0.4	&1.8	&$0.28$		&$0.04$\\

36	&IRAS\,17233		&0.4	&0.3			&4513	&19			&5.3	&1.1				&400
&100	&0				&0.0	&1.4	&$3.14$		&$0.44$\\

37	&IRAS\,17243		&0.7	&0.3			&4581	&19			&5.3	&1.1				&3800
&1100	&200				&0.3	&1.6	&$0.50$		&$0.07$\\

38	&89\,Her		&0.0	&0.3			&4785	&21			&3.3	&1.1		                &2400
&400	&100				&0.4	&1.8	&$0.58$		&$0.08$\\

39	&AI\,Sco		&0.3	&0.3			&3657	&12			&3.7	&1.3				&1000
&300	&100				&0.3	&1.8	&$0.50$		&$0.07$\\

40	&IRAS\,18123		&0.3	&0.3			&3686	&12			&5.3	&1.2				&1500
&700	&100				&0.5	&1.0	&$0.39$		&$0.06$\\

41	&IRAS\,18158		&0.9	&0.3			&4824	&21			&5.5	&1.2				&7000
&1800	&400				&0.6	&1.4	&$1.04$		&$0.15$\\

42	&AC\,Her		&0.1	&0.3			&4036	&15			&3.7	&1.3				&900
&200	&100				&0.3	&0.2	&$0.04$		&$0.01$\\

43	&AD\,Aql		&0.6	&0.3			&4691	&50			&8.0	&2.8				&2000
&300	&100				&0.7	&0.0	&$0.03$		&$0.00$\\

44	&IRAS\,19125		&1.2	&0.3			&5748	&34			&6.2	&1.3				&7700
&1400	&300				&0.5	&1.6	&$0.45$		&$0.06$\\

45	&EP\,Lyr		&0.5	&0.3			&5143	&24			&3.7	&1.3				&10000
&1500	&500				&0.3	&1.2	&$0.01$		&$0.00$\\

46	&IRAS\,19157		&0.8	&0.3			&5740	&33			&6.0	&1.3				&5600
&1000	&200				&0.5	&1.4	&$0.44$		&$0.06$\\

47	&IRAS\,20056		&0.7	&0.3			&4216	&23			&7.4	&1.6				&1400
&400	&100				&0.3	&1.4	&$3.16$		&$0.45$\\

48	&R\,Sge			&0.4	&0.3			&4236	&16			&3.4	&1.2				&1800
&400	&100				&0.4	&1.6	&$0.31$		&$0.04$\\

49	&V\,Vul			&0.1	&0.3			&3902	&13			&3.7	&1.3				&1400
&400	&100				&0.7	&1.4	&$0.31$		&$0.04$\\

50	&HD\,213985		&0.2	&0.3			&6167	&34			&5.5	&1.1				&10000
&1600	&300				&0.6	&1.4	&$0.24$		&$0.03$\\

51	&BD$+$39$^{\circ}$4926	&0.2	&0.3			&	&			&\multicolumn{2}{c}{}			&
&\multicolumn{2}{c}{}			&	&	&\multicolumn{2}{c}{}\\
\end{supertabular}
\end{center}

\vspace{1cm}
\tablecaption{For those stars for which we have a parallax obtained by
Hipparcos, we derive the distance $D$ (pc) and the luminosity $L$
($10^3\,$L$_{\sun}$) based on that parallax. In the second part of the
table, distance and luminosity estimates are given based on the P-L
relation of \citet{98Alcock}. Periods $P$ (days) are from SIMBAD and
bolometric corrections BC$_V$ from \citet{98Bessell}. The luminosity
$L$ (in $10^3\,$L$_{\sun}$) and the distance $D$ (in kpc) are given
together with the estimated errors. Where we don't have a pulsation
period, distances are estimated based on a luminosity of $L =
5000\pm2000\,L_{\sun}$.\label{tab:distances}}
\begin{center}
\tablehead{
\hline
\hline
No	&Name			&\multicolumn{2}{c}{Parallax}	&Distance $D$		&$L$
&Period $P$	&BC$_V$		&\multicolumn{2}{c}{$L$}			&\multicolumn{2}{c}{Distance $D$}\\
	&			&\multicolumn{2}{c}{(mas)}	&(pc)			&($10^3\,$L$_{\sun}$)
&(days)		&		&\multicolumn{2}{c}{($10^3\,$L$_{\sun}$)}	&\multicolumn{2}{c}{(kpc)}\\
\hline}
\tabletail{\hline}
\begin{supertabular}{ll|r@{\,$\pm$\,}lcc|crr@{\,$\pm$\,}lr@{\,$\pm$\,}l}
1	&TW\,Cam		&\multicolumn{2}{c}{}		&			&
&$85.6$		&$-0.36$	&$3.8$		&$2.6$				&$2.8$		&$1.0$\\

2	&RV\,Tau		&\multicolumn{2}{c}{}		&			&
&$78.698$	&$-0.51$	&$3.8$		&$2.6$				&$2.1$		&$0.7$\\

3	&IRAS\,05208		&\multicolumn{2}{c}{}		&			&
&		&$-0.98$	&\multicolumn{2}{c}{}				&$4.1$		&$0.8$\\

4	&DY\,Ori		&\multicolumn{2}{c}{}		&			&
&$60.26$	&$0.00$		&$1.6$		&$1.0$				&$1.9$		&$0.6$\\

5	&CT\,Ori		&\multicolumn{2}{c}{}		&			&
&$135.52$	&$-0.08$	&$6.0$		&$4.5$				&$6.6$		&$2.5$\\

6	&SU\,Gem		&\multicolumn{2}{c}{}		&			&
&$50.12$	&$-0.03$	&$1.2$		&$0.8$				&$2.2$		&$0.7$\\

7	&UY\,CMa		&\multicolumn{2}{c}{}		&			&
&$113.9$	&$-0.08$	&$4.5$		&$3.3$				&$8.5$		&$3.1$\\

8	&HD\,44179		&$2.62$		&$2.37$		&$200$-$382$-$4000$	&$0.01$-$0.03$-$3.81$
&		&$0.11$		&\multicolumn{2}{c}{}				&\multicolumn{2}{c}{$< 4.6$}\\

9	&HD\,46703		&\multicolumn{2}{c}{}		&			&
&		&$0.06$		&\multicolumn{2}{c}{}				&$4.3$		&$0.9$\\

10	&ST\,Pup		&\multicolumn{2}{c}{}		&			&
&$18.8864$	&$-0.01$	&$0.3$		&$0.1$				&$1.3$		&$0.4$\\

11	&HD\,52961		&\multicolumn{2}{c}{}		&			&
&		&$0.04$		&\multicolumn{2}{c}{}				&$2.3$		&$0.5$\\

12	&SAO\,173329		&\multicolumn{2}{c}{}		&			&
&		&$0.11$		&\multicolumn{2}{c}{}				&$7.0$		&$1.5$\\

13	&U\,Mon			&$1.45$		&$0.82$		&$441$-$690$-$1587$	&$1.21$-$2.96$-$15.65$
&$92.26$	&$-0.24$	&$3.8$		&$2.7$				&$0.8$		&$0.3$\\

14	&AR\,Pup		&$2.75$		&$1.16$		&$256$-$364$-$629$	&$0.03$-$0.06$-$0.17$
&$75$		&$0.00$		&$2.2$		&$1.5$				&\multicolumn{2}{c}{$< 1.8$}\\

15	&IRAS\,08544		&\multicolumn{2}{c}{}		&			&
&		&$0.11$		&\multicolumn{2}{c}{}				&$0.8$		&$0.2$\\

16	&IRAS\,09060		&\multicolumn{2}{c}{}		&			&
&		&$0.08$		&\multicolumn{2}{c}{}				&$4.4$		&$0.9$\\

17	&IRAS\,09144		&\multicolumn{2}{c}{}		&			&
&		&$-0.01$	&\multicolumn{2}{c}{}				&$2.7$		&$0.6$\\

18	&IW\,Car		&\multicolumn{2}{c}{}		&			&
&$67.5$		&$0.08$		&$1.7$		&$1.2$				&$0.7$		&$0.2$\\

19	&IRAS\,09400		&\multicolumn{2}{c}{}		&			&
&		&		&\multicolumn{2}{c}{}				&\multicolumn{2}{c}{}\\

20	&IRAS\,09538		&\multicolumn{2}{c}{}		&			&
&		&$-0.08$	&\multicolumn{2}{c}{}				&$7.7$		&$1.6$\\

21	&HR\,4049		&$1.50$		&$0.63$		&$469$-$667$-$1149$	&$1.67$-$3.39$-$10.05$
&		&$0.11$		&\multicolumn{2}{c}{}				&$0.8$		&$0.2$\\

22	&IRAS\,10174		&\multicolumn{2}{c}{}		&			&
&		&		&\multicolumn{2}{c}{}				&\multicolumn{2}{c}{}\\

23	&IRAS\,10456		&$1.18$		&$0.60$		&$562$-$847$-$1724$	&$3.31$-$7.52$-$31.15$
&		&$-0.70$	&\multicolumn{2}{c}{}				&$0.7$		&$0.1$\\

24	&HD\,95767		&\multicolumn{2}{c}{}		&			&
&		&$0.10$		&\multicolumn{2}{c}{}				&$2.0$		&$0.4$\\

25	&GK\,Car		&\multicolumn{2}{c}{}		&			&
&$55.6$		&		&\multicolumn{2}{c}{}				&\multicolumn{2}{c}{}\\

26	&IRAS\,11472		&\multicolumn{2}{c}{}		&			&
&		&$-0.03$	&\multicolumn{2}{c}{}				&$9.7$		&$2.0$\\

27	&RU\,Cen		&\multicolumn{2}{c}{}		&			&
&$64.727$	&$0.00$		&$1.7$		&$1.2$				&$1.6$		&$0.6$\\

28	&SX\,Cen		&\multicolumn{2}{c}{}		&			&
&$32.8642$	&$0.02$		&$0.6$		&$0.4$				&$1.3$		&$0.4$\\

29	&HD\,108015		&\multicolumn{2}{c}{}		&			&
&		&$0.11$		&\multicolumn{2}{c}{}				&$2.5$		&$0.5$\\

30	&IRAS\,13258		&\multicolumn{2}{c}{}		&			&
&		&		&\multicolumn{2}{c}{}				&\multicolumn{2}{c}{}\\

31	&EN\,TrA		&\multicolumn{2}{c}{}		&			&
&$36.9$		&$0.02$		&$0.7$		&$0.4$				&$1.1$		&$0.3$\\

32	&IRAS\,15469		&\multicolumn{2}{c}{}		&			&
&		&$0.11$		&\multicolumn{2}{c}{}				&$1.6$		&$0.3$\\

33	&IRAS\,15556		&\multicolumn{2}{c}{}		&			&
&		&		&\multicolumn{2}{c}{}				&\multicolumn{2}{c}{}\\

34	&IRAS\,16230		&\multicolumn{2}{c}{}		&			&
&		&$0.06$		&\multicolumn{2}{c}{}				&$6.0$		&$1.2$\\

35	&IRAS\,17038		&\multicolumn{2}{c}{}		&			&
&		&$-0.36$	&\multicolumn{2}{c}{}				&$4.6$		&$1.0$\\

36	&IRAS\,17233		&\multicolumn{2}{c}{}		&			&
&		&$0.05$		&\multicolumn{2}{c}{}				&\multicolumn{2}{c}{$< 9.4$}\\

37	&IRAS\,17243		&\multicolumn{2}{c}{}		&			&
&		&$0.07$		&\multicolumn{2}{c}{}				&$3.9$		&$0.8$\\

38	&89\,Her		&$1.02$		&$0.59$		&$621$-$980$-$2326$	&$1.38$-$3.45$-$19.43$
&$70$		&$0.09$		&$1.8$		&$1.2$				&$0.7$		&$0.2$\\

39	&AI\,Sco		&\multicolumn{2}{c}{}		&			&
&$71.0$		&$-0.24$	&$2.5$		&$1.7$				&$1.9$		&$0.7$\\

40	&IRAS\,18123		&\multicolumn{2}{c}{}		&			&
&		&$-0.24$	&\multicolumn{2}{c}{}				&$5.2$		&$1.1$\\

41	&IRAS\,18158		&\multicolumn{2}{c}{}		&			&
&		&$0.08$		&\multicolumn{2}{c}{}				&\multicolumn{2}{c}{$< 10.3$}\\

42	&AC\,Her		&\multicolumn{2}{c}{}		&			&
&$75.4619$	&$-0.07$	&$2.4$		&$1.6$				&$1.3$		&$0.4$\\

43	&AD\,Aql		&\multicolumn{2}{c}{}		&			&
&$65.4$		&$0.06$		&$1.7$		&$1.1$				&$4.5$		&$1.5$\\

44	&IRAS\,19125		&\multicolumn{2}{c}{}		&			&
&		&$0.09$		&\multicolumn{2}{c}{}				&$1.9$		&$0.4$\\
			
45	&EP\,Lyr		&\multicolumn{2}{c}{}		&			&
&$83.315$	&$0.11$ 	&$2.3$		&$1.6$				&$3.2$		&$1.1$\\

46	&IRAS\,19157		&\multicolumn{2}{c}{}		&			&
&		&$0.09$		&\multicolumn{2}{c}{}				&$4.2$		&$0.9$\\

47	&IRAS\,20056		&\multicolumn{2}{c}{}		&			&
&		&$-0.03$	&\multicolumn{2}{c}{}				&\multicolumn{2}{c}{$< 9.8$}\\

48	&R\,Sge			&\multicolumn{2}{c}{}		&			&
&$70.594$	&$0.00$		&$2.0$		&$1.4$				&$1.8$		&$0.6$\\

49	&V\,Vul			&\multicolumn{2}{c}{}		&			&
&$75.72$	&$-0.15$	&$2.6$		&$1.8$				&$2.0$		&$0.7$\\

50	&HD\,213985		&\multicolumn{2}{c}{}		&			&
&		&$0.04$		&\multicolumn{2}{c}{}				&$3.6$		&$0.7$\\

51	&BD$+$39$^{\circ}$4926	&\multicolumn{2}{c}{}		&			&
&		&$0.11$		&\multicolumn{2}{c}{}				&$4.8$		&$1.0$\\
\end{supertabular}
\end{center}
\clearpage

\begin{figure*}
\centering 
\includegraphics[width=0.33\textwidth]{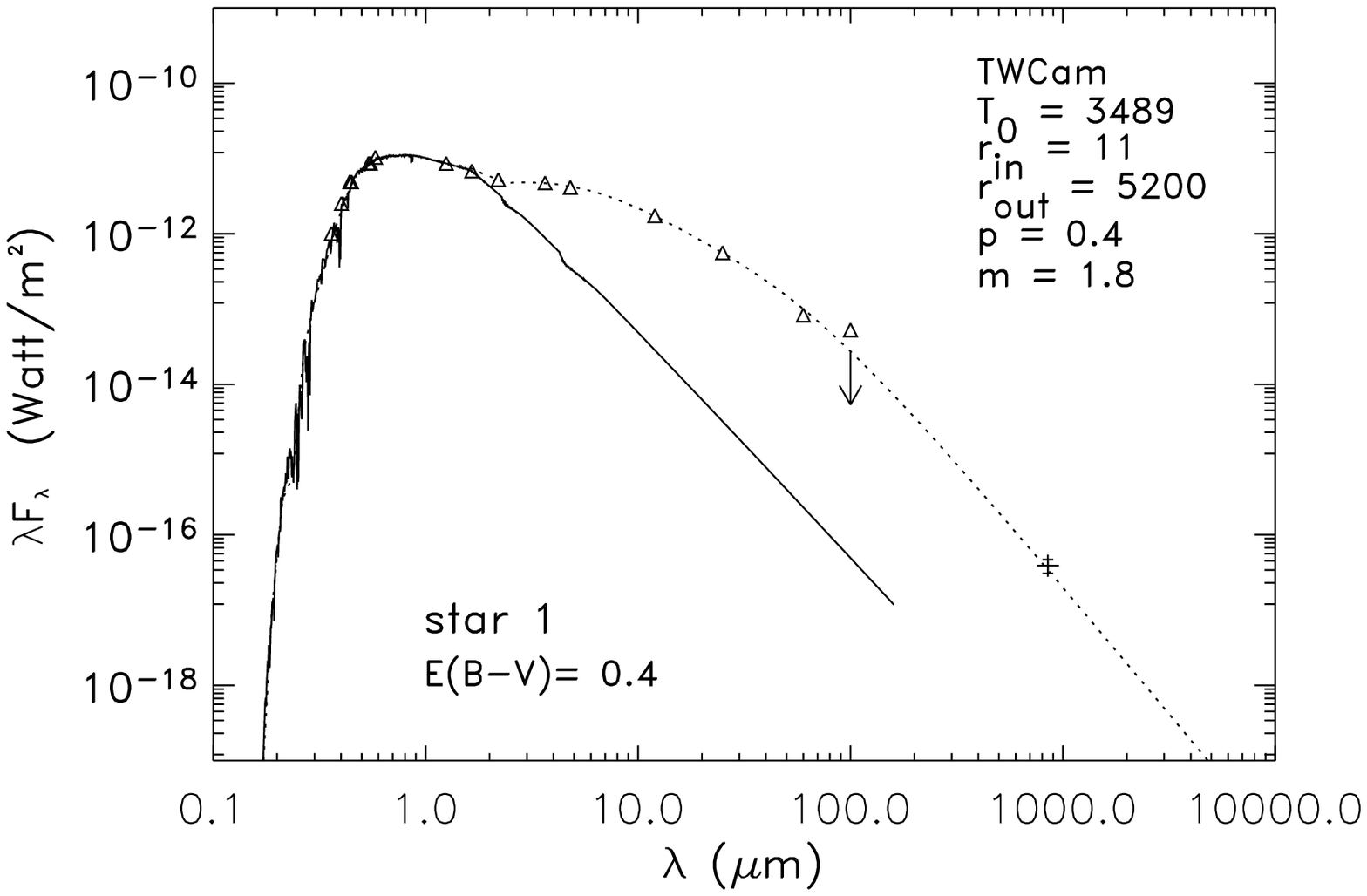}
\includegraphics[width=0.33\textwidth]{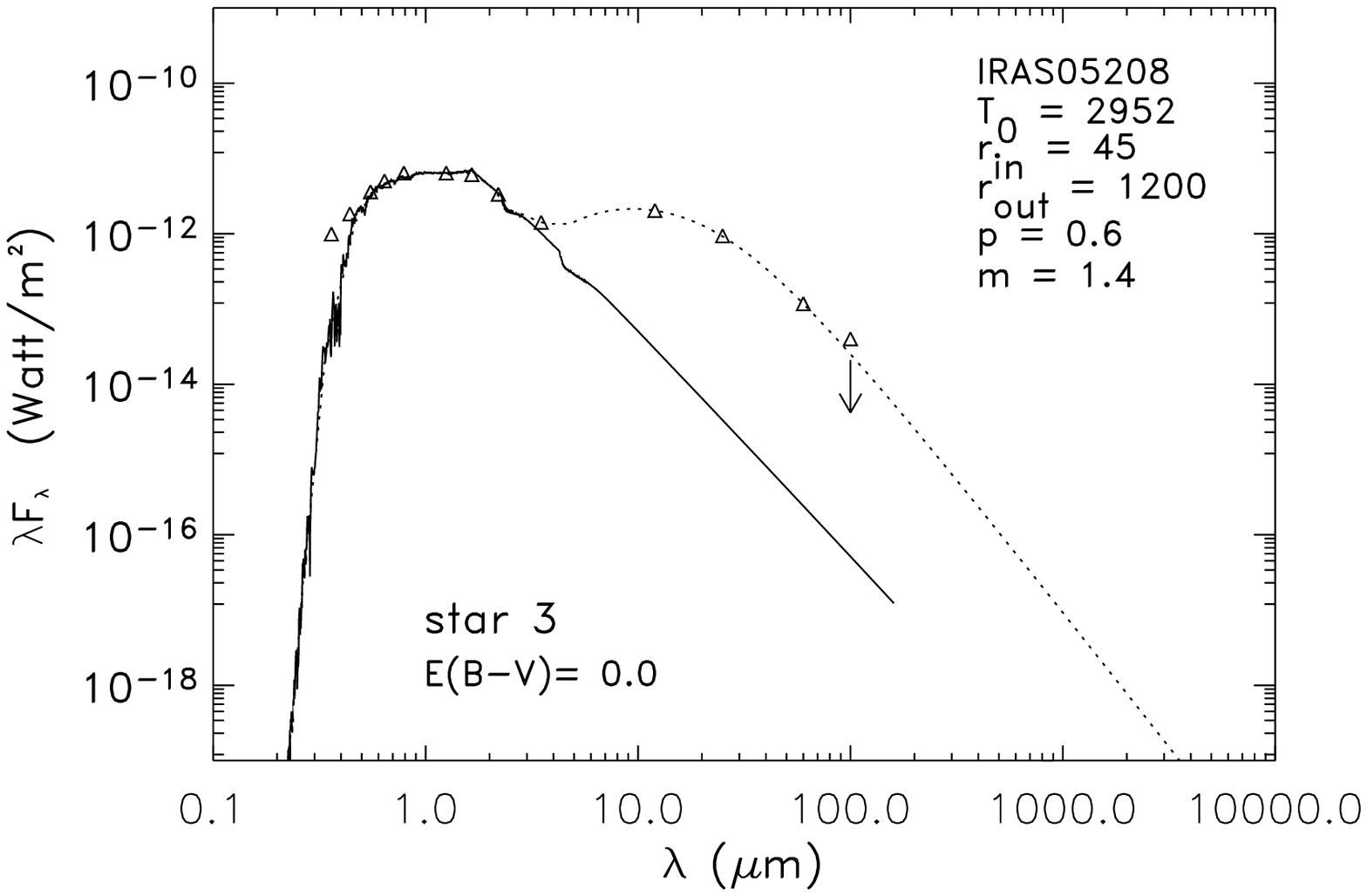}
\includegraphics[width=0.33\textwidth]{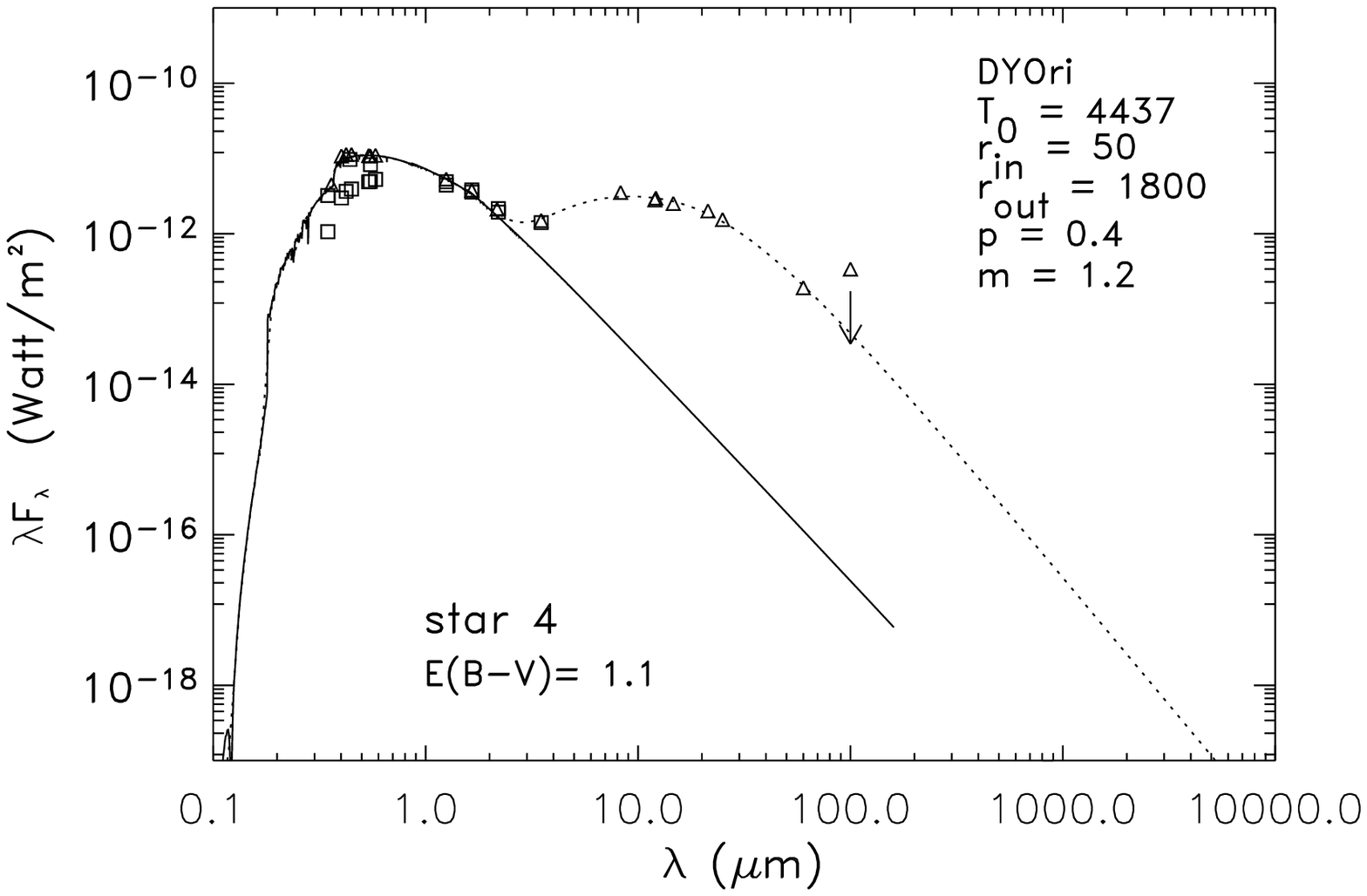}
\\
\includegraphics[width=0.33\textwidth]{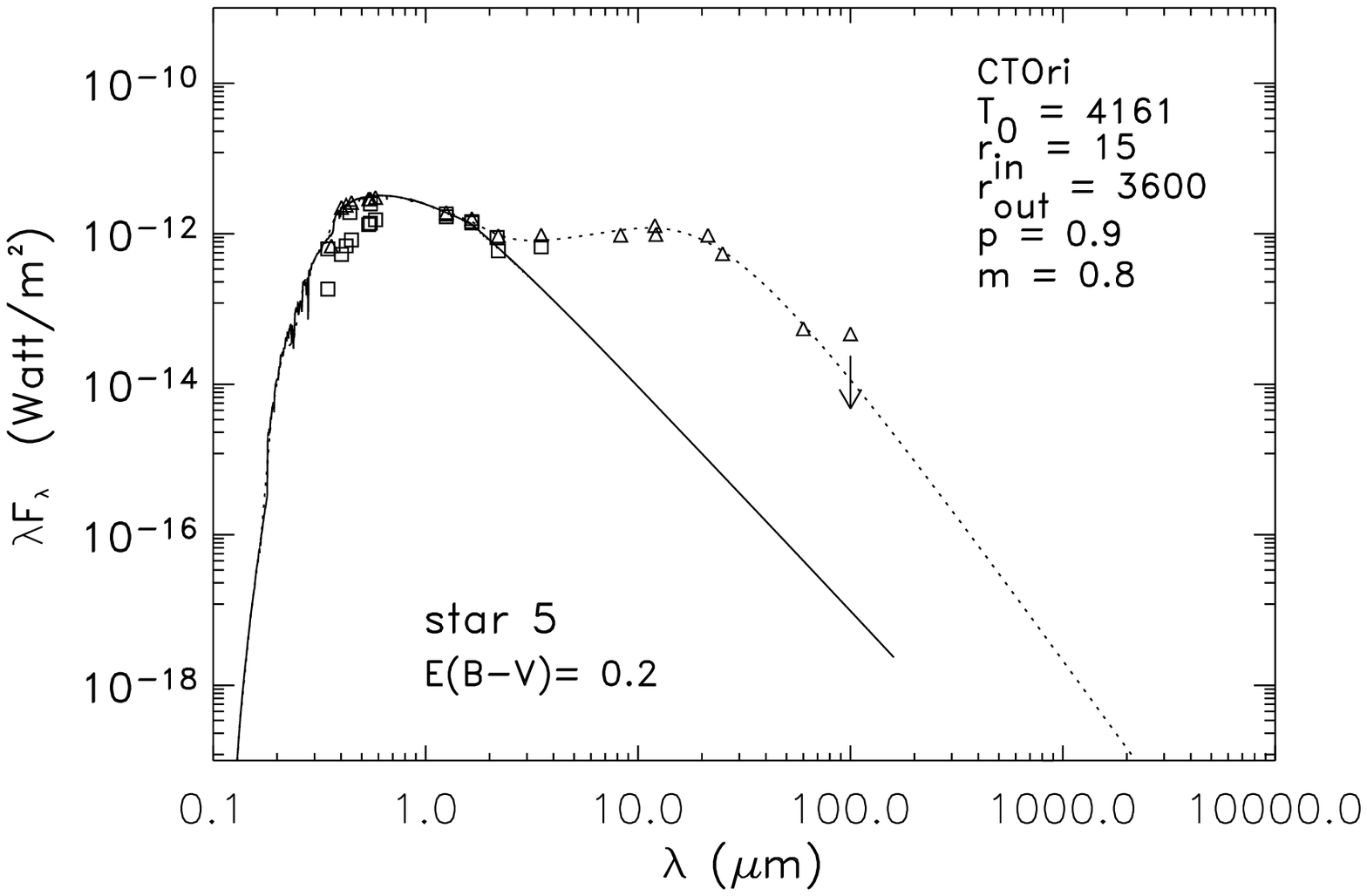}
\includegraphics[width=0.33\textwidth]{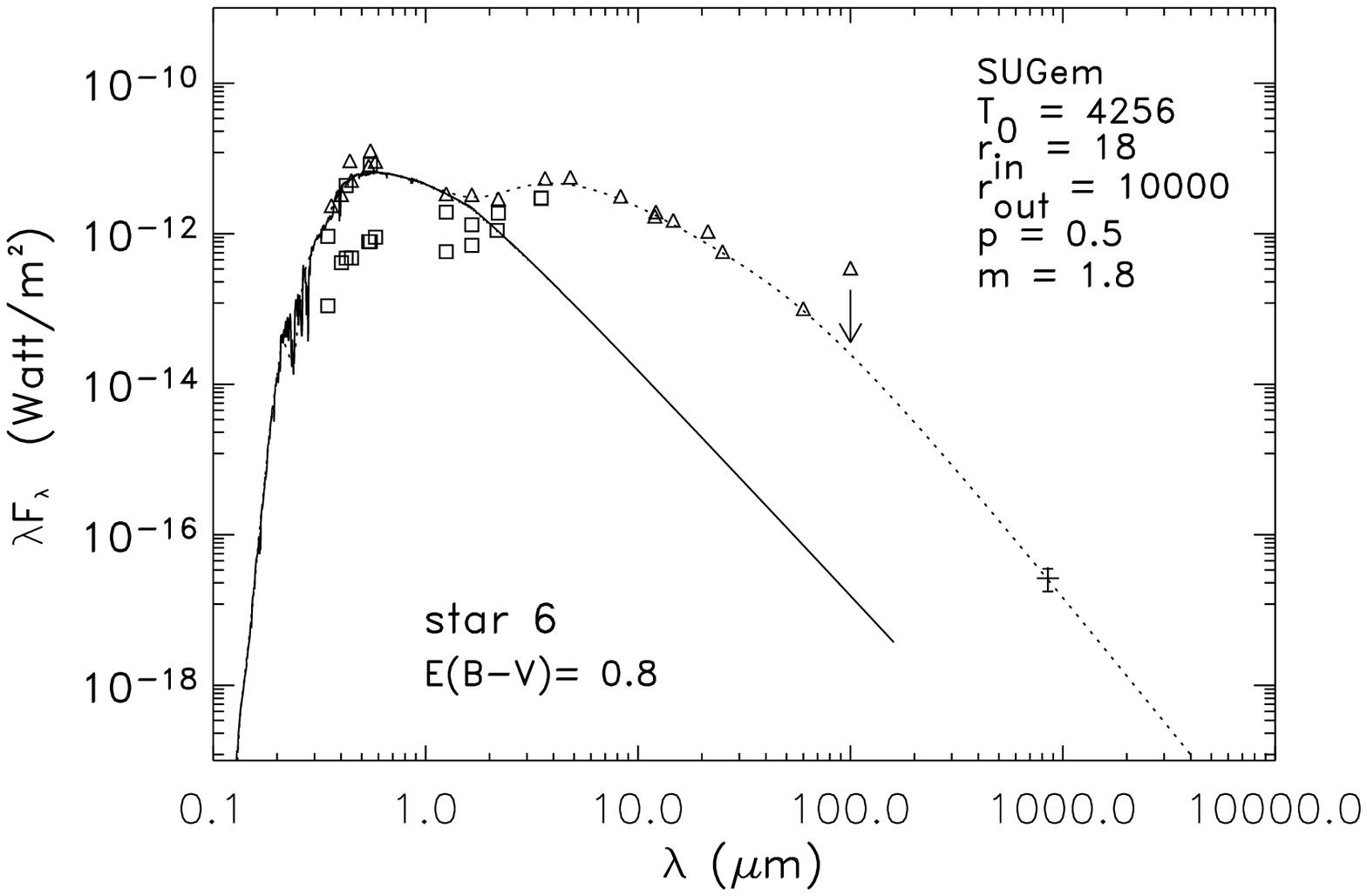}
\includegraphics[width=0.33\textwidth]{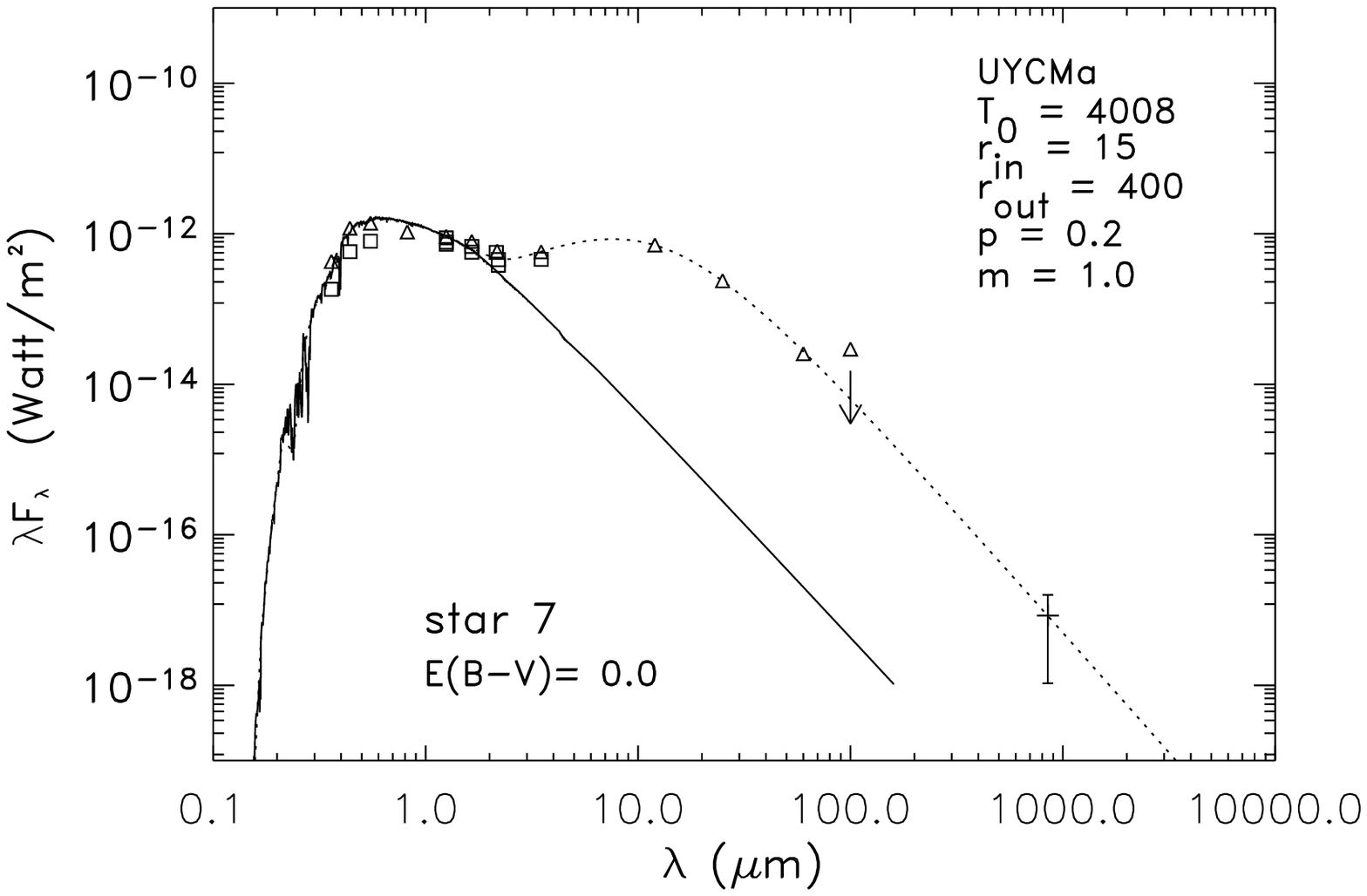}
\\
\includegraphics[width=0.33\textwidth]{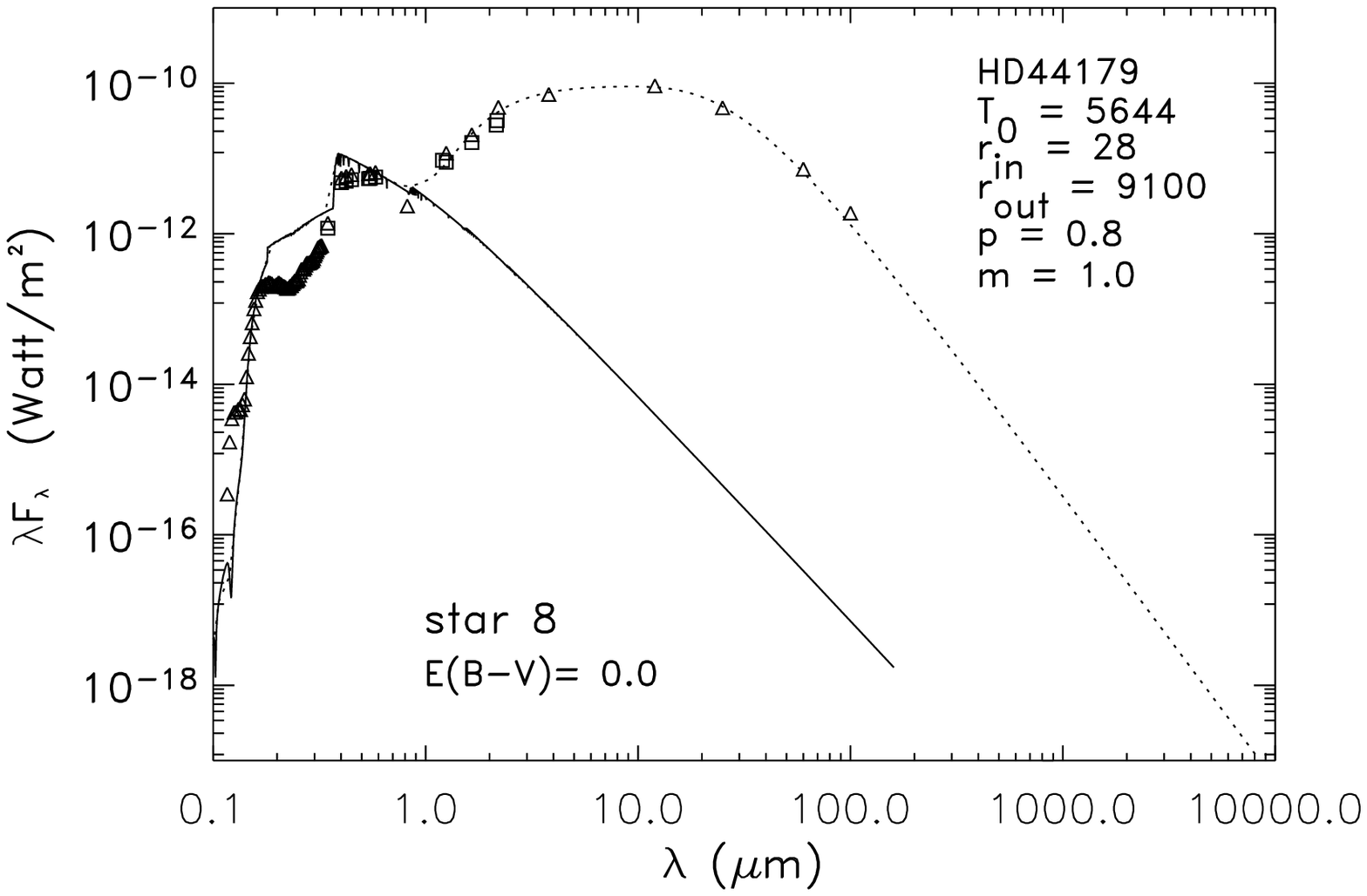}
\includegraphics[width=0.33\textwidth]{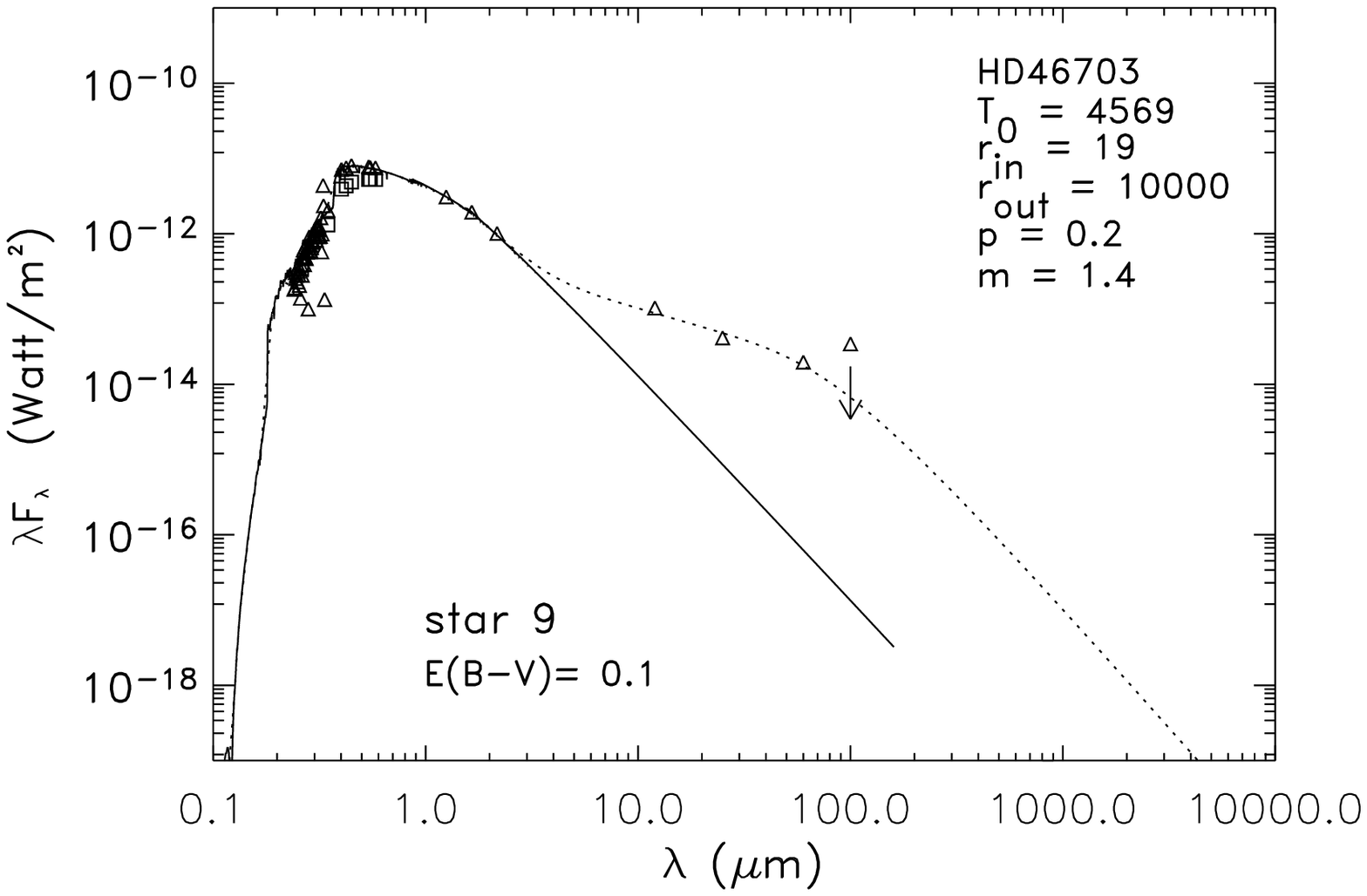}
\includegraphics[width=0.33\textwidth]{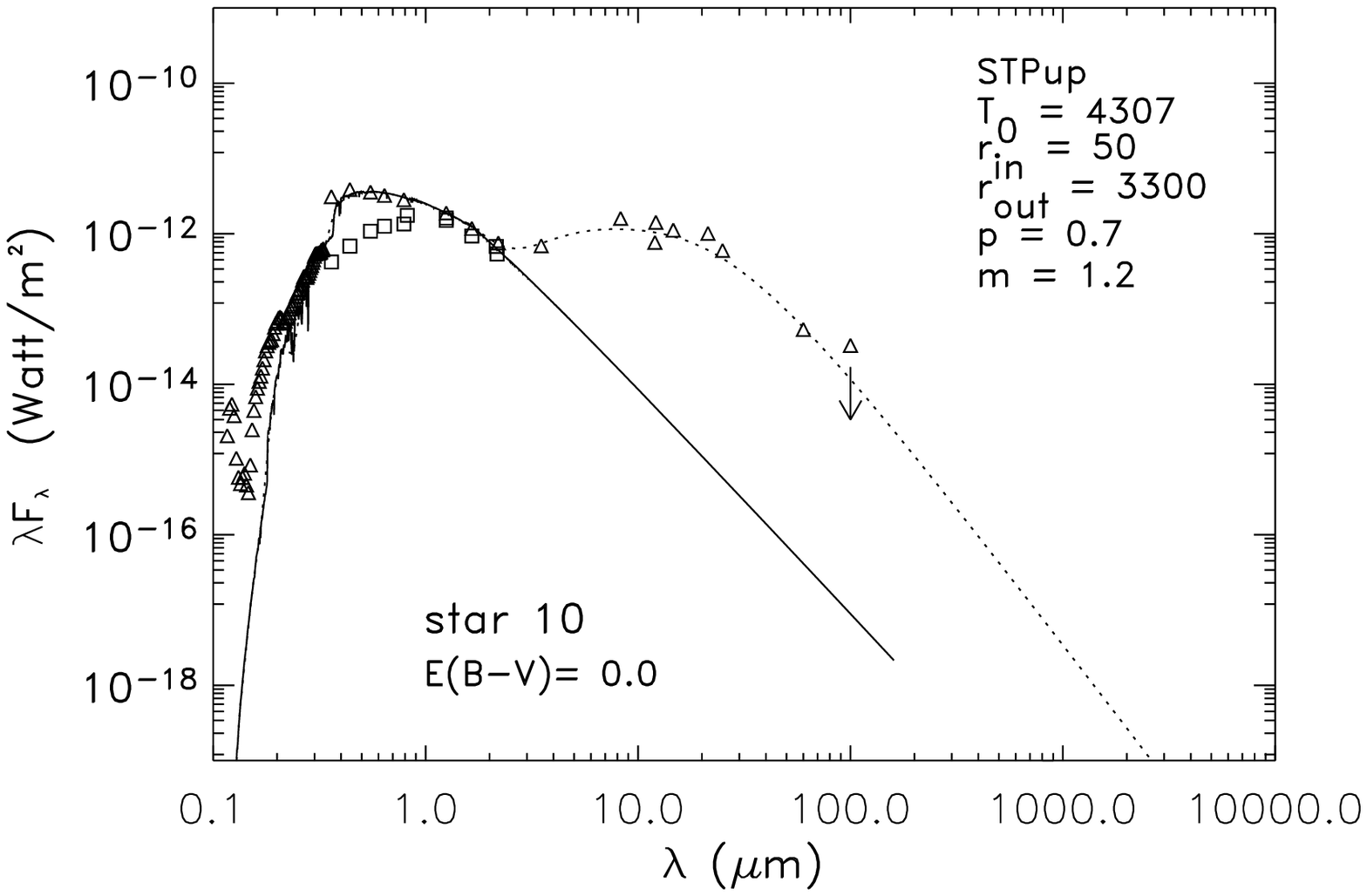}
\\
\includegraphics[width=0.33\textwidth]{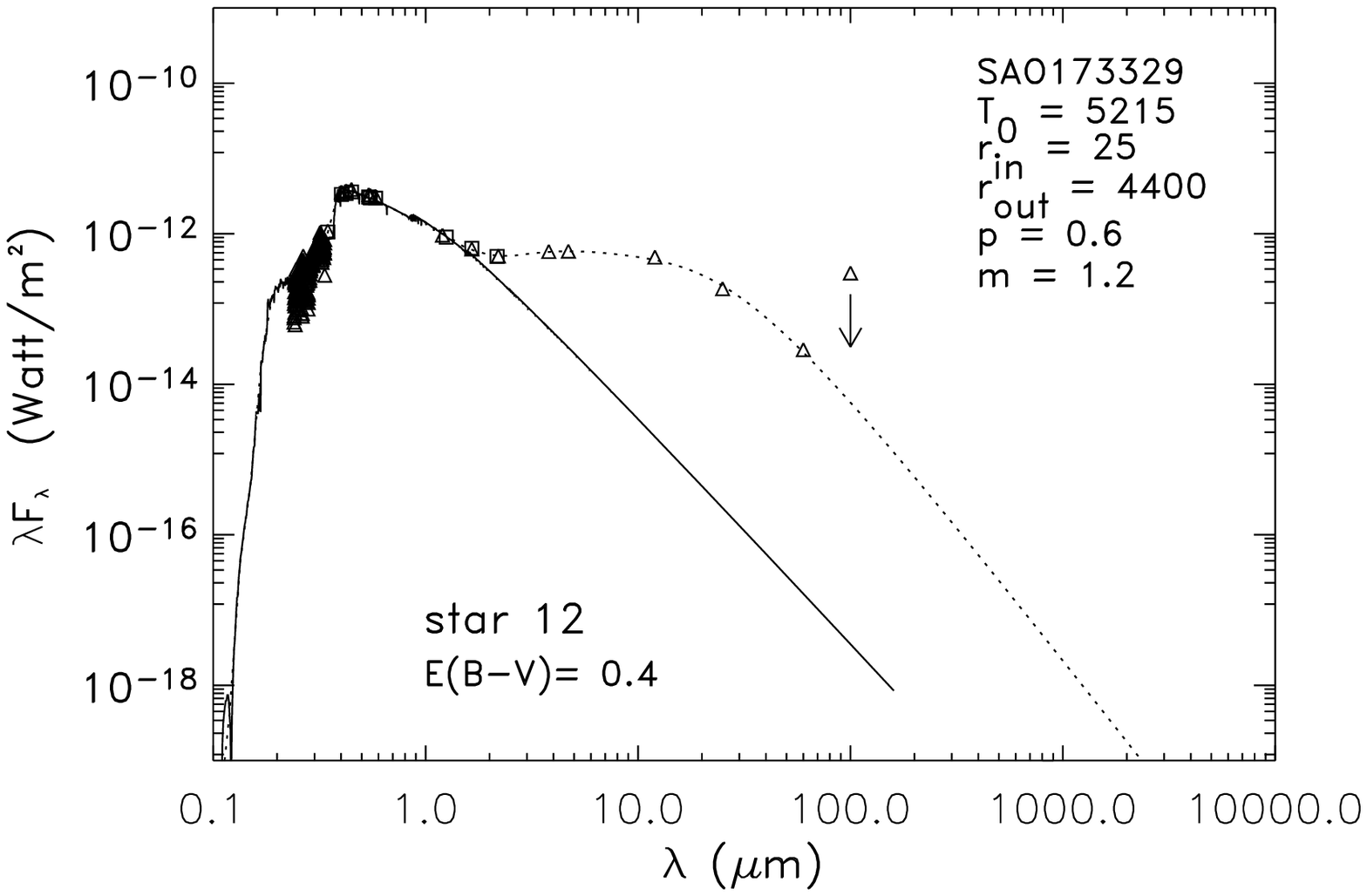}
\includegraphics[width=0.33\textwidth]{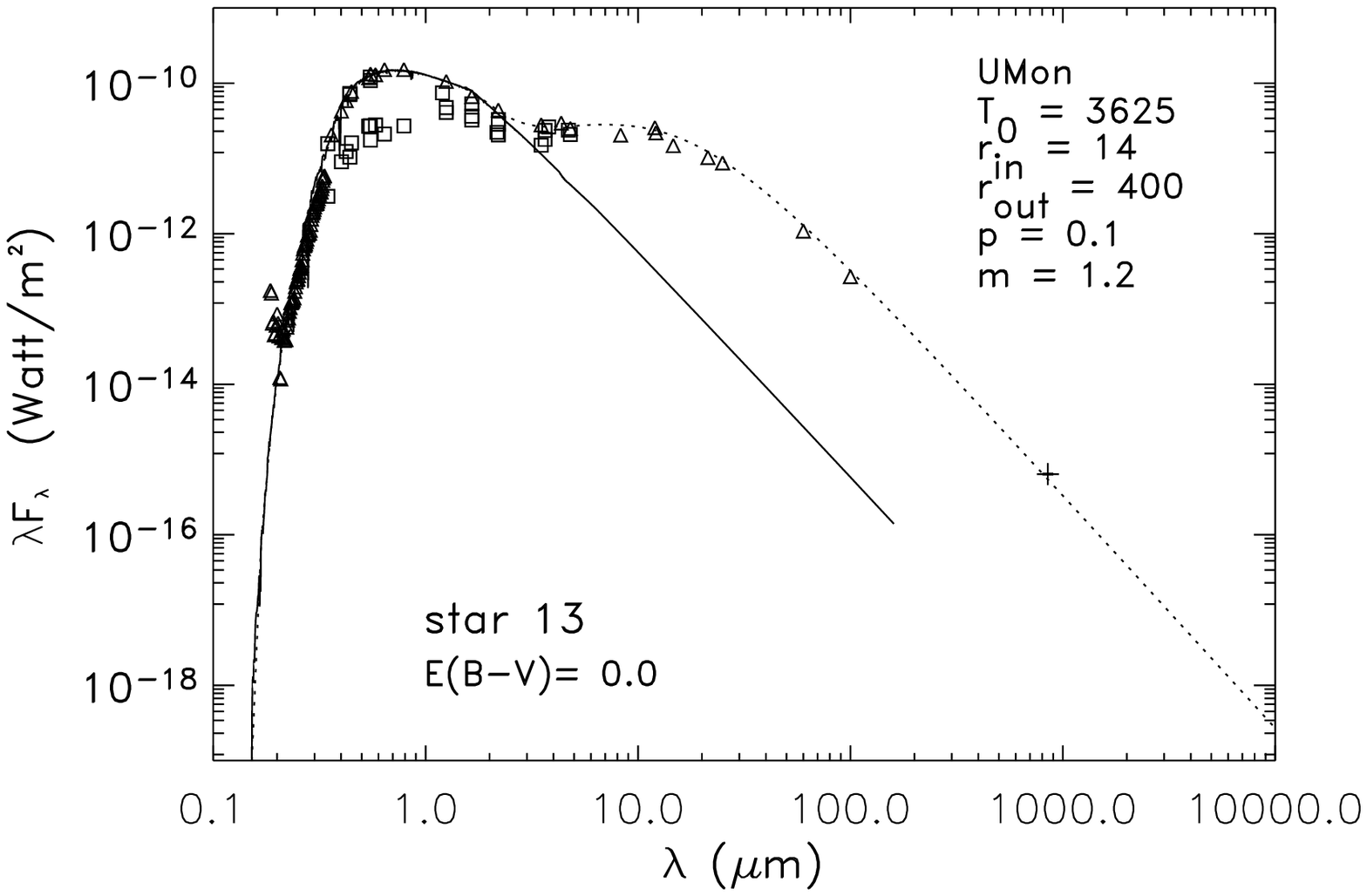}
\includegraphics[width=0.33\textwidth]{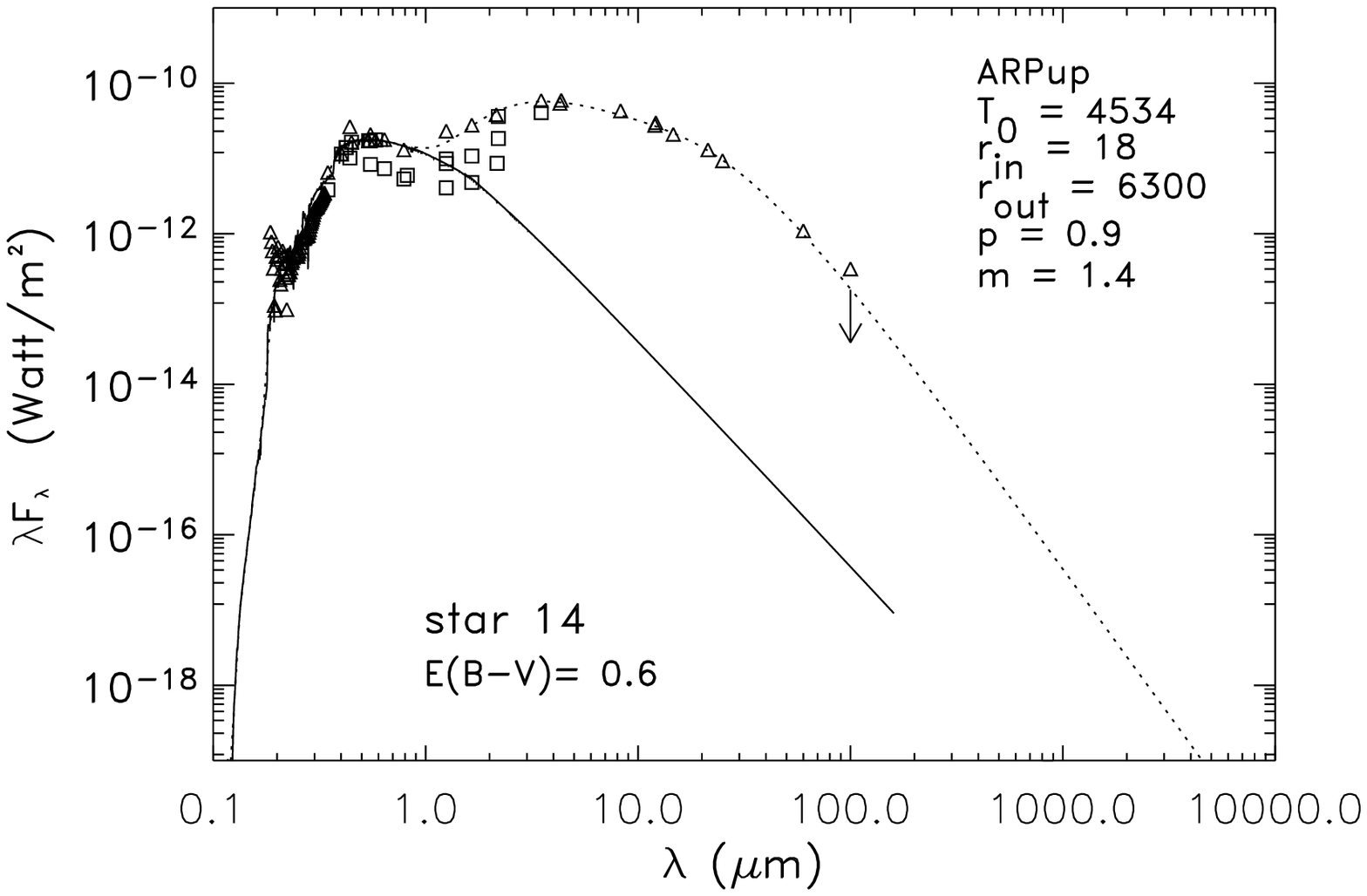}
\\
\includegraphics[width=0.33\textwidth]{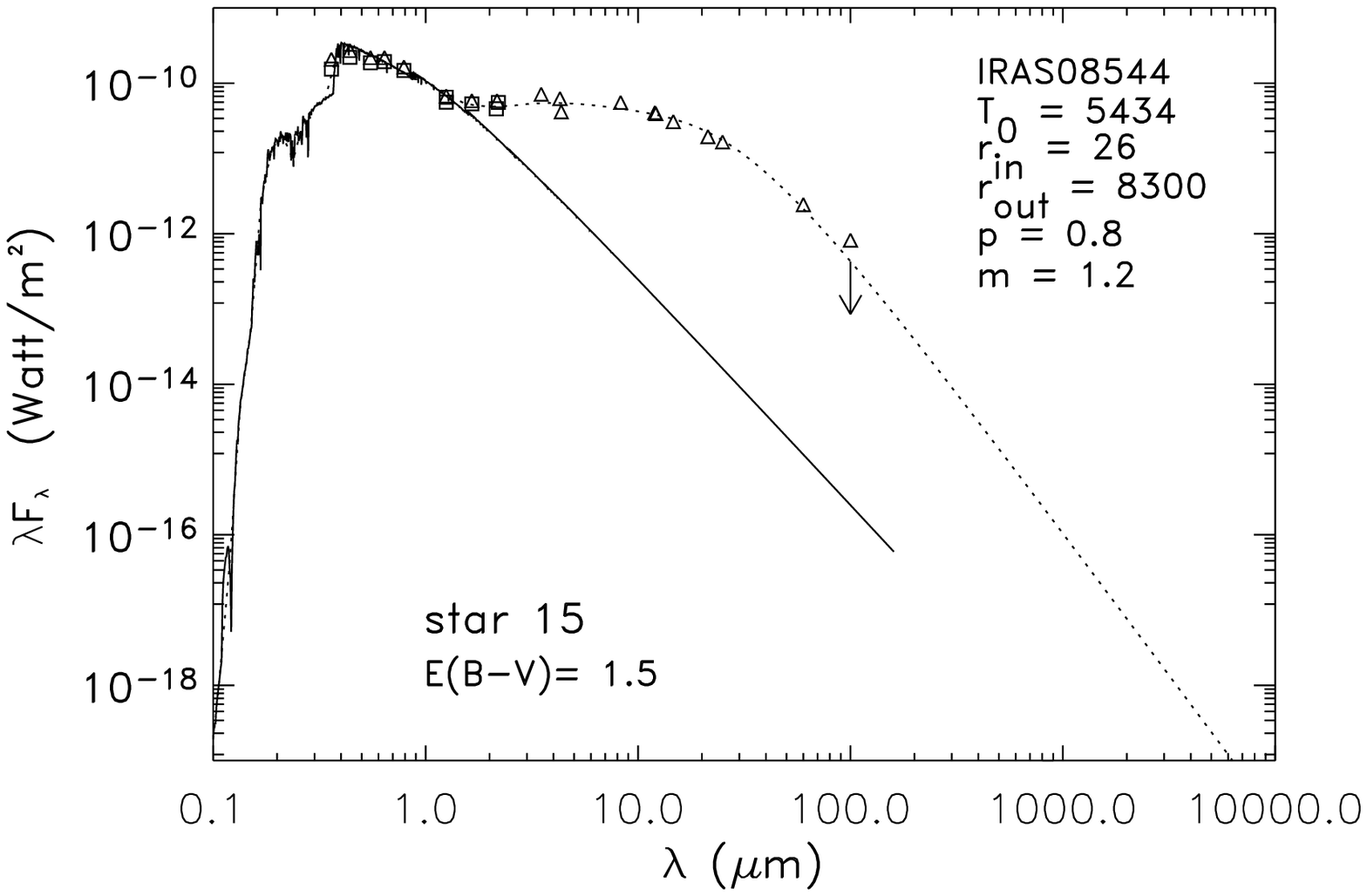}
\includegraphics[width=0.33\textwidth]{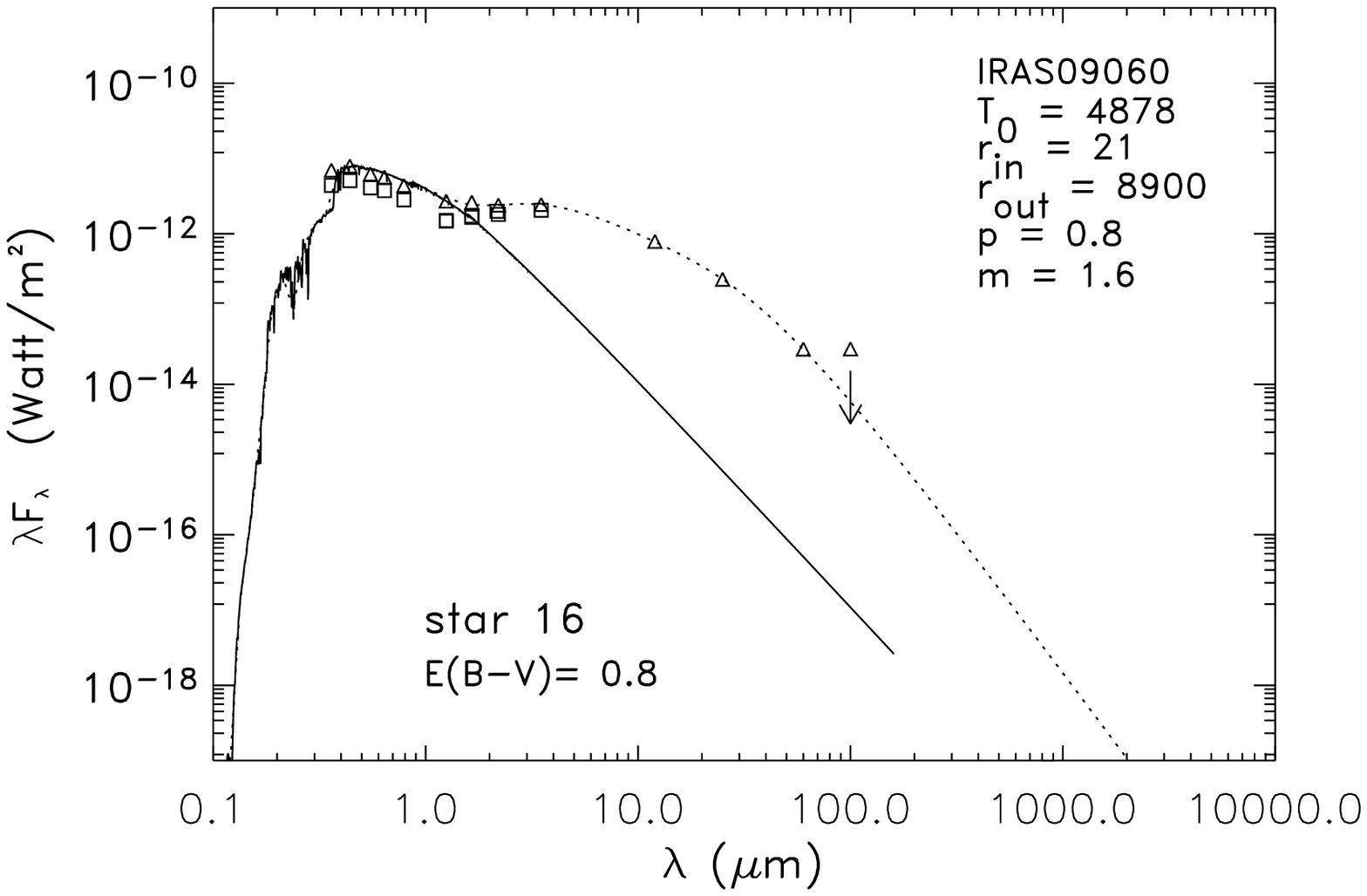}
\includegraphics[width=0.33\textwidth]{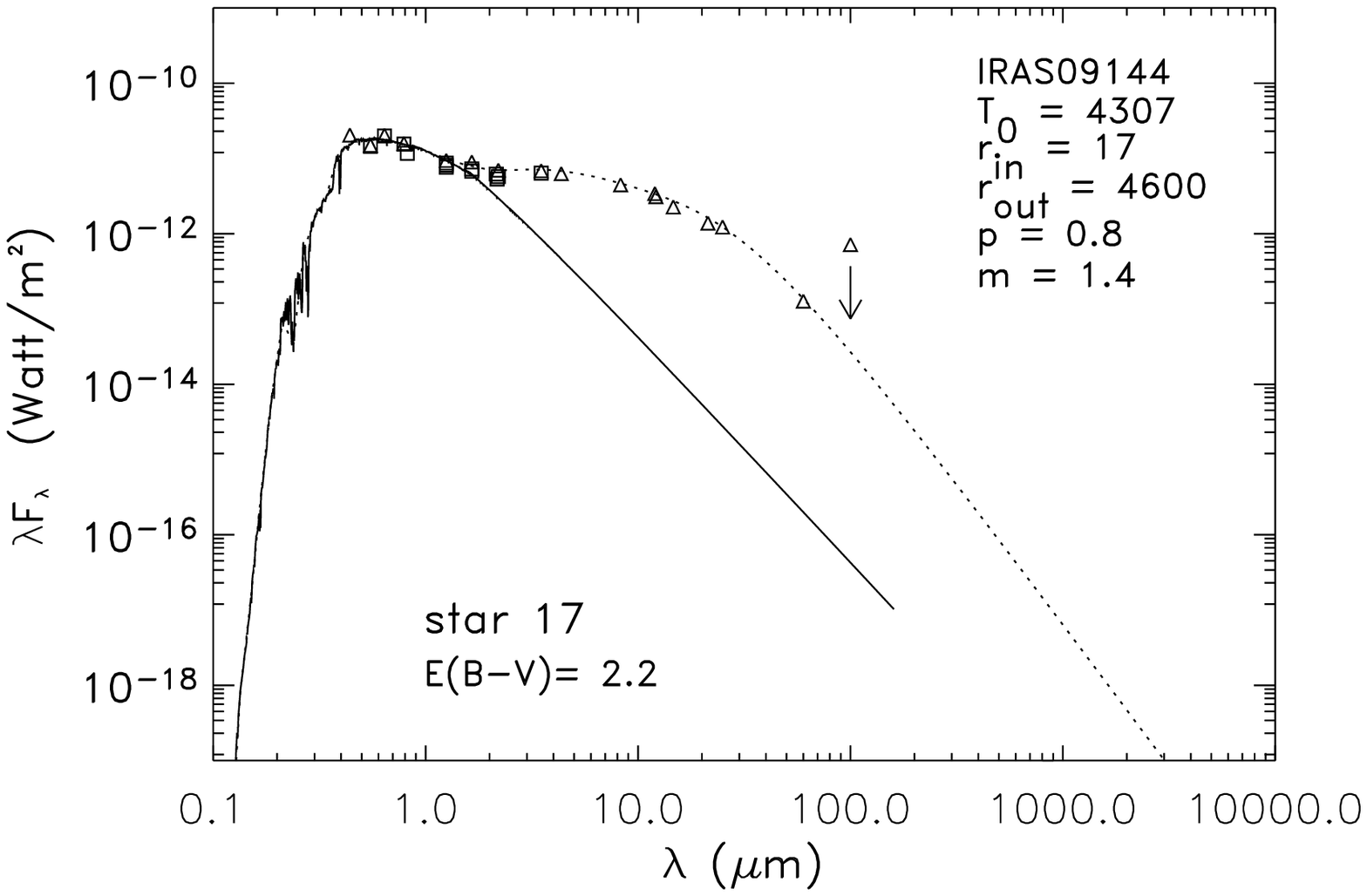}
\caption{The SEDs of all post-AGB stars in our sample for which we
have all the data needed to complete the SEDs, except those of
HD\,52961, HD\,213985, RV\,Tau, IW\,Car, IRAS\,17233 and IRAS\,18123
(see Fig.~\ref{plots}). The dereddened fluxes are given together with
the scaled photospheric Kurucz model representing the unattenuated
stellar photosphere (solid line). An optically thin dust model was
used to fit the IR\,excess (dotted line). Data found in the literature
together with our 7 band Geneva photometry (only the maxima) are
plotted as triangles. The minimal data points (squares) were not used
for the determination of $E(B-V)$. They are shown to give an
indication of the amplitude of the pulsations. Where available,
crosses represent our $850\,\mu$m SCUBA data point. Error bars on the
$850\,\mu$m SCUBA data point are plotted as well, for some objects
though smaller than the symbols. The arrow at the $100\,\mu$m flux
point signifies an upper limit. \label{app:plots}}
\end{figure*}

\addtocounter{figure}{-1}

\begin{figure*}
\includegraphics[width=0.33\textwidth]{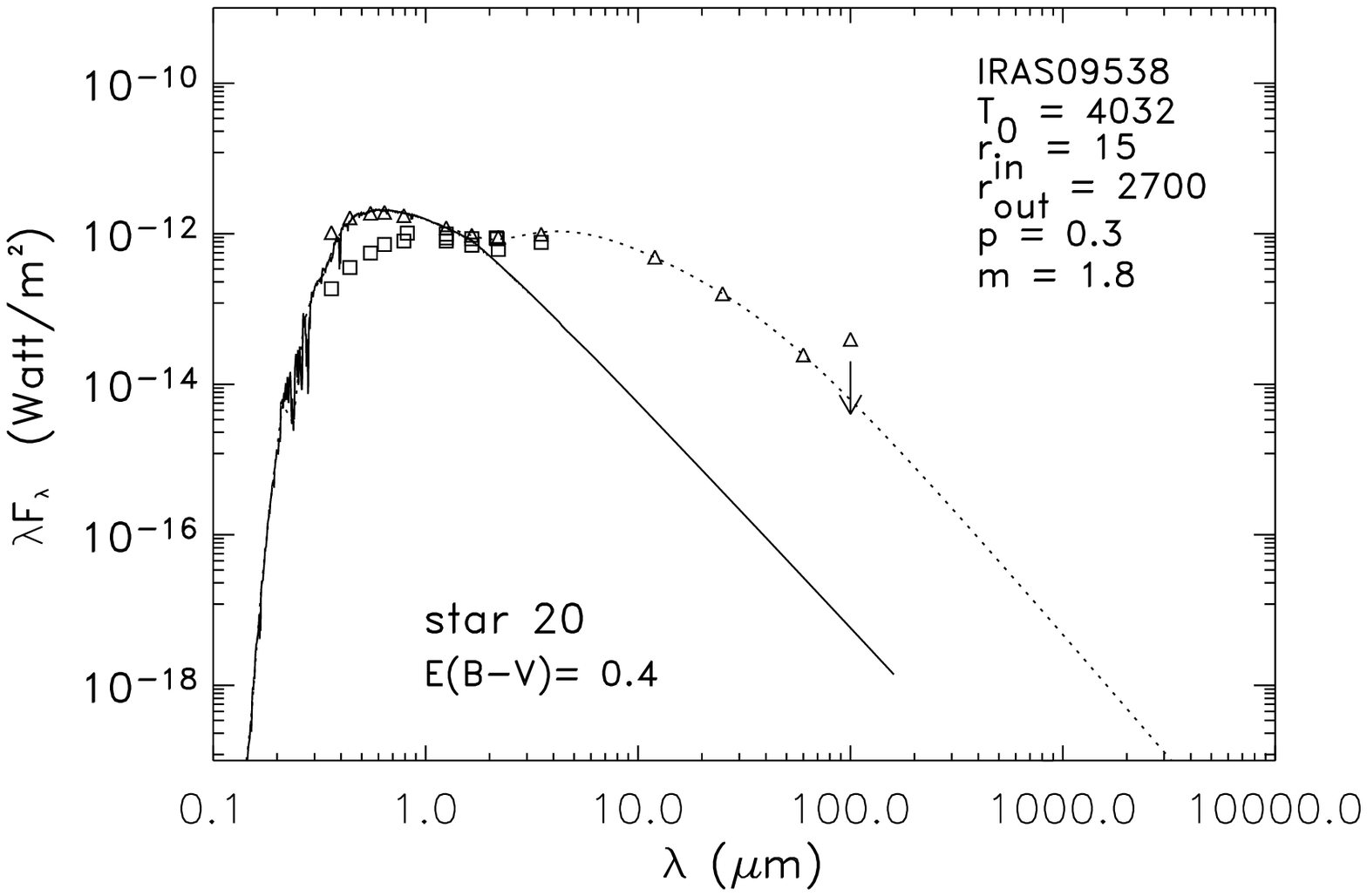}
\includegraphics[width=0.33\textwidth]{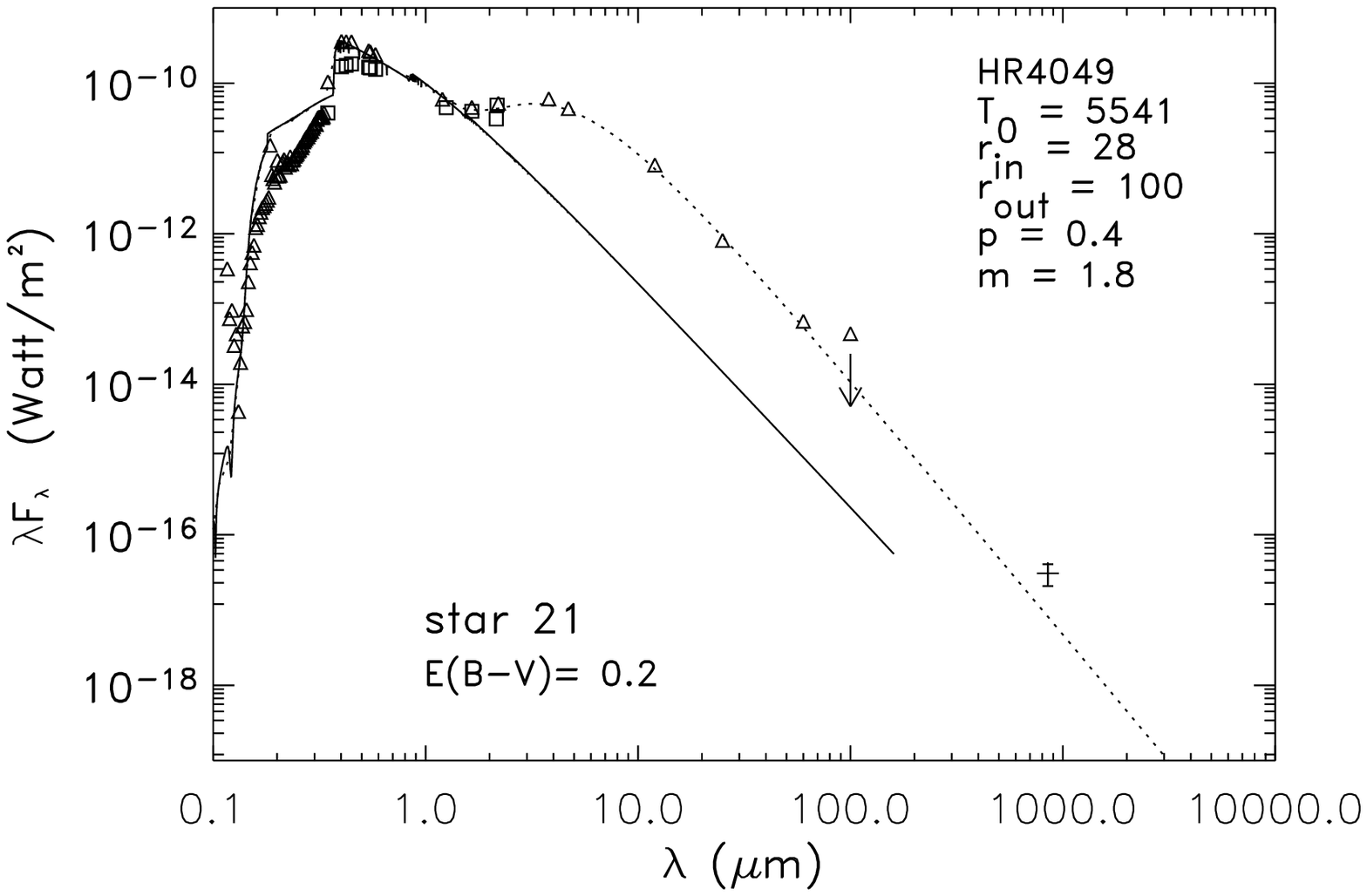}
\includegraphics[width=0.33\textwidth]{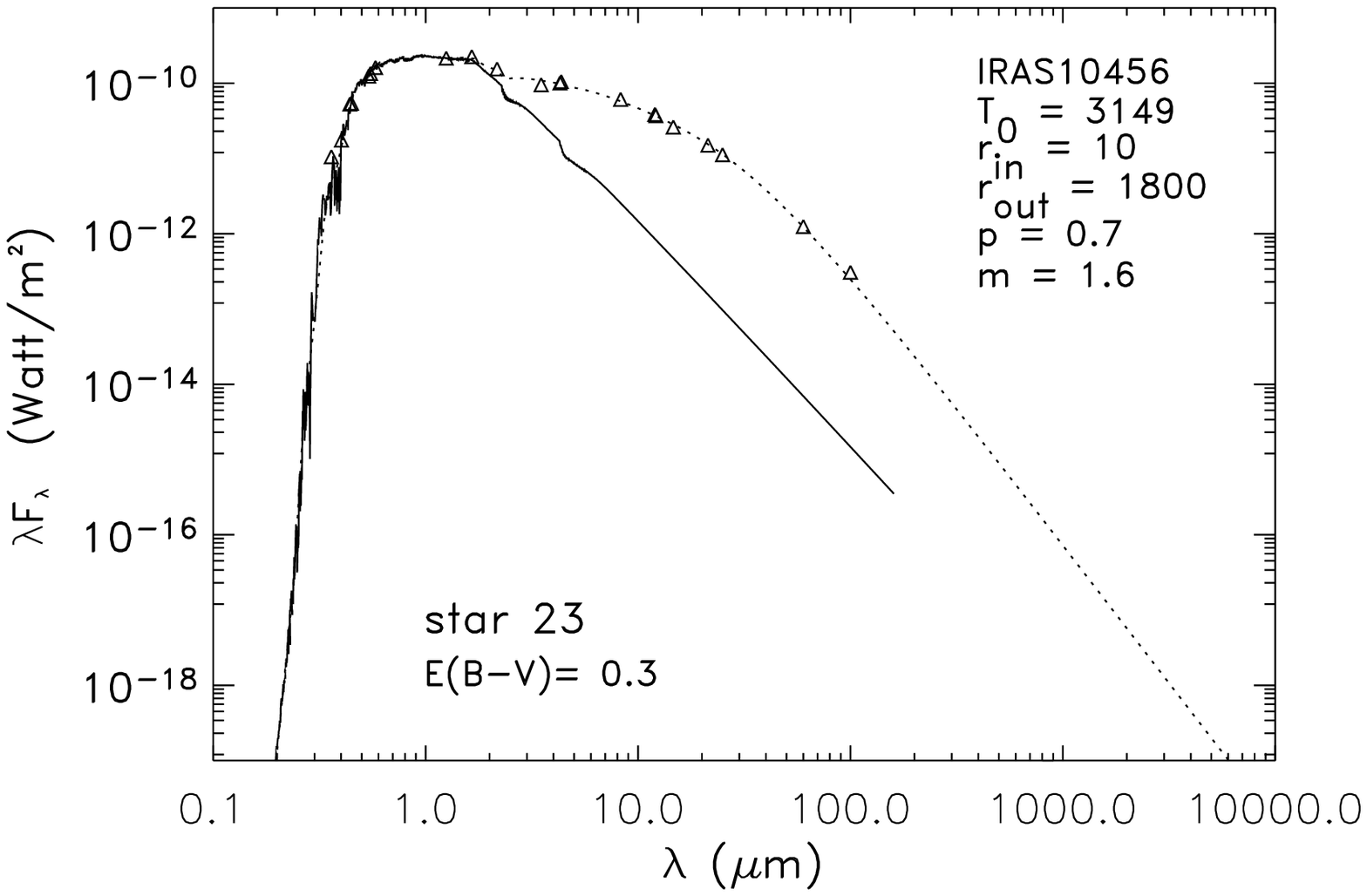}
\\
\includegraphics[width=0.33\textwidth]{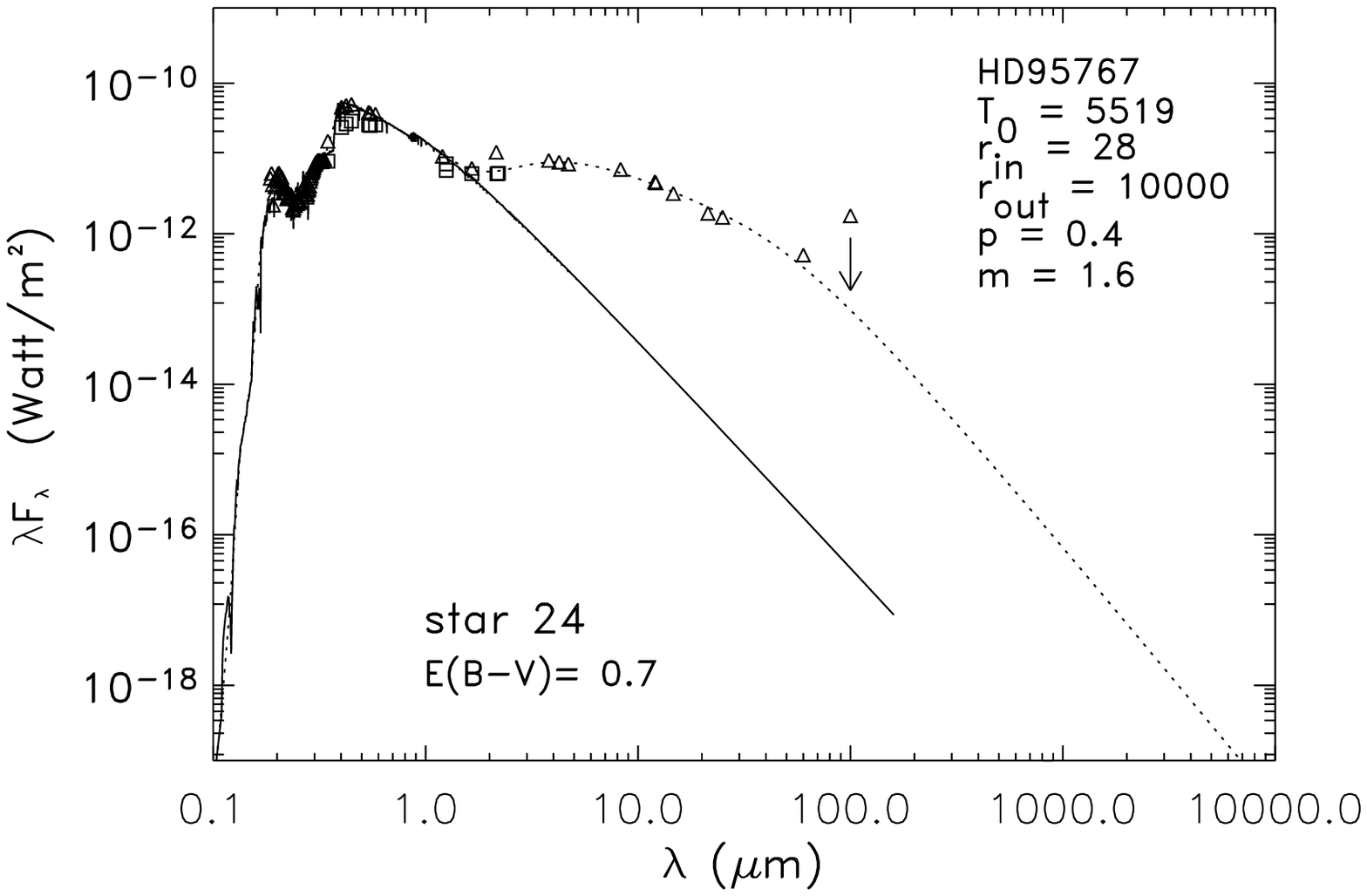}
\includegraphics[width=0.33\textwidth]{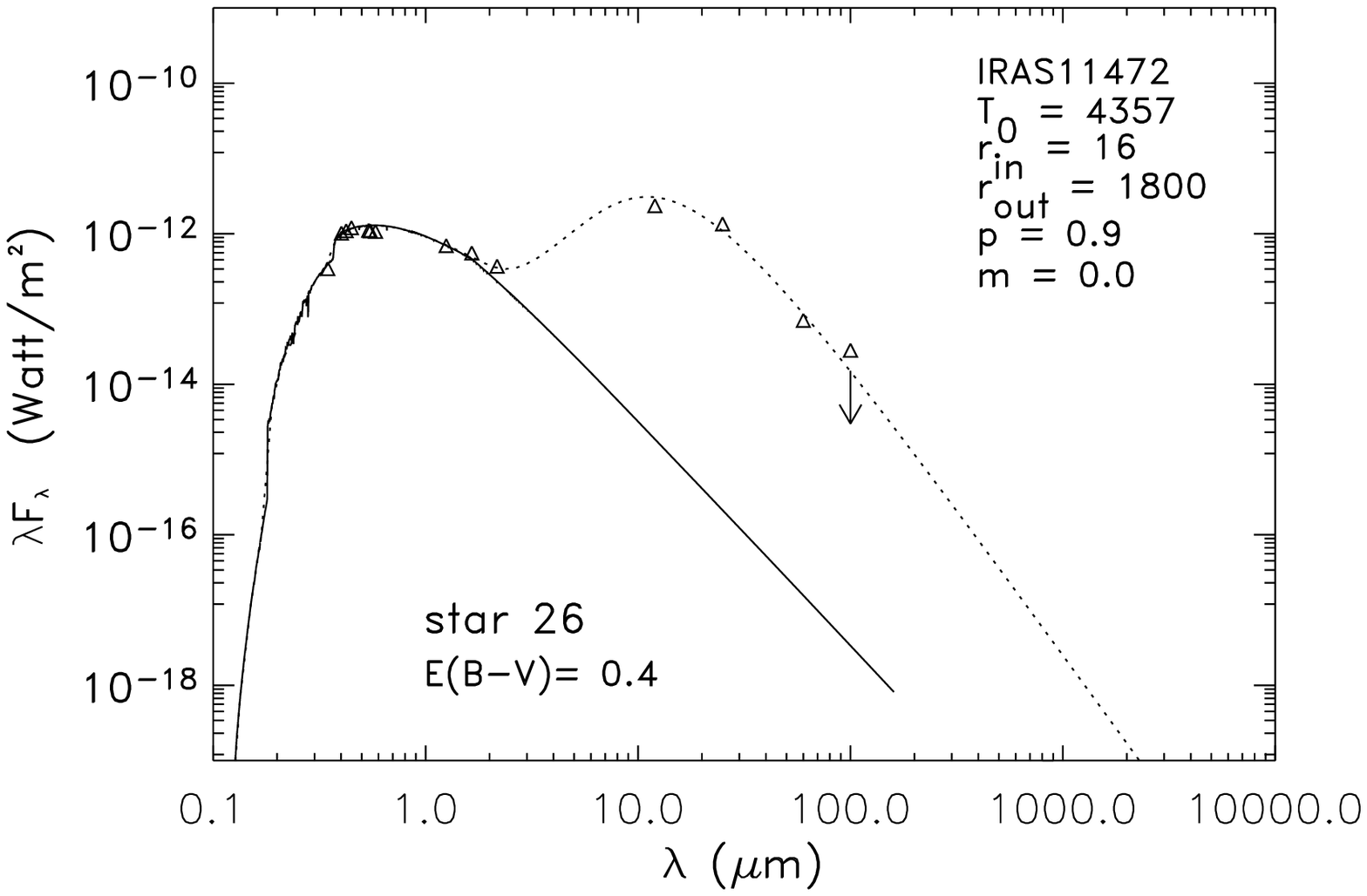}
\includegraphics[width=0.33\textwidth]{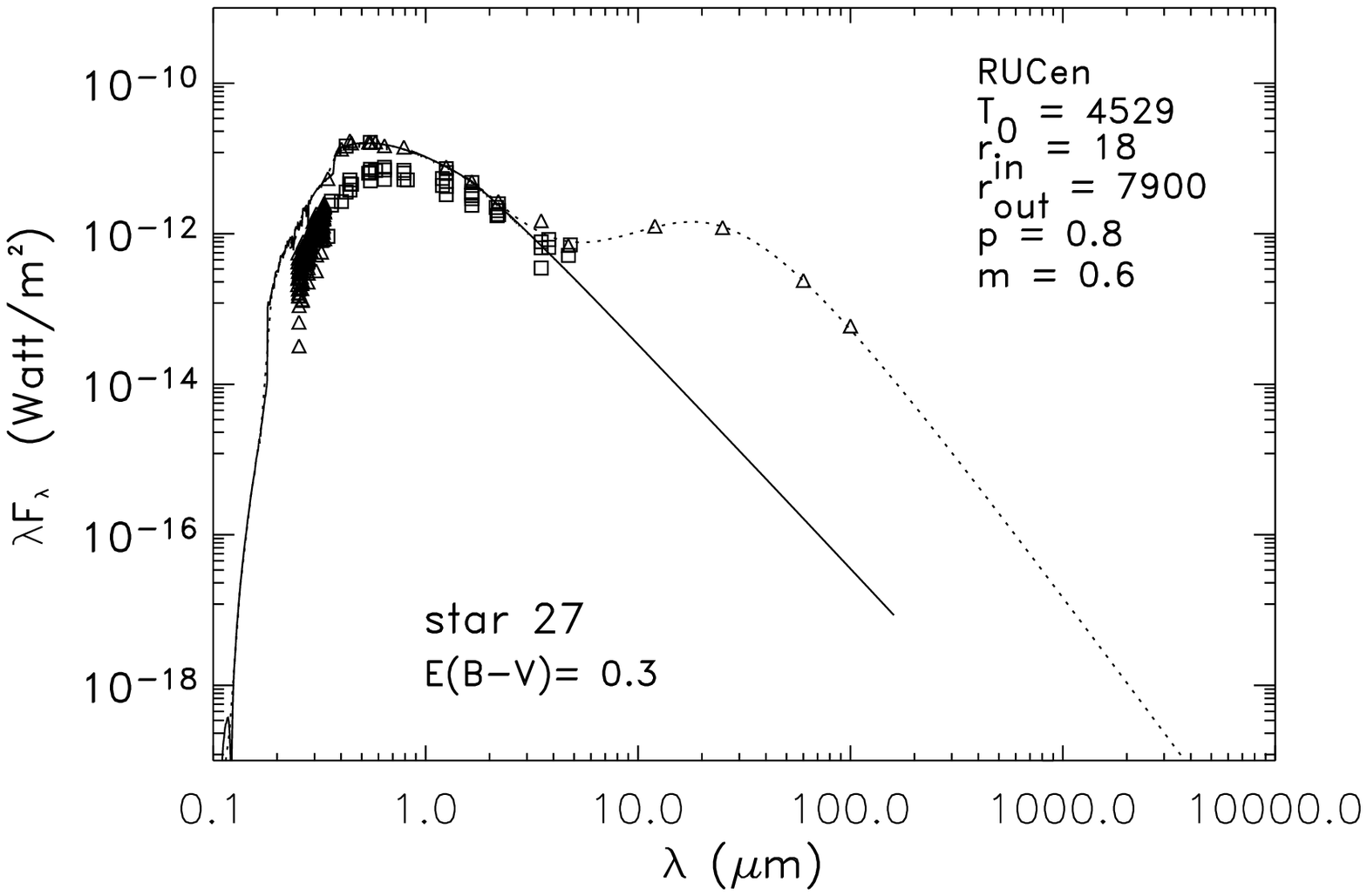}
\\
\includegraphics[width=0.33\textwidth]{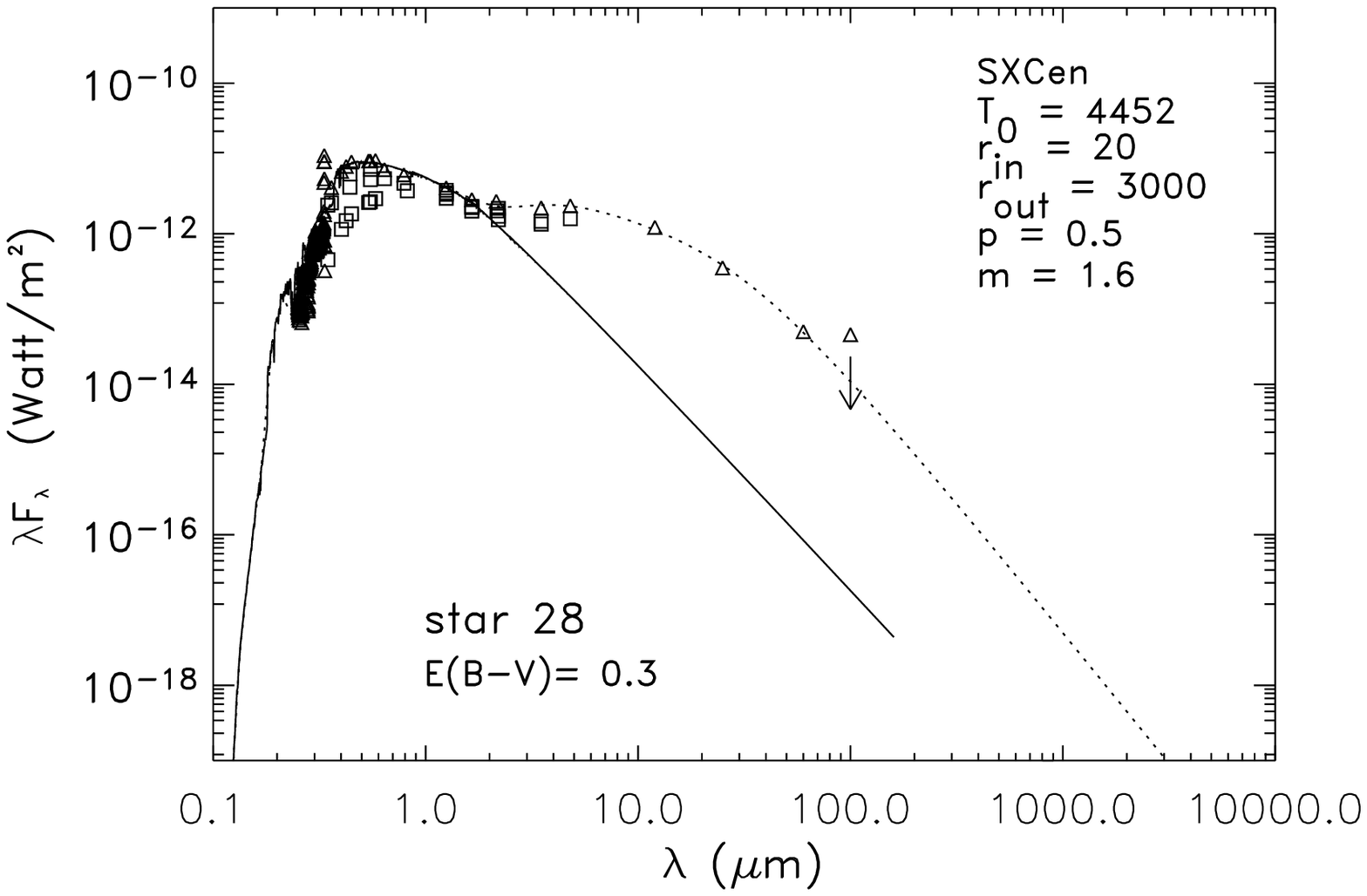}
\includegraphics[width=0.33\textwidth]{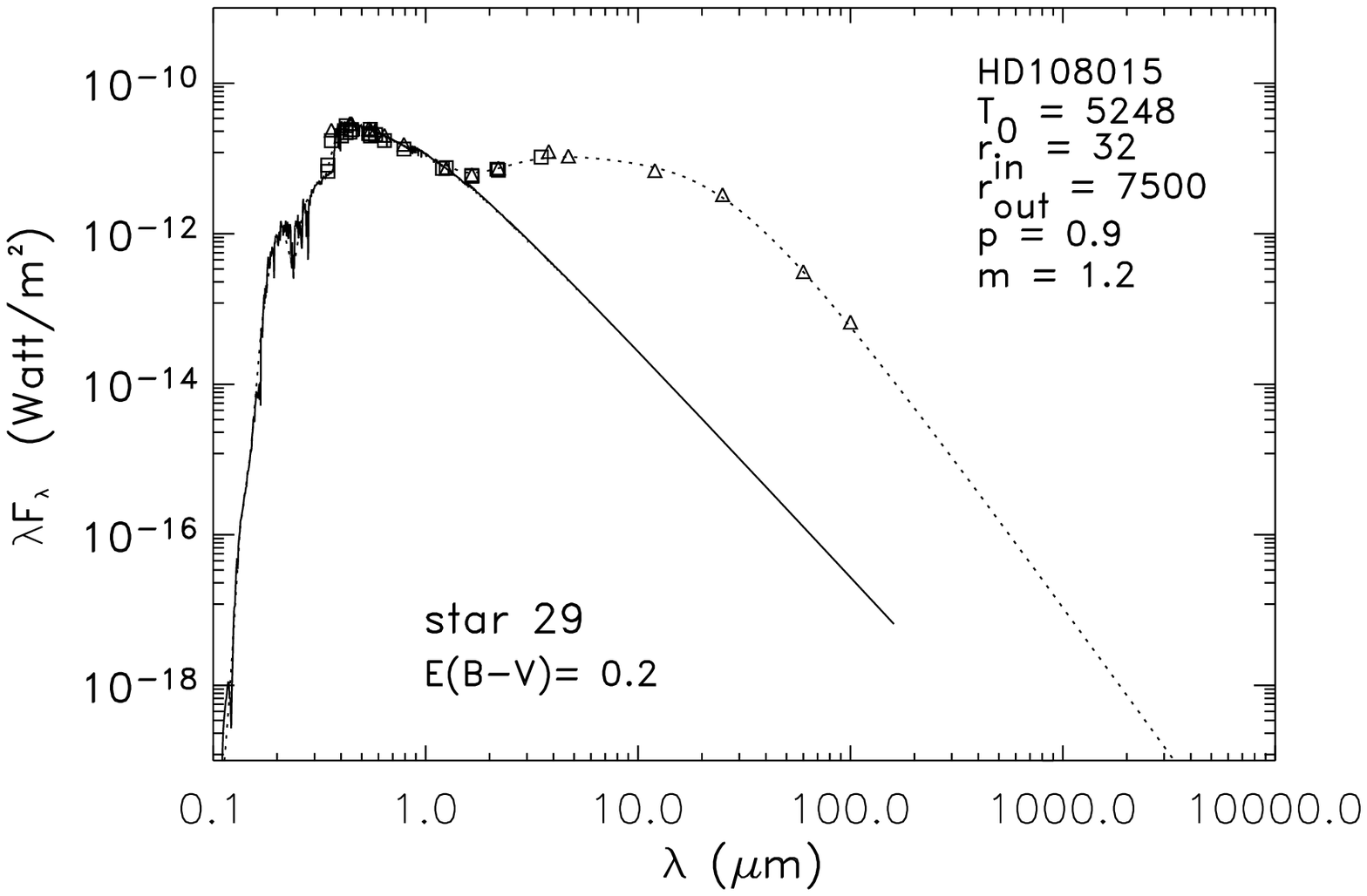}
\includegraphics[width=0.33\textwidth]{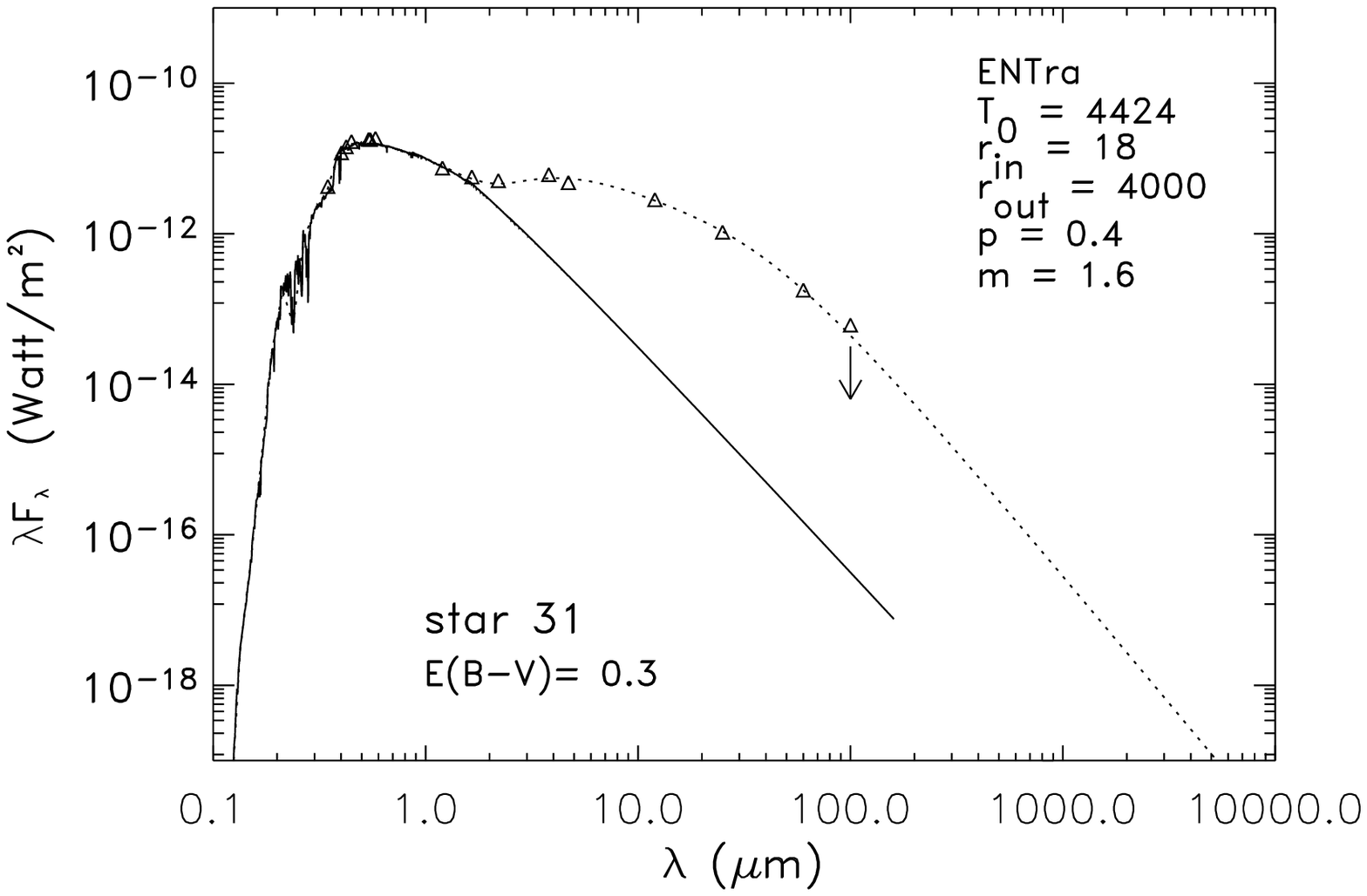}
\\
\includegraphics[width=0.33\textwidth]{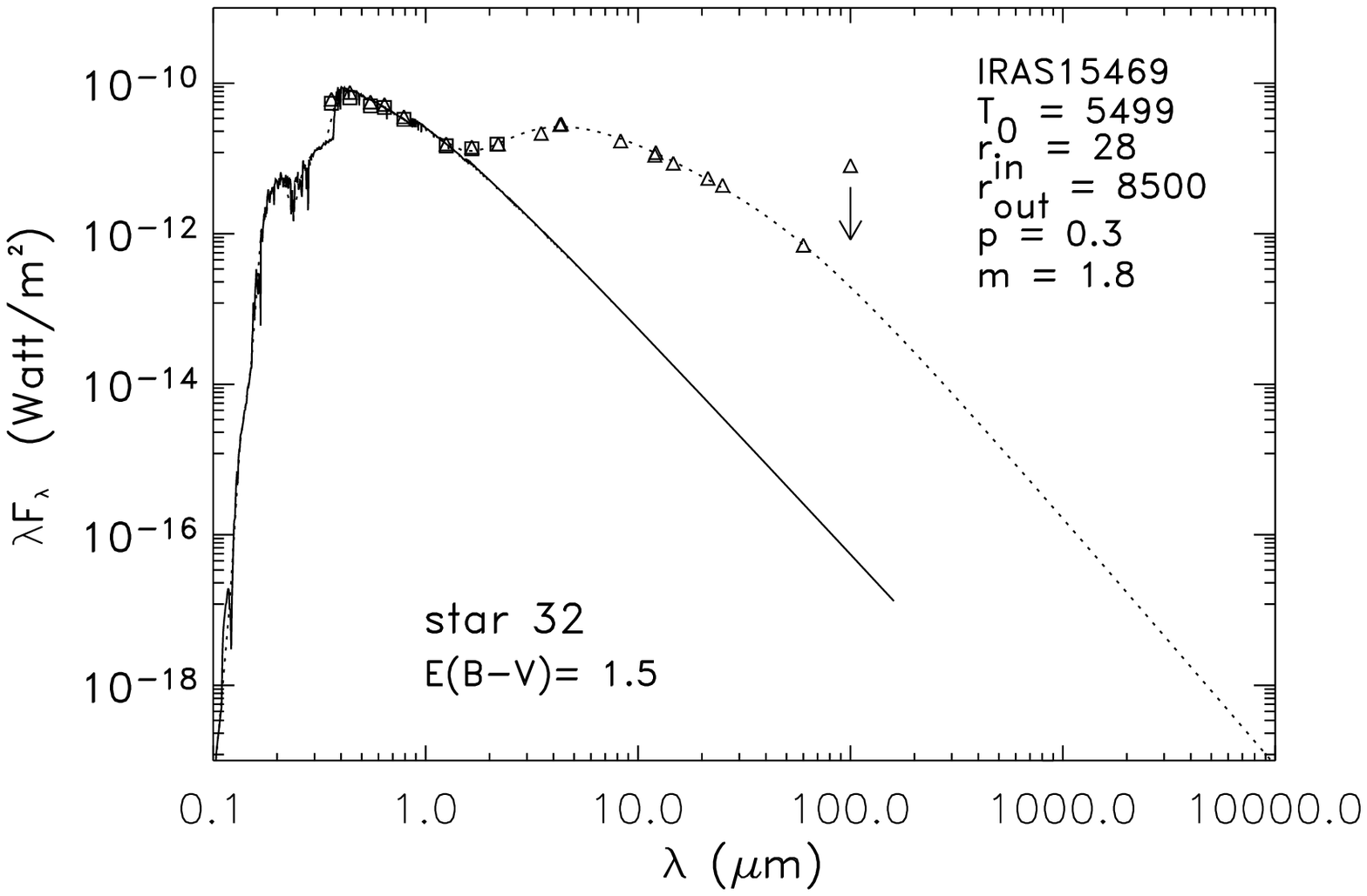}
\includegraphics[width=0.33\textwidth]{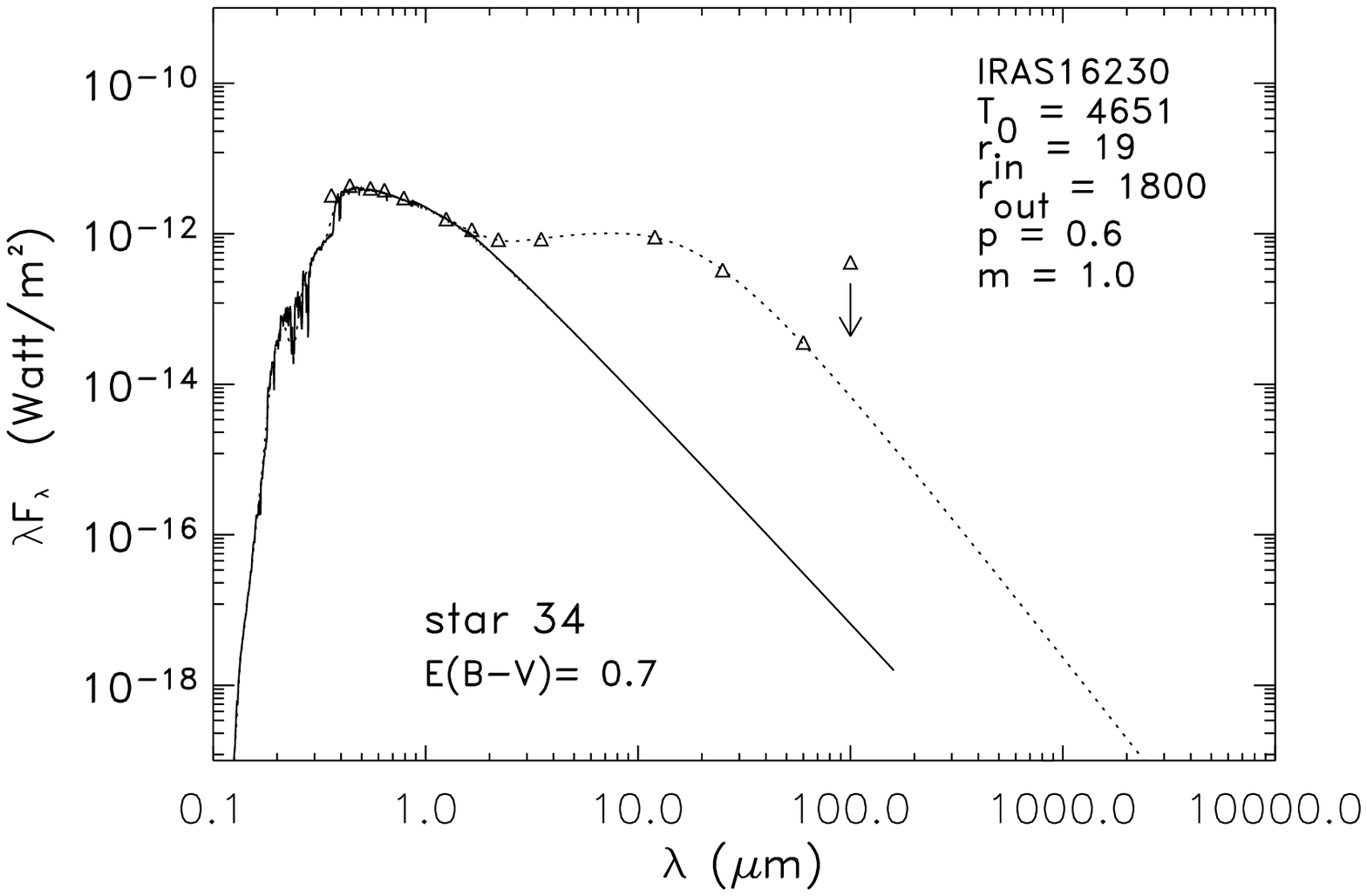}
\includegraphics[width=0.33\textwidth]{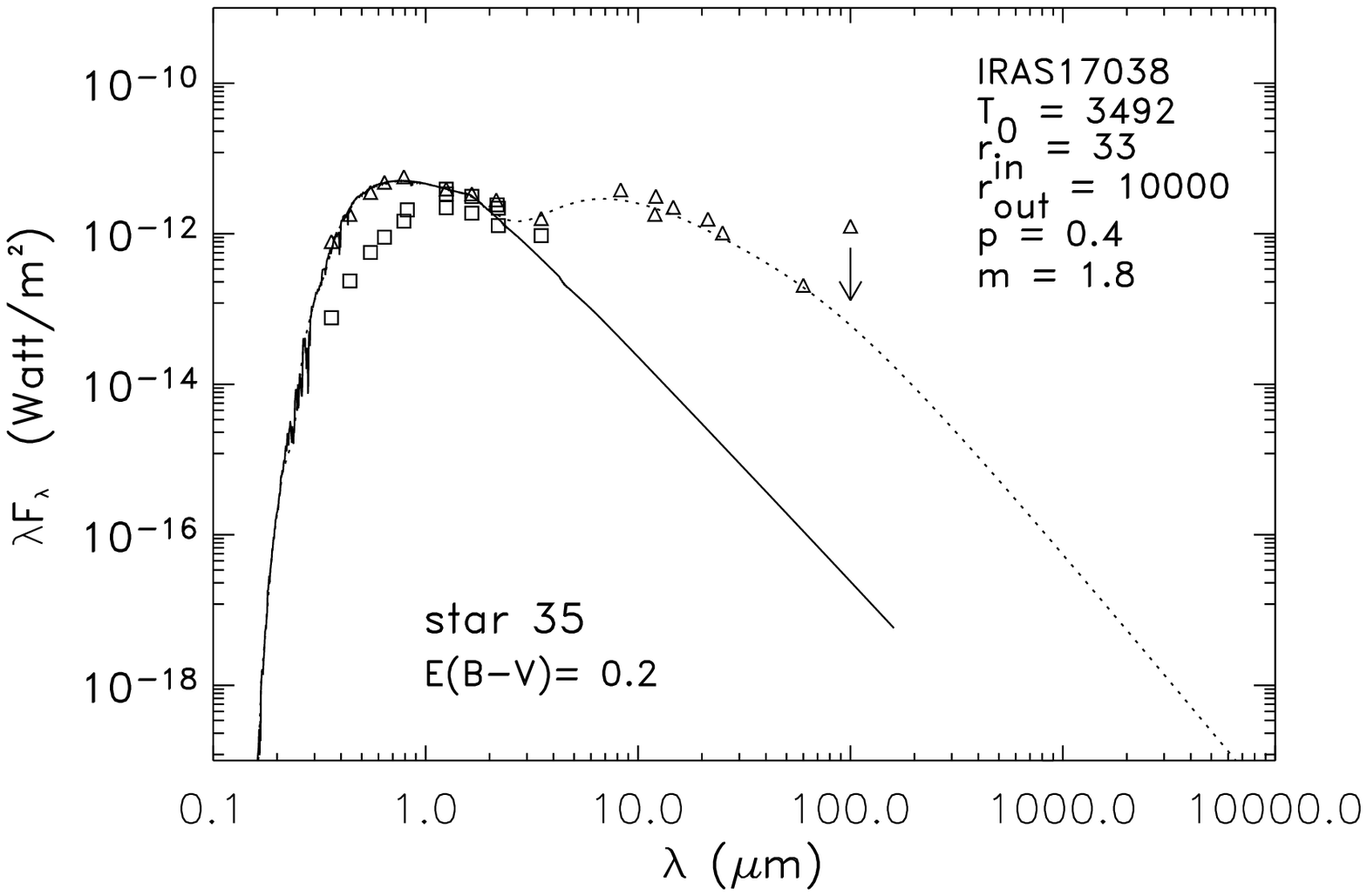}
\\
\includegraphics[width=0.33\textwidth]{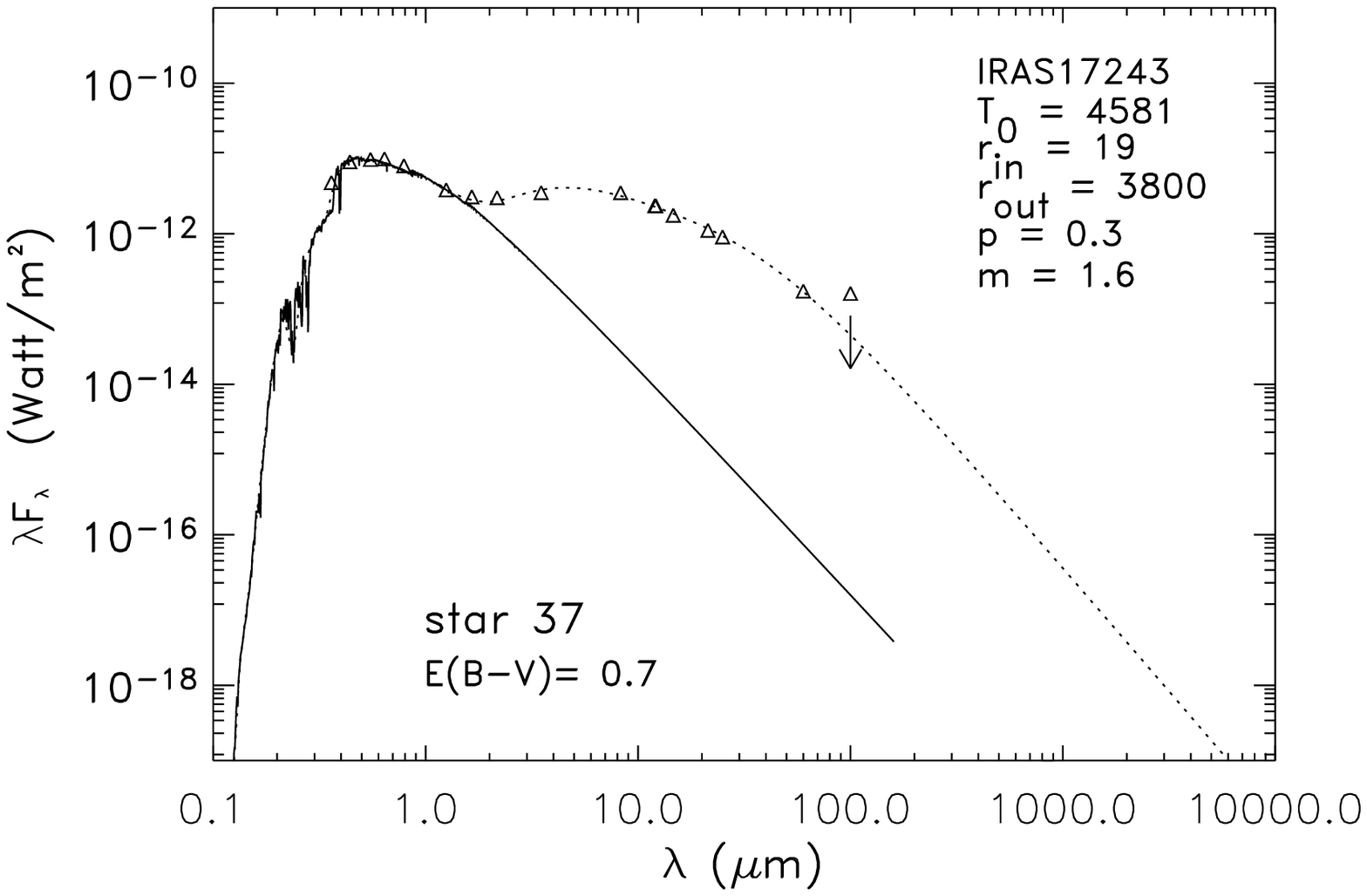}
\includegraphics[width=0.33\textwidth]{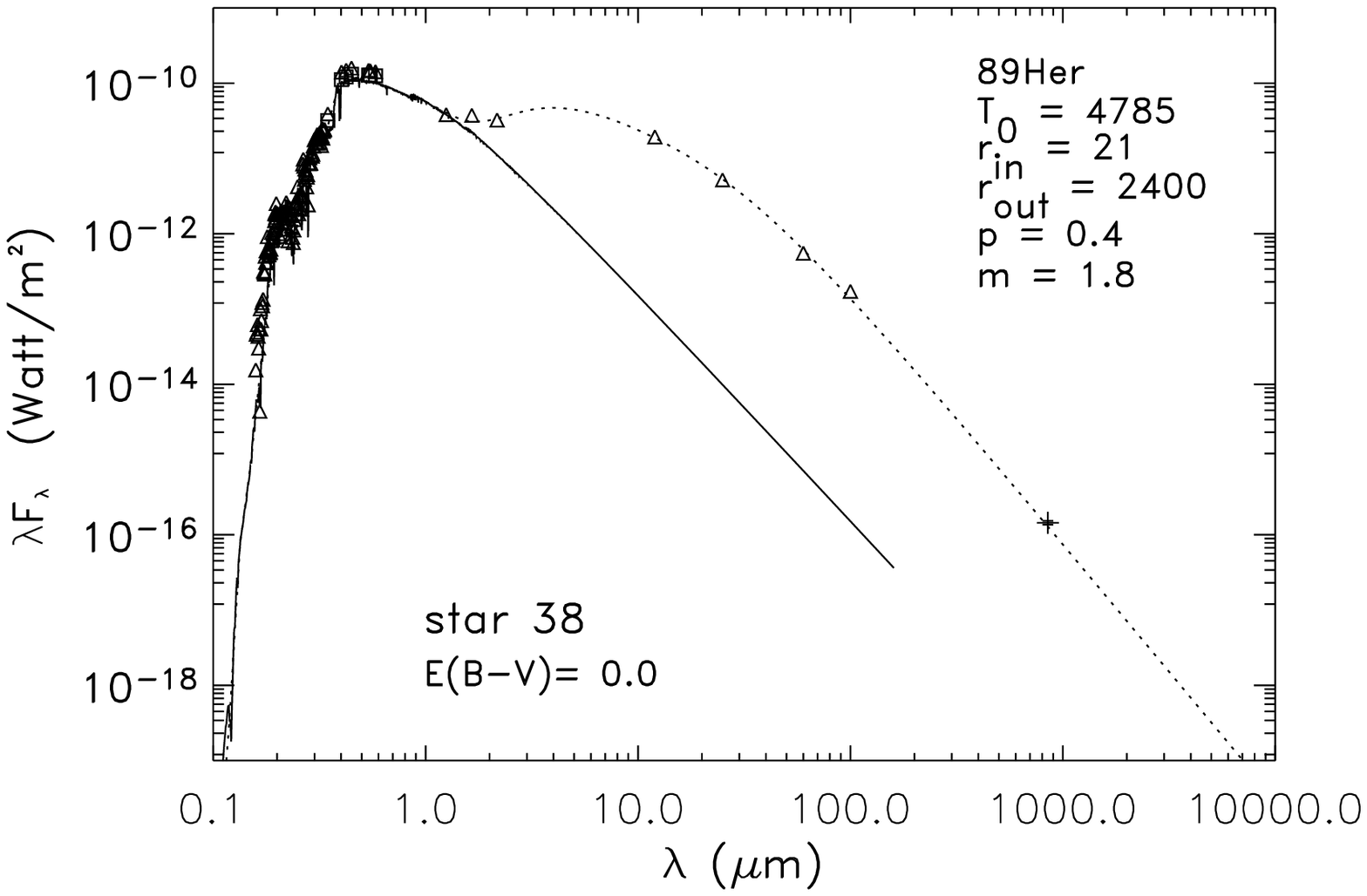}
\includegraphics[width=0.33\textwidth]{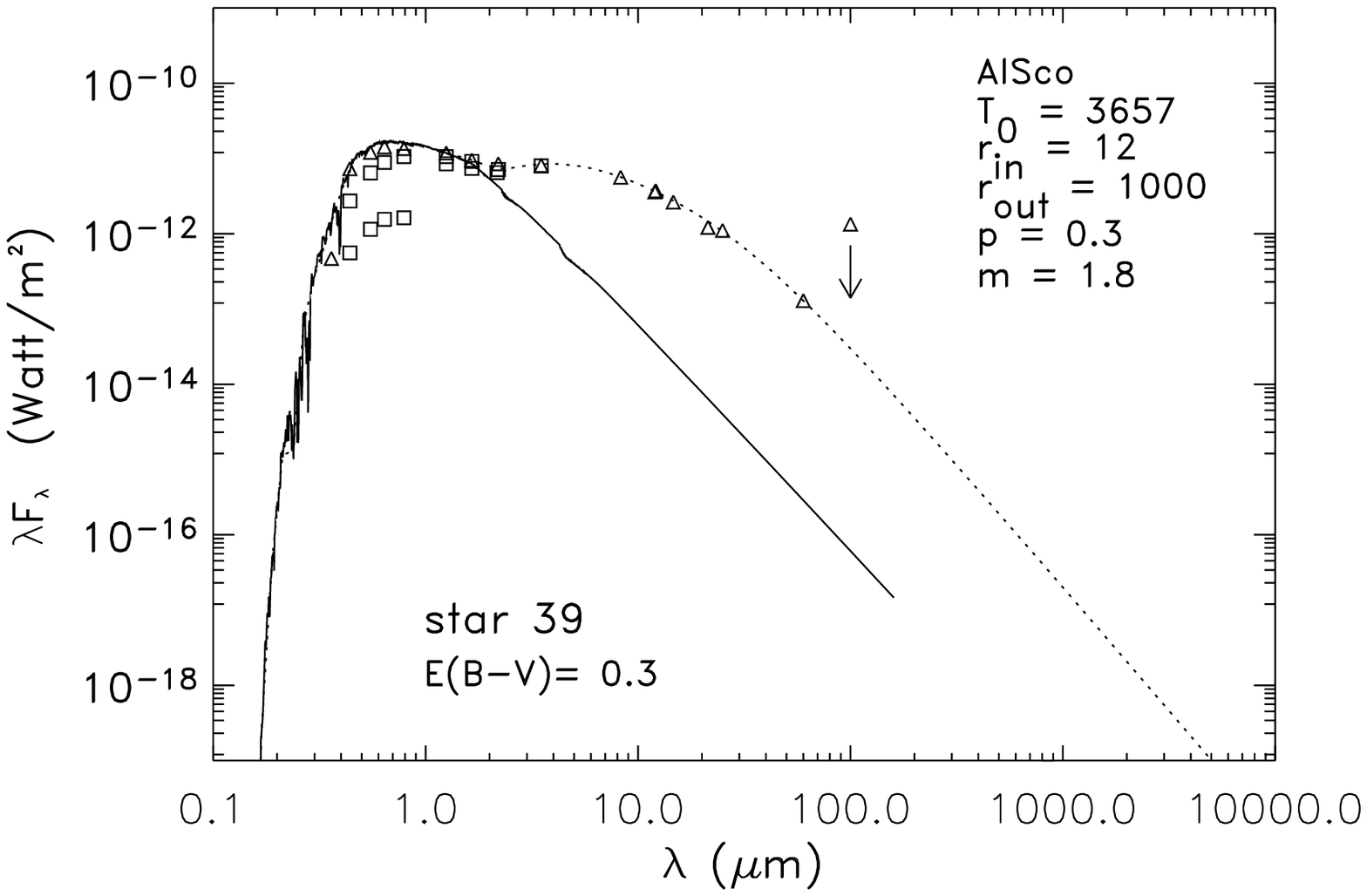}
\caption{The SEDs of all post-AGB objects from our sample. Continued.}
\end{figure*}

\addtocounter{figure}{-1}

\begin{figure*}
\includegraphics[width=0.33\textwidth]{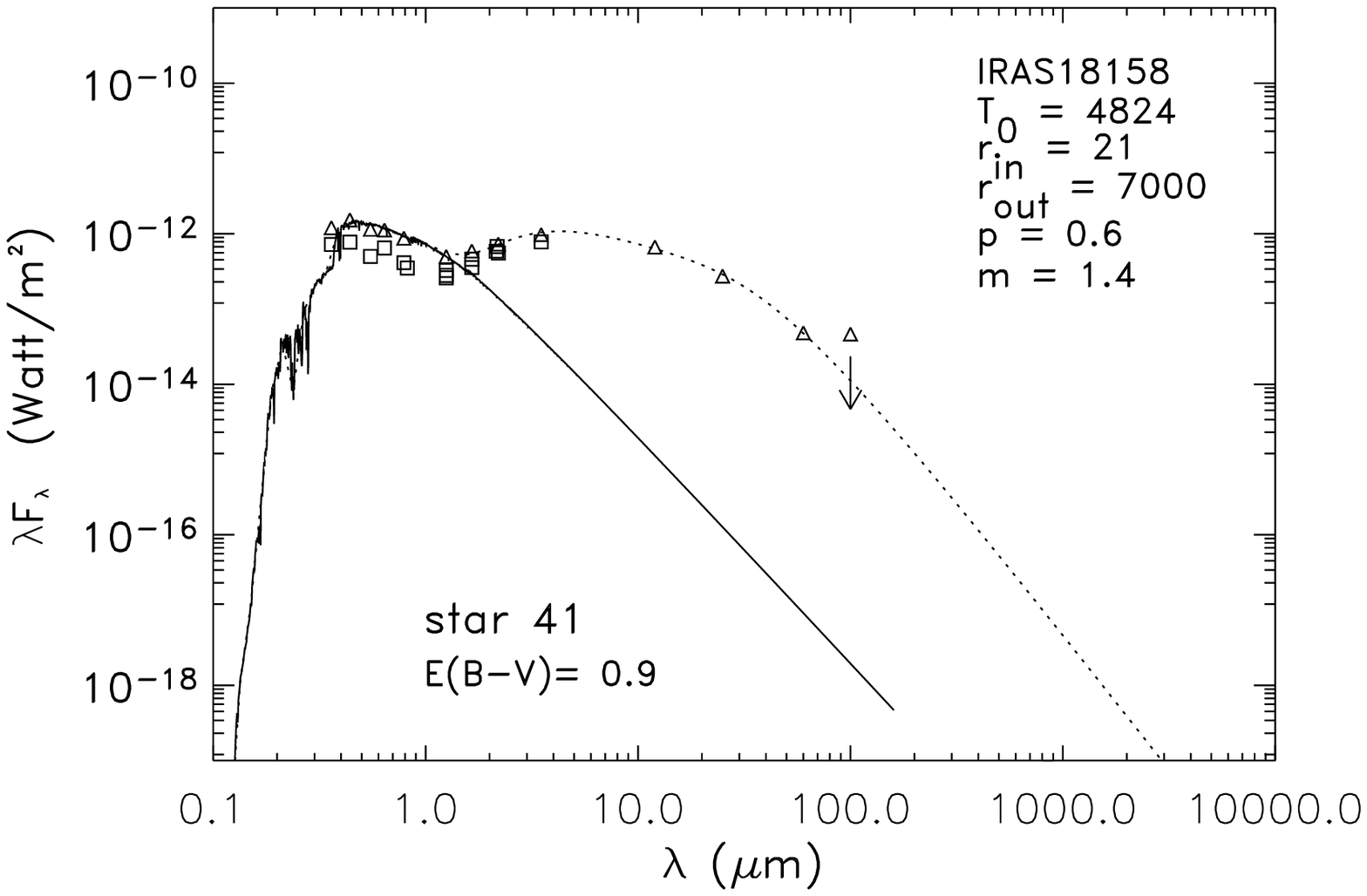}
\includegraphics[width=0.33\textwidth]{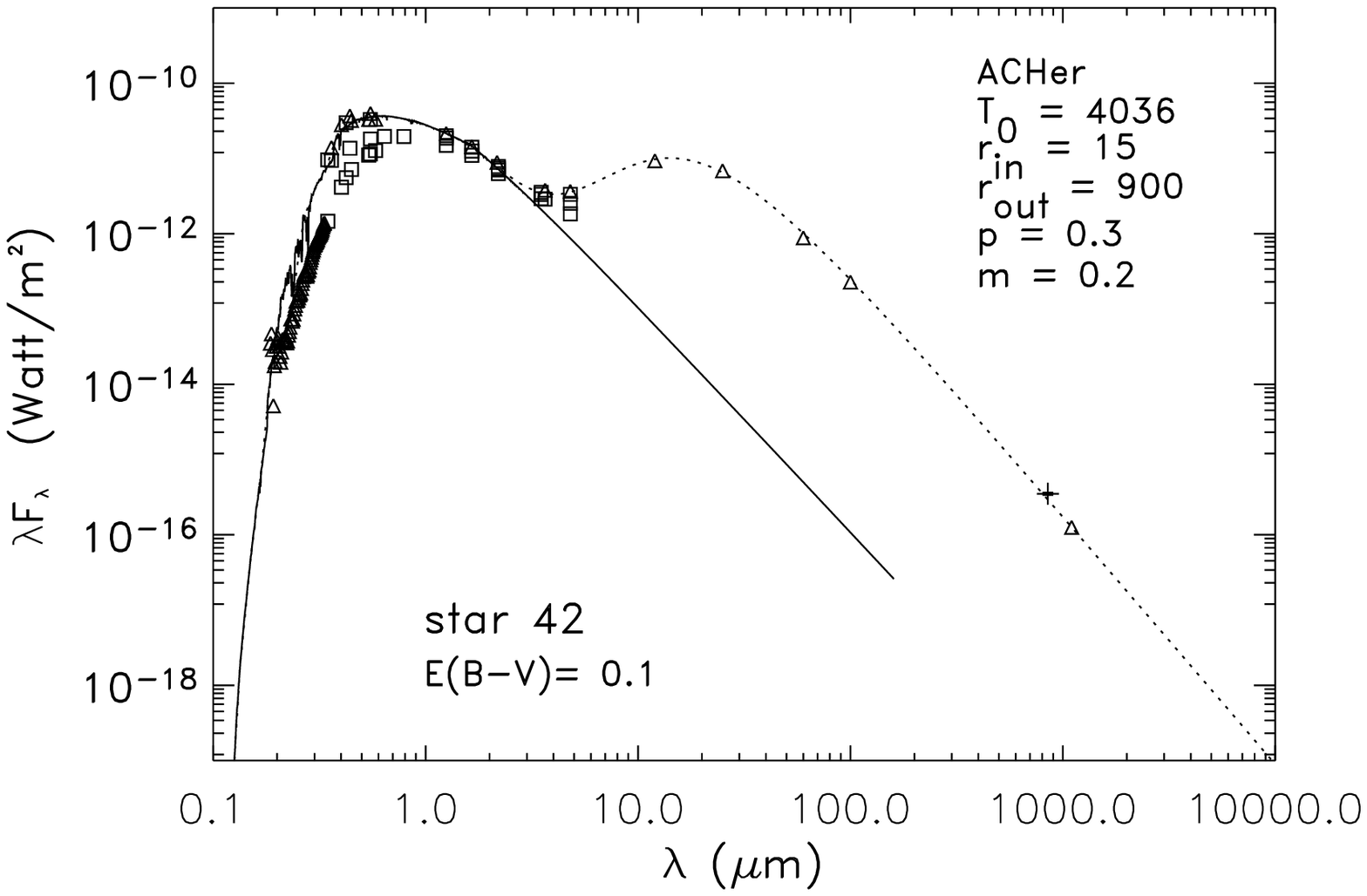}
\includegraphics[width=0.33\textwidth]{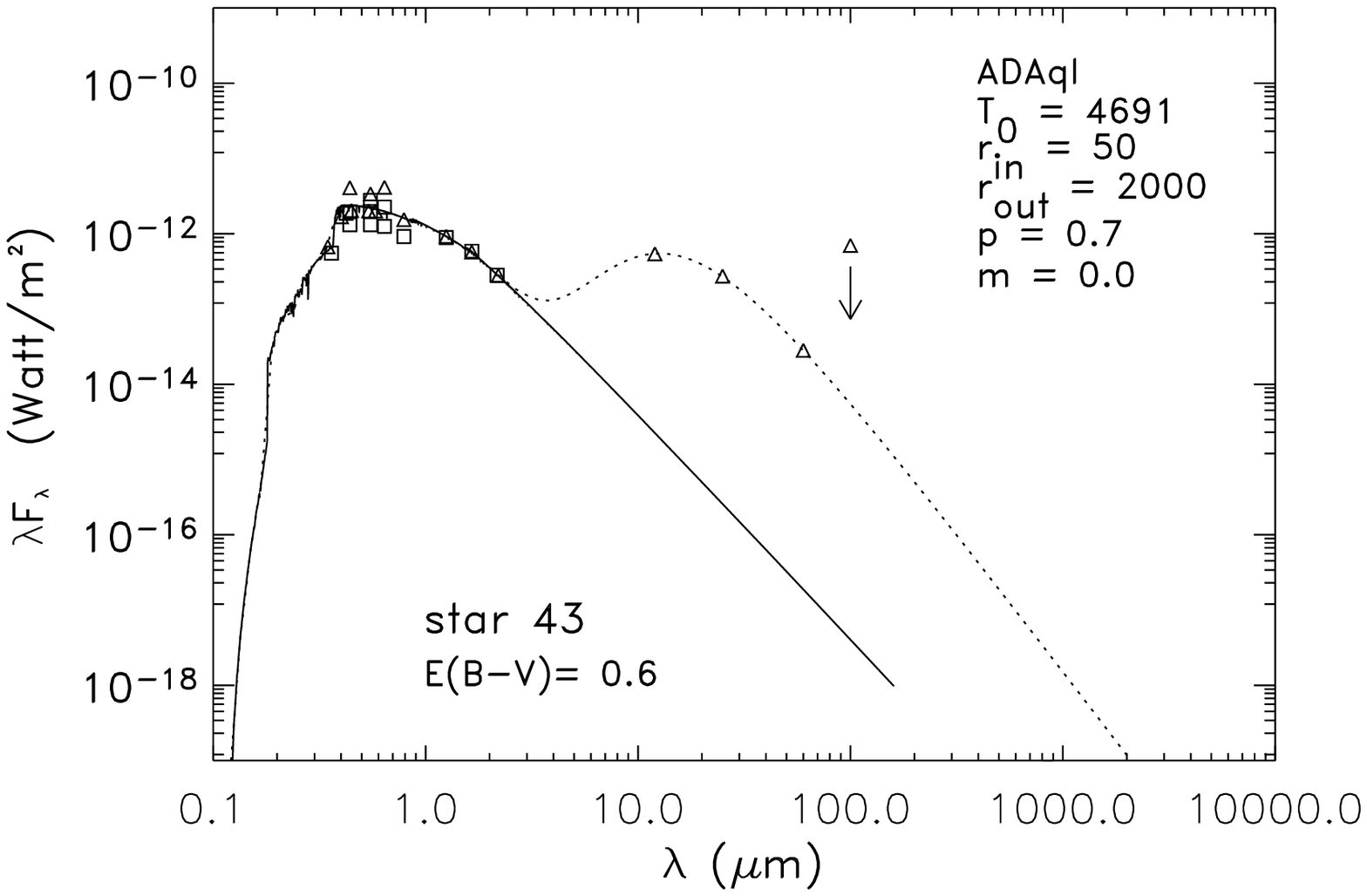}
\\
\includegraphics[width=0.33\textwidth]{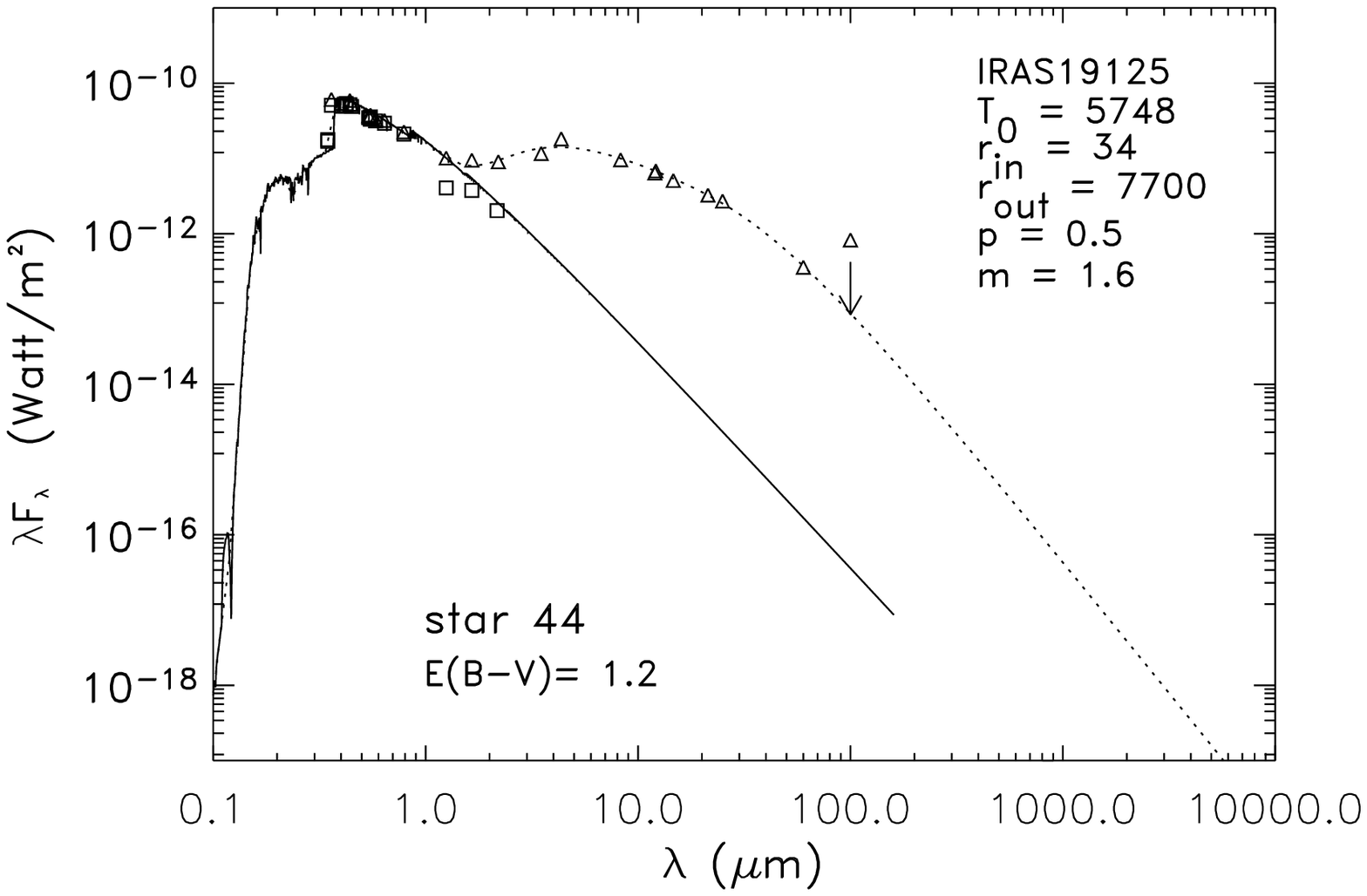}
\includegraphics[width=0.33\textwidth]{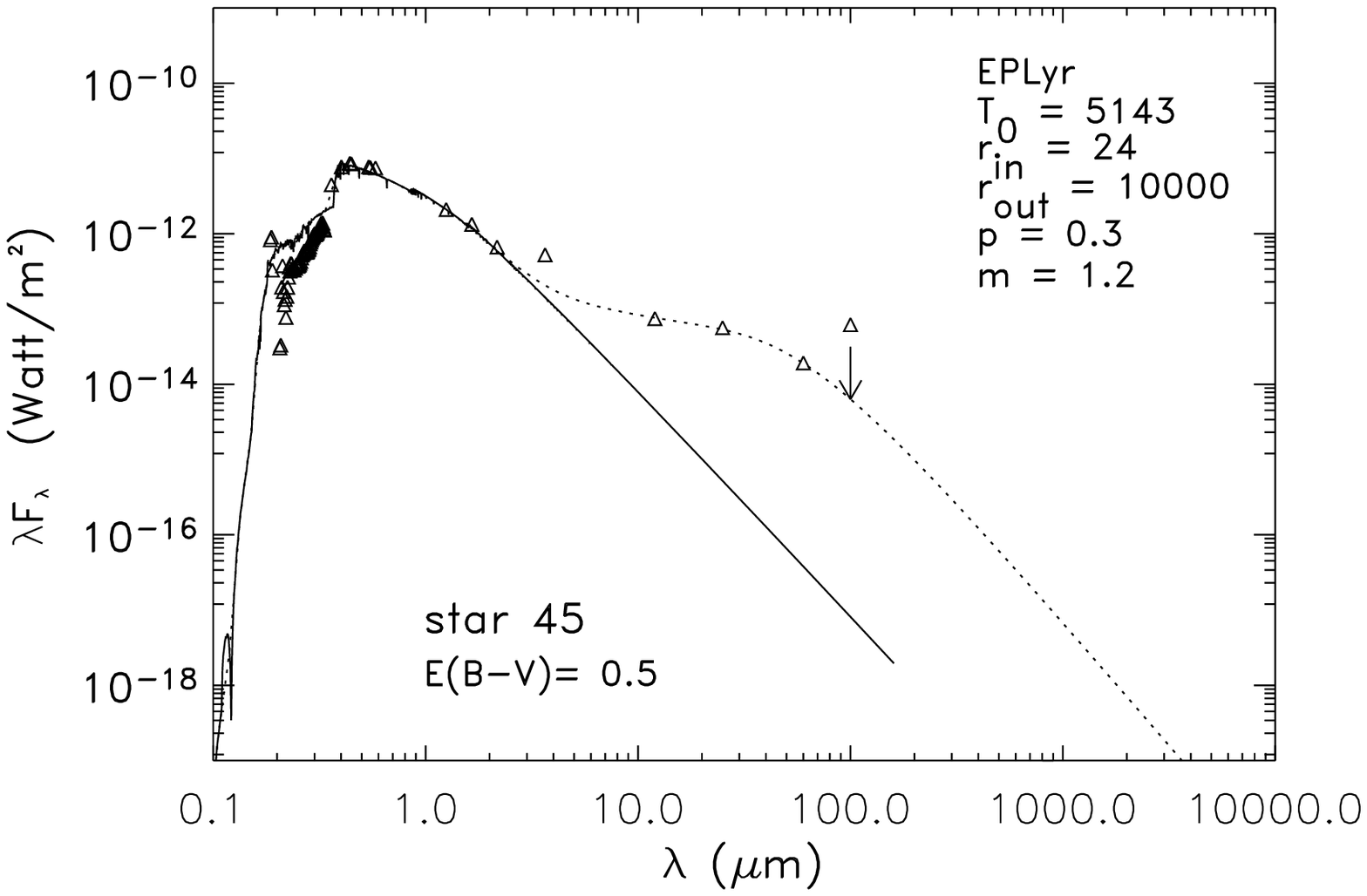}
\includegraphics[width=0.33\textwidth]{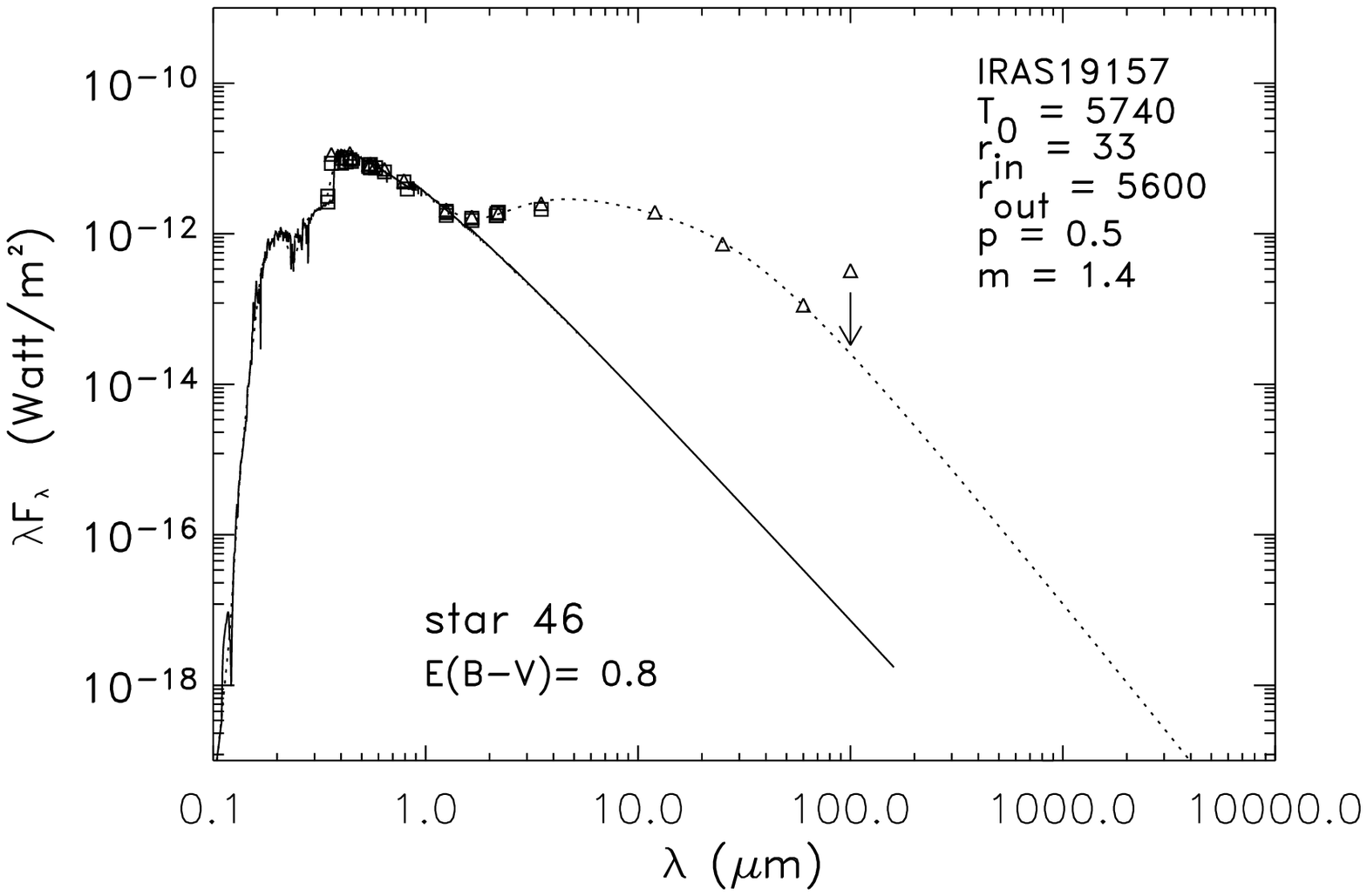}
\\
\includegraphics[width=0.33\textwidth]{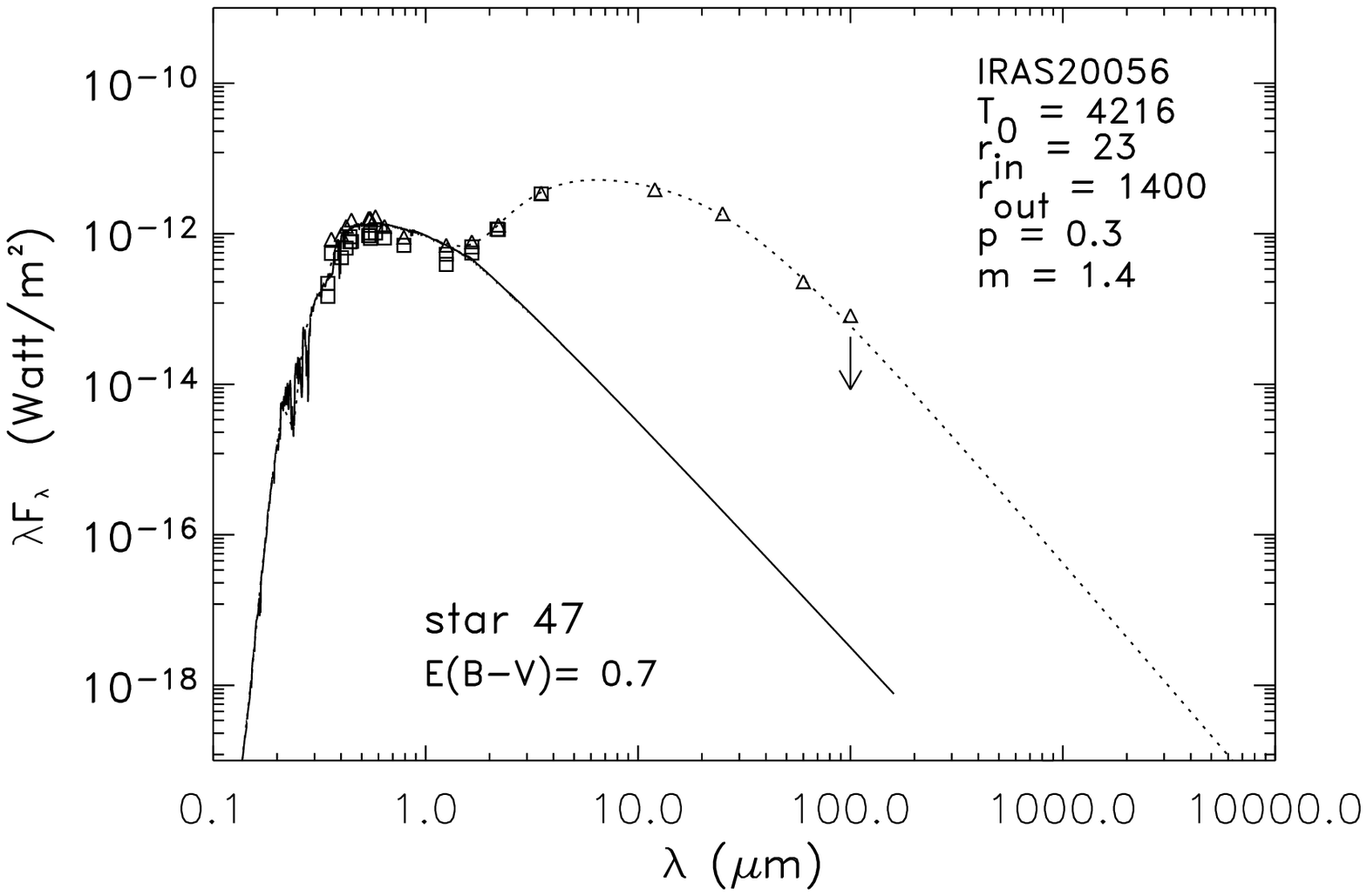}
\includegraphics[width=0.33\textwidth]{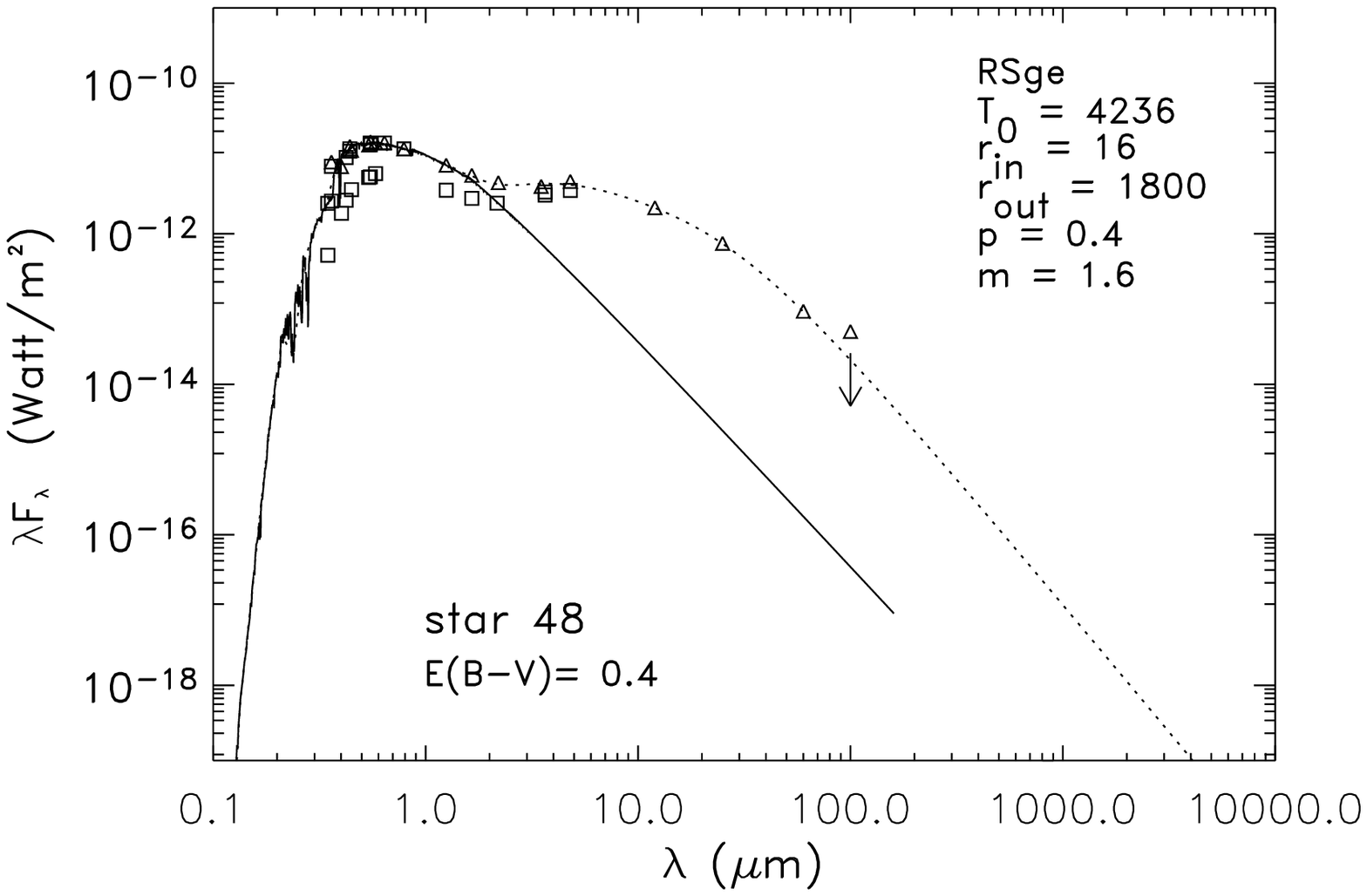}
\includegraphics[width=0.33\textwidth]{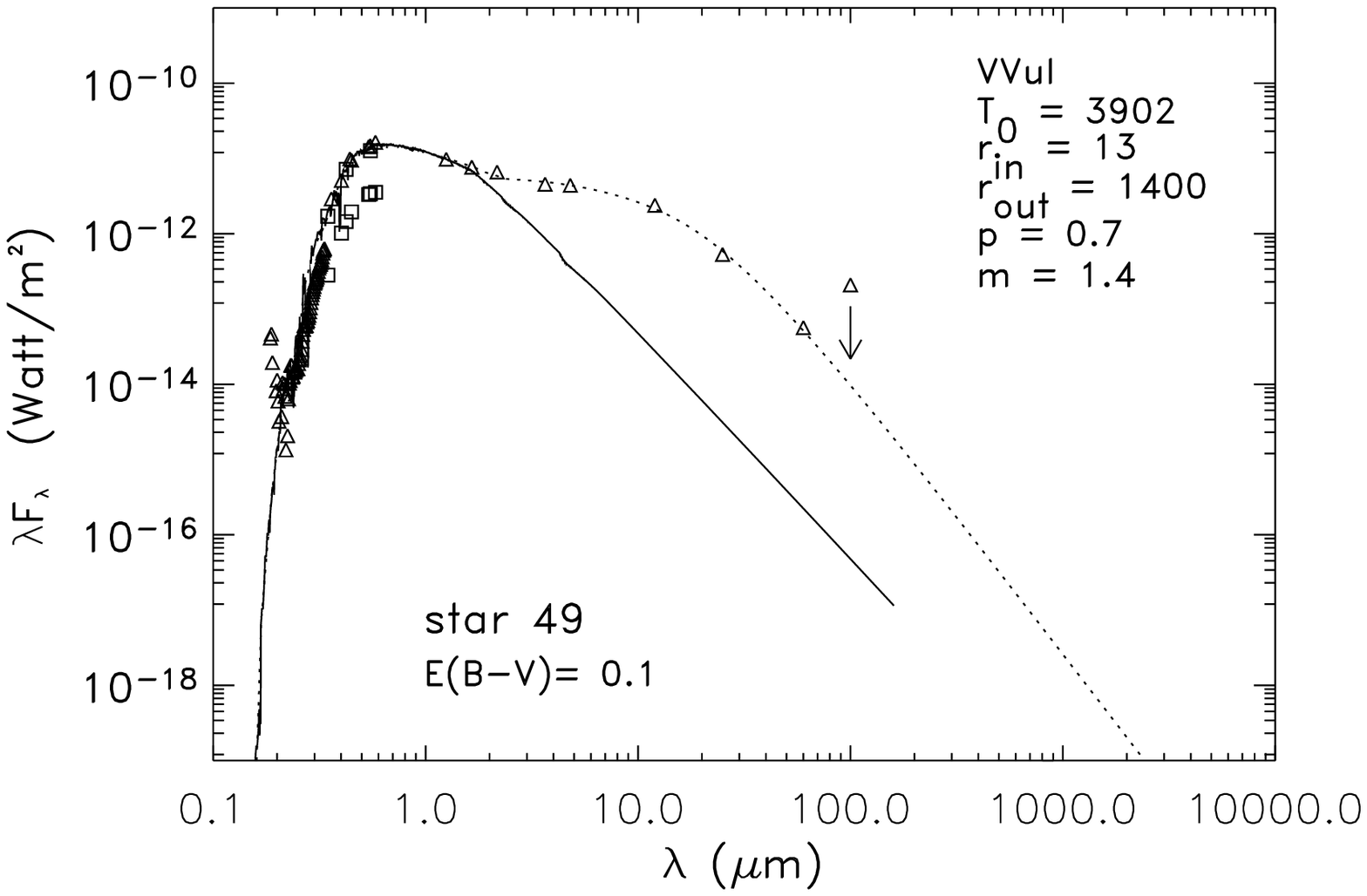}
\\
\includegraphics[width=0.33\textwidth]{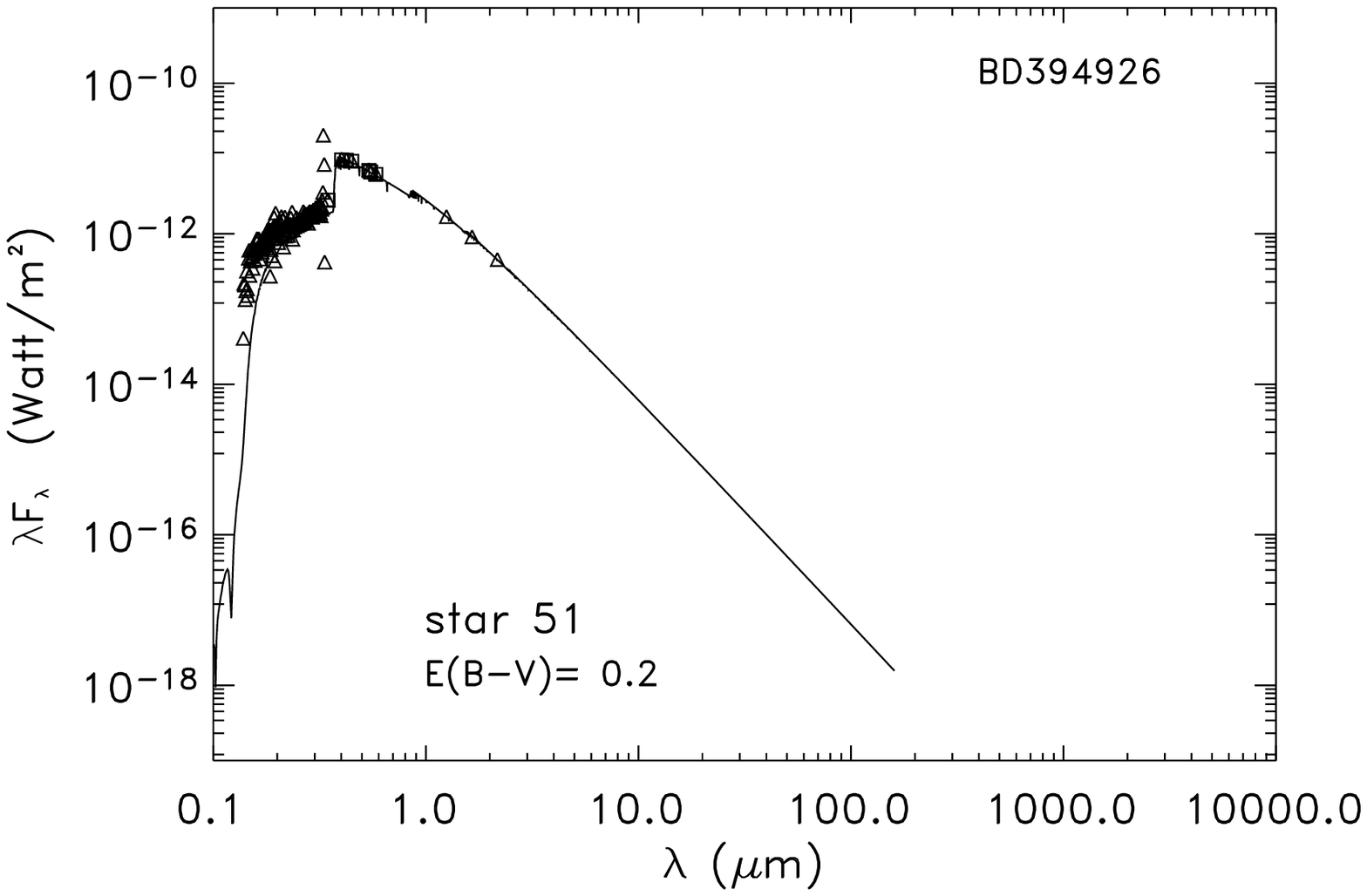}
\caption{The SEDs of all post-AGB objects from our sample. Continued.}
\end{figure*}

\clearpage
\section{Orbital Periods}
\label{app:Porbit}
\vspace{0.5cm}
\tablecaption{The orbital period is given. Note that we included further refinements of our own based on longer time series. The original reference is
given.\label{tab:Porbit}}
\begin{center}
\tablehead{
\hline
\hline
No	&Name			&\multicolumn{2}{c}{$P_{\mathrm{orbit}}$}	&$a \sin i$		&Reference\\
	&			&\multicolumn{2}{c}{(days)}			&AU			&\\
\hline}
\tabletail{\hline}
\begin{supertabular}{llr@{\,$\pm$\,}lll}
8	&HD\,44179		&$319$					&$2$	&$0.34$			&\citet{95VanWinckel}\\

9	&HD\,46703		&$600$					&$2$	&$0.92$			&\citet{93Waelkens}\\

10	&ST\,Pup		&$410.4$				&$2.9$	&$0.65$			&\citet{96Gonzalez}\\

11	&HD\,52961		&$1310$					&$8$	&$1.60$			&\citet{99VanWinckel}\\

12	&SAO\,173329		&$115.9$				&$1$	&$0.14$			&\citet{00VanWinckel}\\

13	&U\,Mon			&\multicolumn{2}{c}{$\pm\,2597$}		&$3.6$			&\citet{95Pollard}\\

15	&IRAS\,08544		&$503$					&$2$	&$0.38$			&\citet{03Maas}\\

21	&HR\,4049		&$430.7$				&$1$	&$0.60$			&\citet{91bWaelkens}\\

24	&HD\,95767		&\multicolumn{2}{c}{$\pm\,2050$}		&			&\citet{00VanWinckel}\\

28	&SX\,Cen		&$595$					&$7$	&$1.12$			&\citet{02Maas}\\

31	&EN\,TrA		&$1534$					&$21$	&$2.07$			&\citet{99VanWinckel}\\

38	&89\,Her		&$288.4$				&$4$	&$0.08$			&\citet{84Arellano}\\

42	&AC\,Her		&$1194$					&$6$	&$1.39$			&\citet{98VanWinckel}\\

50	&HD\,213985		&$259$					&$1$	&$0.78$			&\citet{00VanWinckel}\\

51	&BD$+$39$^{\circ}$4926	&$775$					&$5$	&$0.10$			&\citet{70Kodaira}\\
\end{supertabular}
\end{center}
\clearpage

\section{Data}
\label{app:data}
\vspace{0.5cm}
\tablecaption{Geneva data were acquired with the $70\,$cm Swiss
Telescope at La Silla (Chile) and with the Flemish Mercator Telescope
at La Palma (Spain), using the refurbished Geneva photometer P7
\citep{04Raskin}. Our total dataset was scanned for the maximum and
minimum magnitudes. Observation dates, number of measurements and
total timebases of these maxima and minima are given as
well. Additional data were found in the Geneva database: the General
Catalogue of Photometric Data (GCPD, 
\texttt{http://obswww.unige.ch/gcpd/gcpd.html}).\label{DATA:Geneva}}
\begin{footnotesize}
\begin{center}
\tablehead{
\hline
\hline
No	&Name			&JD		&Number		&Timebase	&$U$	&$B$	&$V$	&$B1$	&$B2$	&$V1$	&$G$	&Reference\\
	&			&2400000	&		&(days)		&	&	&	&	&	&	&	&\\
\hline}
\tabletail{\hline}
\begin{supertabular}{lllccllllllll}
1	&TW\,Cam		&52580.598	&11		&375		&12.907	&10.288	&9.490	&11.637	&11.424	&10.305 &10.360 &Mercator\\
	&			&52295.400	&		&		&13.000	&10.499	&9.716	&11.807	&11.642	&10.533	&10.579	&Mercator\\

2	&RV\,Tau		&52305.412	&37		&827		&12.511	&9.957	&9.036	&11.323	&11.077	&9.859	&9.886	&Mercator\\
	&			&52609.487	&		&		&14.867	&12.091	&10.822	&13.507	&13.130	&11.650	&11.655	&Mercator\\

4	&DY\,Ori		&52229.717	&27		&880		&14.242	&11.709	&11.229	&12.781	&13.019	&12.035	&12.119	&Mercator\\
	&			&52986.649	&		&		&15.447	&12.933	&12.078	&14.177	&14.167	&12.908	&12.921	&Mercator\\

5	&CT\,Ori		&53021.482	&31		&884		&12.119	&10.020	&10.174	&10.993	&11.397	&10.924	&11.175	&Mercator\\
	&			&52247.539	&		&		&13.454	&11.368	&10.983	&12.548	&12.637	&11.776	&11.914	&Mercator\\

6	&SU\,Gem		&52543.715	&18		&781		&14.301	&11.616	&10.697	&12.902	&12.808	&11.539	&11.540	&Mercator\\
	&			&52307.508	&		&		&16.610	&14.028	&13.248	&15.160	&15.386	&14.042	&14.057	&Mercator\\

8	&HD\,44179		&		&		&		&10.465	&8.362	&8.832	&9.285	&9.806	&9.578	&9.861	&GCPD\\
	&			&50403.809	&96		&2230		&10.398	&8.307	&8.777	&9.234	&9.751	&9.518	&9.815	&$70\,$cm Swiss\\
	&			&48705.546	&		&		&10.565	&8.470	&8.938	&9.390	&9.921	&9.684	&9.972	&$70\,$cm Swiss\\

9	&HD\,46703		&		&		&		&10.588	&8.668	&9.022	&9.633	&10.064	&9.742	&10.083	&GCPD\\
	&			&50100.497	&104		&6347		&10.402	&8.398	&8.855	&9.340	&9.812	&9.574	&9.927	&$70\,$cm Swiss\\
	&			&49300.642	&		&		&10.889	&8.998	&9.251	&9.995	&10.361	&9.989	&10.309	&$70\,$cm Swiss\\

11	&HD\,52961		&		&		&		&8.818	&6.914	&7.376	&7.845	&8.348	&8.105	&8.453	&GCPD\\
	&			&48575.657	&177		&5631		&8.667	&6.717	&7.237	&7.637	&8.161	&7.959	&8.326	&$70\,$cm Swiss\\
	&			&52612.675	&		&		&8.955	&7.080	&7.522	&8.013	&8.504	&8.253	&8.592	&Mercator\\

12	&SAO\,173329		&		&		&		&12.420	&10.320	&10.642	&11.297	&11.721	&11.382	&11.687	&GCPD\\
	&			&47894.684	&75		&2917		&12.392	&10.290	&10.609	&11.256	&11.674	&11.352	&11.653	&$70\,$cm Swiss\\
	&			&49846.473	&		&		&12.425	&10.377	&10.690	&11.339	&11.759	&11.427	&11.716	&$70\,$cm Swiss\\

13	&U\,Mon			&		&		&		&8.428	&6.330	&5.980	&7.596	&7.508	&6.752	&6.948	&GCPD\\
	&			&47545.766	&110		&1217		&7.759	&5.792	&5.575	&7.013	&6.990	&6.334	&6.570	&$70\,$cm Swiss\\
	&			&47853.822	&		&		&9.507	&7.484	&7.194	&8.697	&8.698	&7.941	&8.245	&$70\,$cm Swiss\\

14	&AR\,Pup		&		&		&		&11.415	&9.274	&9.226	&10.360	&10.568	&9.988	&10.207	&GCPD\\
	&			&43605.517	&2		&416		&11.313	&9.229	&9.179	&10.318	&10.511	&9.935	&10.165	&$70\,$cm Swiss\\
	&			&43189.707	&		&		&11.889	&9.603	&9.353	&10.772	&10.822	&10.125	&10.300	&$70\,$cm Swiss\\

21	&HR\,4049		&		&		&		&7.010	&4.896	&5.504	&5.808	&6.349	&6.221	&6.617	&GCPD\\
	&			&45271.870	&371		&6943		&6.570	&4.584	&5.284	&5.471	&6.056	&5.997	&6.421	&$70\,$cm Swiss\\
	&			&47233.683	&		&		&7.605	&5.373	&5.827	&6.317	&6.791	&6.550	&6.918	&$70\,$cm Swiss\\

23	&IRAS\,10456		&		&		&		&10.460	&7.480	&6.292	&9.144	&8.455	&7.130	&7.117	&GCPD\\

24	&HD\,95767		&		&		&		&10.865	&8.722	&8.782	&9.779	&10.048	&9.542	&9.767	&GCPD\\
	&			&49086.620	&102		&6148		&10.707	&8.576	&8.706	&9.601	&9.921	&9.461	&9.708	&$70\,$cm Swiss\\
	&			&47636.567	&		&		&10.941	&8.816	&8.840	&9.887	&10.125	&9.596	&9.808	&$70\,$cm Swiss\\

26	&IRAS\,11472		&53548.427	&1		&		&13.645	&11.619	&11.790	&12.622	&12.949	&12.533	&12.824	&Mercator\\

27	&RU\,Cen		&		&		&		&10.766	&8.950	&9.007	&10.000	&10.276	&9.761	&10.017	&GCPD\\
	&			&44778.506	&23		&751		&10.219	&8.429	&8.574	&9.441	&9.773	&9.320	&9.591	&$70\,$cm Swiss\\
	&			&44304.731	&		&		&12.135	&9.953	&9.567	&11.184	&11.150	&10.353	&10.538	&$70\,$cm Swiss\\

28	&SX\,Cen		&		&		&		&11.732	&9.735	&9.636	&10.858	&11.005	&10.400	&10.616	&GCPD\\
	&			&44778.526	&15		&751		&11.099	&9.101	&9.190	&10.183	&10.406	&9.943	&10.203	&$70\,$cm Swiss\\
	&			&44669.855	&		&		&12.912	&10.907	&10.545	&12.109	&12.132	&11.329	&11.472	&$70\,$cm Swiss\\

29	&HD\,108015		&		&		&		&9.433	&7.479	&7.934	&8.471	&8.864	&8.667	&9.006	&GCPD\\
	&			&49386.836	&111		&2867		&9.337	&7.378	&7.865	&8.356	&8.772	&8.592	&8.937	&$70\,$cm Swiss\\
	&			&50115.801	&		&		&9.541	&7.606	&8.020	&8.613	&8.975	&8.749	&9.082	&$70\,$cm Swiss\\

31	&EN\,TrA		&		&		&		&10.687	&8.648	&8.677	&9.754	&9.932	&9.435	&9.671	&GCPD\\
	&			&47636.730	&61		&2245		&10.478	&8.449	&8.477	&9.565	&9.728	&9.229	&9.470	&$70\,$cm Swiss\\
	&			&48133.582	&		&		&11.293	&9.181	&9.088	&10.350	&10.439	&9.852	&10.056	&$70\,$cm Swiss\\

38	&89\,Her		&		&		&		&6.860	&4.869	&5.425	&5.815	&6.296	&6.148	&6.526	&GCPD\\
	&			&41401.613	&14		&3507		&6.769	&4.779	&5.364	&5.714	&6.212	&6.080	&6.471	&$70\,$cm Swiss\\
	&			&41434.593	&		&		&6.987	&5.006	&5.513	&5.965	&6.419	&6.241	&6.609	&$70\,$cm Swiss\\

40	&IRAS\,18123		&52415.660	&31		&441		&14.010	&10.882	&9.945	&12.385	&11.927	&10.764	&10.805	&Mercator\\
	&			&52508.430	&		&		&14.719	&11.420	&10.267	&13.023	&12.428	&11.090	&11.119	&Mercator\\

42	&AC\,Her		&		&		&		&9.419	&7.450	&7.456	&8.536	&8.755	&8.221	&8.460	&GCPD\\
	&			&52367.729	&46		&5038		&8.726	&6.900	&7.261	&7.854	&8.311	&8.002	&8.310	&Mercator\\
	&			&52097.548	&		&		&10.770	&8.729	&8.408	&9.927	&9.941	&9.187	&9.354	&Mercator\\

43	&AD\,Aql		&52112.508	&1		&		&13.780	&11.789	&11.697	&12.853	&13.092	&12.460	&12.672	&Mercator\\

44	&IRAS\,19125		&52476.526	&33		&441		&12.822	&10.395	&10.214	&11.425	&11.707	&11.003	&11.181	&Mercator\\
	&			&52428.678	&		&		&12.875	&10.466	&10.289	&11.516	&11.784	&11.064	&11.237	&Mercator\\

45	&EP\,Lyr		&52491.424	&35		&481		&11.808	&9.818	&9.954	&10.806	&11.181	&10.707	&10.974	&Mercator\\
	&			&52136.422	&		&		&13.144	&11.134	&10.880	&12.267	&12.371	&11.648	&11.841	&Mercator\\

46	&IRAS\,19157		&52561.390	&4		&60		&12.968	&10.690	&10.688	&11.694	&12.058	&11.468	&11.659	&Mercator\\
	&			&52505.446	&		&		&13.163	&10.831	&10.775	&11.857	&12.177	&11.547	&11.752	&Mercator\\

47	&IRAS\,20056		&52580.350	&33		&444		&14.999	&12.594	&12.196	&13.828	&13.764	&12.985	&13.111	&Mercator\\
	&			&52854.611	&		&		&15.429	&12.943	&12.505	&14.213	&14.122	&13.274	&13.392	&Mercator\\

48	&R\,Sge			&52089.601	&41		&1164		&11.471	&9.183	&8.925	&10.413	&10.376	&9.700	&9.869	&Mercator\\
	&			&52428.658	&		&		&13.199	&10.603	&9.994	&11.965	&11.685	&10.777	&10.903	&Mercator\\

49	&V\,Vul			&52131.599	&39		&1167		&10.603	&8.449	&8.128	&9.702	&9.632	&8.909	&9.076	&Mercator\\
	&			&52419.684	&		&		&12.555	&10.187	&9.728	&11.454	&11.350	&10.494	&10.740	&Mercator\\

50	&HD\,213985		&		&		&		&9.975	&8.053	&8.827	&8.940	&9.536	&9.536	&9.968	&GCPD\\
	&			&47397.749	&156		&5871		&9.789	&7.904	&8.706	&8.781	&9.395	&9.420	&9.853	&$70\,$cm Swiss\\
	&			&50051.540	&		&		&10.618	&8.556	&9.209	&9.478	&10.008	&9.922	&10.327	&$70\,$cm Swiss\\
		
51	&BD$+$39$^{\circ}$4926	&		&		&		&10.536	&8.512	&9.263	&9.386	&10.009	&9.975	&10.399	&GCPD\\
	&			&53250.653	&70		&14211		&10.481	&8.507	&9.241	&9.380	&9.996	&9.952	&10.364	&Mercator\\
	&			&52561.442	&		&		&10.502	&8.528	&9.278	&9.408	&10.026	&9.979	&10.407	&Mercator\\
\end{supertabular}
\end{center}
\end{footnotesize}
\vspace{0.5cm}
\tablecaption{Ground-based optical ($UBVRI$) data, acquired over a
long period, found in the literature. If possible the maximum and
minimum magnitudes of a dataset are taken into
account.\label{DATA:optical}}
\begin{center}
\tablehead{
\hline
\hline
No	&Name			&JD		&$U$	&$B$	&$V$	&$R$	&$I$	&System		&Reference\\
	&			&2400000	&	&	&	&	&	&		&\\
\hline}
\tabletail{\hline}
\begin{supertabular*}{\textwidth}{llllllllll}
1	&TW\,Cam		&		&12.12	&10.94	&9.51	&	&	&Johnson	&\citet{79Dawson}\\

2	&RV\,Tau		&		&12.52	&10.93	&9.19	&	&	&Johnson	&\citet{79Dawson}\\

3	&IRAS\,05208		&52294.367	&11.132	&10.498	&9.352	&8.643	&7.947	&Cousins	&SAAO\\
	&			&52253.552	&11.185	&10.665	&9.608	&8.915	&8.139	&Cousins	&SAAO\\

4	&DY\,Ori		&		&13.45	&12.75	&11.48	&	&	&Johnson	&\citet{79Dawson}\\

5	&CT\,Ori		&		&11.69	&11.24	&10.31	&	&	&Johnson	&\citet{79Dawson}\\

6	&SU\,Gem		&		&12.88	&11.70	&10.19	&	&	&Johnson	&\citet{79Dawson}\\

7	&UY\,CMa		&		&11.36	&11.03	&10.40	&	&	&Johnson	&\citet{79Dawson}\\
	&			&		&12.28	&11.81	&11.00	&	&	&Johnson	&\citet{79Dawson}\\

10	&ST\,Pup		&48363.260	&9.890	&9.686	&9.362	&9.126	&8.835	&Cousins	&\citet{93Kilkenny}\\
	&			&48672.365	&12.058	&11.568	&10.661	&10.151	&9.642	&Cousins	&\citet{93Kilkenny}\\

13	&U\,Mon			&		&7.15	&6.59	&5.66	&	&	&Johnson	&\citet{79Dawson}\\
	&			&49258.126	&	&6.497	&5.436	&4.946	&4.511	&Cousins	&\citet{96Pollard}\\
	&			&48499.199	&	&8.614	&7.632	&7.088	&6.387	&Cousins	&\citet{96Pollard}\\

14	&AR\,Pup		&48622.972	&	&9.782	&9.093	&8.672	&8.216	&Cousins	&\citet{96Pollard}\\
	&			&48854.238	&	&10.823	&10.092	&9.638	&9.186	&Cousins	&\citet{96Pollard}\\

15	&IRAS\,08544		&51609.372	&11.660	&10.533	&9.017	&8.052	&7.024	&Cousins	&SAAO\\
	&			&51300.233	&11.982	&10.756	&9.192	&8.186	&7.139	&Cousins	&SAAO\\

16	&IRAS\,09060		&50957.232	&12.390	&11.817	&10.959	&10.383	&9.749	&Cousins	&SAAO\\
	&			&49399.455	&12.893	&12.305	&11.417	&10.839	&10.227	&Cousins	&SAAO\\

17	&IRAS\,09144		&51625.329	&	&	&13.857	&12.267	&10.777	&Cousins	&SAAO\\
	&			&51626.333	&	&15.886	&13.890	&12.293	&10.797	&Cousins	&SAAO\\

18	&IW\,Car		&48650.035	&	&8.657	&7.852	&7.379	&6.772	&Cousins	&\citet{96Pollard}\\
	&			&48911.056	&	&9.755	&8.758	&8.148	&7.454	&Cousins	&\citet{96Pollard}\\

20	&IRAS\,09538		&52030.329	&12.772	&12.083	&11.161	&10.623	&10.056	&Cousins	&SAAO\\
	&			&51576.506	&14.638	&13.733	&12.485	&11.691	&10.901	&Cousins	&SAAO\\

23	&IRAS\,10456		&49053.470	&9.829	&7.942	&6.246	&	&	&Cousins	&SAAO\\
	&			&49078.387	&9.869	&7.963	&6.267	&	&	&Cousins	&SAAO\\

27	&RU\,Cen		&46244.28	&11.31	&10.46	&9.43	&8.89	&8.38	&Cousins	&\citet{87Goldsmith}\\
	&			&46245.34	&11.44	&10.59	&9.53	&8.98	&8.47	&Cousins	&\citet{87Goldsmith}\\
	&			&48269.101	&	&9.158	&8.529	&8.178	&7.606	&Cousins	&\citet{96Pollard}\\
	&			&48319.041	&	&10.798	&9.811	&9.294	&8.688	&Cousins	&\citet{96Pollard}\\

28	&SX\,Cen		&46243.26	&10.85	&10.26	&9.43	&8.97	&8.53	&Cousins	&\citet{87Goldsmith}\\
	&			&46248.30	&11.36	&10.70	&9.77	&9.27	&8.81	&Cousins	&\citet{87Goldsmith}\\

29	&HD\,108015		&52734.477	&8.524	&8.225	&7.834	&7.598	&7.343	&Cousins	&SAAO\\
	&			&53139.350	&8.873	&8.506	&8.048	&7.771	&7.498	&Cousins	&SAAO\\

32	&IRAS\,15469		&49489.459	&12.985	&11.921	&10.493	&9.609	&8.676	&Cousins	&SAAO\\
	&			&49442.626	&13.063	&12.090	&10.622	&9.718	&8.769	&Cousins	&SAAO\\

34	&IRAS\,16230		&49585.254	&12.800	&12.102	&11.157	&10.585	&9.985	&Cousins	&SAAO\\
	&			&53129.518	&13.324	&12.760	&11.628	&10.915	&10.272	&Cousins	&SAAO\\

35	&IRAS\,17038		&52411.554	&12.229	&11.241	&9.922	&9.158	&8.420	&Cousins	&SAAO\\
	&			&51786.255	&14.756	&13.453	&11.916	&10.982	&9.895	&Cousins	&SAAO\\

36	&IRAS\,17233		&51663.588	&12.719	&12.312	&11.705	&11.287	&10.830	&Cousins	&SAAO\\
	&			&49590.265	&14.219	&13.564	&12.663	&12.172	&11.722	&Cousins	&SAAO\\

37	&IRAS\,17243		&52837.410	&12.378	&11.316	&10.200	&9.544	&8.919	&Cousins	&SAAO\\
	&			&52507.388	&13.479	&12.216	&10.915	&10.186	&9.507	&Cousins	&SAAO\\

39	&AI\,Sco		&46245.48	&13.21	&11.16	&9.55	&8.73	&7.92	&Cousins	&\citet{87Goldsmith}\\
	&			&49238.995	&	&10.088	&8.868	&8.215	&7.656	&Cousins	&\citet{96Pollard}\\
	&			&4823.095	&	&12.884	&11.422	&10.623	&9.954	&Cousins	&\citet{96Pollard}\\

40	&IRAS\,18123		&49537.405	&13.033	&11.335	&9.876	&9.146	&8.524	&Cousins	&SAAO\\
	&			&49970.295	&13.907	&12.062	&10.537	&9.781	&9.030	&Cousins	&SAAO\\

41	&IRAS\,18158		&48799.539	&14.720	&13.970	&13.070	&12.380	&11.670	&Cousins	&SAAO\\
	&			&49172.542	&14.857	&14.349	&13.699	&12.742	&12.312	&Cousins	&SAAO\\

42	&AC\,Her		&		&8.00	&7.66	&7.03	&	&	&Johnson	&\citet{79Dawson}\\
	&			&46248.47	&9.11	&8.68	&7.87	&7.40	&6.91	&Cousins	&\citet{87Goldsmith}\\

43	&AD\,Aql		&46253.53	&14.30	&12.61	&11.07	&10.26	&	&Cousins	&\citet{87Goldsmith}\\
	&			&48524.899	&	&11.811	&11.285	&10.917	&10.534	&Cousins	&\citet{96Pollard}\\
	&			&48152.985	&	&13.039	&12.097	&11.558	&11.069	&Cousins	&\citet{96Pollard}\\

44	&IRAS\,19125		&49563.415	&11.724	&11.097	&10.100	&9.438	&8.661	&Cousins	&SAAO\\
	&			&49510.555	&11.921	&11.237	&10.215	&9.532	&8.731	&Cousins	&SAAO\\

45	&EP\,Lyr		&		&10.92	&10.69	&10.12	&	&	&Johnson	&\citet{79Dawson}\\
		
46	&IRAS\,19157		&49207.444	&11.862	&11.398	&10.624	&10.126	&9.588	&Cousins	&SAAO\\
	&			&51761.419	&12.160	&11.605	&10.753	&10.213	&9.632	&Cousins	&SAAO\\

47	&IRAS\,20056		&46654.41	&14.25	&13.44	&12.37	&11.79	&11.29	&Cousins	&\citet{88Menzies}\\
	&			&46627.46	&14.31	&13.61	&12.55	&11.94	&11.39	&Cousins	&\citet{88Menzies}\\

48	&R\,Sge			&		&11.04	&9.93	&8.84	&	&	&Johnson	&\citet{79Dawson}\\
	&			&46242.57	&10.43	&9.71	&8.79	&8.31	&7.84	&Cousins   	&\citet{87Goldsmith}\\
	&			&46243.58	&10.57	&9.79	&8.83	&8.32	&7.84	&Cousins   	&\citet{87Goldsmith}\\

49	&V\,Vul			&		&9.70	&9.09	&8.27	&	&	&Johnson	&\citet{79Dawson}\\
\end{supertabular*}
\end{center}
\clearpage
\tablecaption{Ground-based near-infrared ($JHKLM$) photometry found in
the literature. If there was more than one measurement we used both
the maximum and minimum datapoints. For the data from 2\,MASS and
DENIS we made use of the catalogues found in VIZIER
(\texttt{http://vizier.u-strasbg.fr/viz-bin/VizieR}).\label{DATA:NIR}}
\begin{center}
\begin{footnotesize}
\tablehead{
\hline
\hline
No	&Name			&JD		&$I$	&$J$	&$H$	&$K$	&$L$	&$M$	&System		&Reference\\
	&			&2400000	&	&	&	&	&	&	&		&\\
\hline}
\tabletail{\hline}
\begin{supertabular*}{\textwidth}{lllllllllll}
1	&TW\,Cam		&		&	&	&	&	&4.2	&3.5	&Mt Lemmon	&\citet{72bGehrz}\\
	&			&		&	&	&	&	&	&	&Arizona system	&\\
	&			&42445.45	&	&7.04	&6.36	&5.70	&4.40	&	&Tenerife	&\citet{85LloydEvans}\\
	&			&		&	&7.035	&6.364	&5.750	&	&	&2\,MASS	&VIZIER\\

2	&RV\,Tau		&		&	&	&	&5.1	&3.3	&2.3	&Mt Lemmon	&\citet{72bGehrz}\\
	&			&		&	&	&	&	&	&	&Arizona system	&\\
	&			&42445.48	&	&6.39	&5.67	&5.03	&3.64	&	&Tenerife	&\citet{85LloydEvans}\\
	&			&		&	&6.183	&5.488	&4.777	&	&	&2\,MASS	&VIZIER\\

3	&IRAS\,05208		&48637.35	&	&7.05	&6.24	&6.05	&5.64	&	&SAAO		&T.~Lloyd Evans\\
	&			&49313.47	&	&7.20	&6.38	&6.13	&5.64	&	&SAAO		&T.~Lloyd Evans\\
	&			&		&8.705	&6.408	&	&5.953	&	&	&DENIS		&VIZIER\\
	&			&		&	&7.044	&6.314	&6.063	&	&	&2\,MASS	&VIZIER\\

4	&DY\,Ori		&43189.30	&	&8.10	&7.47	&7.02	&5.73	&	&SAAO		&\citet{85LloydEvans}\\
	&			&43158.44	&	&8.30	&7.40	&6.90	&5.80	&	&SAAO		&\citet{85LloydEvans}\\
	&			&		&	&8.073	&7.381	&6.983	&	&	&2\,MASS	&VIZIER\\

5	&CT\,Ori		&43184.37	&	&8.57	&7.82	&8.00	&6.48	&	&SAAO		&\citet{85LloydEvans}\\
	&			&43158.38	&	&8.66	&7.92	&7.50	&6.09	&	&SAAO		&\citet{85LloydEvans}\\
	&			&		&	&8.404	&7.931	&7.607	&	&	&2\,MASS	&VIZIER\\

6	&SU\,Gem		&		&	&	&	&	&4.1	&3.2	&Mt Lemmon	&\citet{72bGehrz}\\
	&			&		&	&	&	&	&	&	&Arizona system	&\\
	&			&43160.37	&	&8.36	&7.38	&6.47	&4.94	&	&SAAO		&\citet{85LloydEvans}\\
	&			&43188.32	&	&8.97	&8.38	&6.95	&4.96	&	&SAAO		&\citet{85LloydEvans}\\
	&			&		&	&10.167	&9.033	&7.584	&	&	&2\,MASS	&VIZIER\\

7	&UY\,CMa		&46107.32	&	&9.35	&8.82	&8.22	&6.62	&	&SAAO		&\citet{85LloydEvans}\\
	&			&43159.36	&	&9.42	&8.46	&8.41	&6.87	&	&SAAO		&\citet{85LloydEvans}\\
	&			&		&9.845	&9.093	&	&8.038	&	&	&DENIS		&VIZIER\\
	&			&		&	&9.042	&8.597	&7.986	&	&	&2\,MASS	&VIZIER\\

8	&HD\,44179		&		&	&6.59	&4.95	&3.19	&1.22	&	&ESO		&C.~Waelkens\\
	&			&		&8.981	&6.280	&3.807	&	&	&	&DENIS		&VIZIER\\
	&			&		&	&6.577	&5.145	&3.655	&	&	&2\,MASS	&VIZIER\\

9	&HD\,46703		&		&	&7.805	&7.513	&7.435	&	&	&2MASS		&VIZIER\\

10	&ST\,Pup		&43159.44	&	&8.38	&8.03	&7.67	&6.42	&	&SAAO		&\citet{85LloydEvans}\\
	&			&		&9.292	&8.437	&	&7.839	&	&	&DENIS		&VIZIER\\
	&			&		&	&8.506	&8.250	&8.082	&	&	&2\,MASS	&VIZIER\\

11	&HD\,52961		&		&	&6.36	&5.98	&5.54	&4.25	&3.97	&ESO		&C.~Waelkens\\
	&			&		&	&6.323	&5.978	&5.526	&	&	&2\,MASS	&VIZIER\\

12	&SAO\,173329		&		&	&9.42	&8.96	&8.27	&6.49	&5.67	&ESO		&C.~Waelkens\\
	&			&		&	&9.375	&8.917	&8.308	&	&	&2\,MASS	&VIZIER\\

13	&U\,Mon			&		&	&	&	&	&2.4	&1.5	&Kitt Peak NO	&\citet{70Gehrz}\\
	&			&		&	&	&	&	&	&	&Arizona system	&\\
	&			&		&	&	&	&3.7	&2.7	&1.7	&Mt Lemmon	&\citet{72bGehrz}\\
	&			&		&	&	&	&	&	&	&Arizona system	&\\
	&			&42684.65	&	&4.01	&3.65	&3.25	&2.40	&	&SAAO		&\citet{85LloydEvans}\\
	&			&43160.45	&	&4.88	&4.43	&4.08	&3.08	&	&SAAO		&\citet{85LloydEvans}\\
	&			&		&	&4.35	&3.93	&3.60	&2.30	&1.61	&ESO		&C.~Waelkens\\
	&			&		&	&4.925	&4.269	&4.042	&	&	&2\,MASS	&VIZIER\\

14	&AR\,Pup		&43184.46	&	&6.12	&4.95	&3.67	&1.68	&	&SAAO		&\citet{85LloydEvans}\\
	&			&42737.58	&	&7.20	&5.98	&4.40	&2.08	&	&SAAO		&\citet{85LloydEvans}\\
	&			&		&8.937	&6.931	&	&3.682	&	&	&DENIS		&VIZIER\\
	&			&		&	&7.891	&6.824	&5.285	&	&	&2\,MASS	&VIZIER\\

15	&IRAS\,08544		&		&	&5.65	&4.67	&3.51	&1.59	&	&SAAO		&T.~Lloyd Evans (mean)\\
	&			&		&9.043	&5.741	&	&3.813	&	&	&DENIS		&VIZIER\\
	&			&		&	&5.575	&4.743	&3.523	&	&	&2\,MASS	&VIZIER\\

16	&IRAS\,09060		&50263.20	&	&8.60	&7.63	&6.68	&5.16	&	&SAAO		&T.~Lloyd Evans\\
	&			&50511.38	&	&9.27	&8.13	&6.99	&5.35	&	&SAAO		&T.~Lloyd Evans\\
	&			&		&	&9.130	&8.060	&6.949	&	&	&2\,MASS	&VIZIER\\

17	&IRAS\,09144		&49125.31	&	&8.33	&7.12	&6.00	&4.23	&	&SAAO		&T.~Lloyd Evans\\
	&			&49475.28	&	&8.55	&7.35	&6.21	&4.31	&	&SAAO		&T.~Lloyd Evans\\
	&			&		&10.801	&8.305	&	&6.235	&	&	&DENIS		&VIZIER\\
	&			&		&	&8.372	&7.391	&6.369	&	&	&2\,MASS	&VIZIER\\

18	&IW\,Car		&43184.48	&	&5.86	&5.18	&4.24	&2.41	&	&SAAO		&\citet{85LloydEvans}\\
	&			&43155.56	&	&5.94	&5.20	&4.31	&2.44	&	&SAAO		&\citet{85LloydEvans}\\
	&			&		&	&5.875	&5.151	&4.367	&	&	&2\,MASS	&VIZIER	\\

19	&IRAS\,09400		&49122.29	&	&6.57	&5.58	&5.16	&4.33	&	&SAAO		&T.~Lloyd Evans\\
	&			&50511.40	&	&6.61	&5.59	&5.20	&4.35	&	&SAAO		&T.~Lloyd Evans\\
	&			&		&	&6.547	&5.626	&5.159	&	&	&2\,MASS	&VIZIER\\

20	&IRAS\,09538		&49741.57	&	&9.18	&8.49	&7.61	&6.09	&	&SAAO		&T.~Lloyd Evans\\
	&			&49031.52	&	&9.62	&8.82	&8.02	&6.36	&	&SAAO		&T.~Lloyd Evans\\
	&			&		&10.530	&9.278	&	&7.703	&	&	&DENIS		&VIZIER\\
	&			&		&	&9.402	&8.574	&7.721	&	&	&2\,MASS	&VIZIER\\

21	&HR\,4049		&		&	&4.73	&4.16	&3.13	&1.39	&0.91	&ESO		&C.~Waelkens\\
	&			&		&	&	&	&3.677	&	&	&DENIS		&VIZIER\\
	&			&		&	&4.921	&4.225	&3.208	&	&	&2\,MASS	&VIZIER\\

22	&IRAS\,10174		&49474.35	&	&5.20	&4.07	&3.54	&2.87	&	&SAAO		&T.~Lloyd Evans\\
	&			&50262.22	&	&5.43	&4.29	&3.74	&3.04	&	&SAAO		&T.~Lloyd Evans\\
	&			&		&8.644	&5.583	&	&3.481	&	&	&DENIS		&VIZIER\\
	&			&		&	&5.313	&4.490	&3.931	&	&	&2\,MASS	&VIZIER\\

23	&IRAS\,10456		&		&	&3.45	&2.55	&2.15	&1.11	&	&SAAO		&T.~Lloyd Evans (mean)\\
	&			&		&	&3.354	&2.469	&2.045	&	&	&2\,MASS	&VIZIER\\

24	&HD\,95767		&		&	&7.05	&6.48	&5.59	&3.50	&2.80	&ESO		&C.~Waelkens\\
	&			&		&8.786	&7.313	&	&4.964	&	&	&DENIS		&VIZIER\\
	&			&		&	&7.098	&6.532	&5.636	&	&	&2\,MASS	&VIZIER\\

25	&GK\,Car		&43184.56	&	&9.05	&8.26	&8.14	&6.42	&	&SAAO		&\citet{85LloydEvans}\\
	&			&43159.52	&	&9.27	&8.46	&7.87	&6.64	&	&SAAO		&\citet{85LloydEvans}\\
	&			&		&10.150	&9.123	&	&7.970	&	&	&DENIS		&VIZIER\\
	&			&		&	&9.328	&8.716	&8.171	&	&	&2\,MASS	&VIZIER	\\

26	&IRAS\,11472		&		&	&9.657	&9.047	&8.630	&	&	&2\,MASS	&VIZIER\\

27	&RU\,Cen		&43188.51	&	&7.08	&6.65	&6.47	&6.53	&	&SAAO		&\citet{85LloydEvans}\\
	&			&43158.58	&	&7.14	&6.68	&6.39	&5.63	&	&SAAO		&\citet{85LloydEvans}\\
	&			&46243.25	&	&7.56	&7.04	&6.83	&6.33	&5.60	&SAAO		&\citet{87Goldsmith}\\
	&			&46248.28	&	&8.00	&7.44	&7.20	&	&	&SAAO		&\citet{87Goldsmith}\\
	&			&		&	&7.45	&7.02	&6.73	&6.08	&5.47	&ESO		&C.~Waelkens\\
	&			&		&	&7.68	&7.15	&6.86	&6.34	&5.80	&ESO		&\citet{97Garcia-Lario}\\
	&			&		&8.600	&7.151	&	&6.654	&	&	&DENIS		&VIZIER\\
	&			&		&	&7.616	&7.186	&6.918	&	&	&2\,MASS	&VIZIER\\

28	&SX\,Cen		&43159.57	&	&7.94	&7.63	&6.88	&5.74	&	&SAAO		&\citet{85LloydEvans}\\
	&			&46243.26	&	&7.86	&7.25	&6.62	&5.20	&4.30	&SAAO		&\citet{87Goldsmith}\\
	&			&46248.30	&	&8.12	&7.51	&7.00	&5.65	&4.73	&SAAO		&\citet{87Goldsmith}\\
	&			&		&8.962	&7.660	&	&6.448	&	&	&DENIS		&VIZIER\\
	&			&		&	&7.875	&7.443	&6.759	&	&	&2\,MASS	&VIZIER\\

29	&HD\,108015		&		&	&7.03	&6.39	&5.32	&3.50	&	&SAAO		&T.~Lloyd Evans (mean)\\
	&			&		&	&7.03	&6.40	&5.28	&3.14	&2.49	&ESO		&C.~Waelkens\\
	&			&		&	&6.941	&6.380	&5.338	&	&	&2\,MASS	&VIZIER\\

30	&IRAS\,13258		&50508.64	&	&11.65	&9.87	&8.29	&6.52	&	&SAAO		&T.~Lloyd Evans\\
	&			&50204.46	&	&12.11	&10.05	&8.48	&6.56	&	&SAAO		&T.~Lloyd Evans\\
	&			&		&14.474	&12.569	&	&	&	&	&DENIS		&VIZIER\\
	&			&		&	&12.160	&10.630	&9.303	&	&	&2\,MASS	&VIZIER\\

31	&EN\,TrA		&42945.38	&	&7.19	&6.67	&6.03	&4.69	&	&SAAO		&\citet{85LloydEvans}\\
	&			&		&	&7.10	&6.53	&5.73	&3.92	&3.38	&ESO		&C.~Waelkens\\
	&			&		&	&7.092	&6.578	&5.931	&	&	&2\,MASS	&VIZIER	\\

32	&IRAS\,15469		&		&	&7.24	&6.21	&4.89	&2.89	&	&SAAO		&T.~Lloyd Evans (mean)\\
	&			&		&	&7.190	&6.235	&4.967	&	&	&2\,MASS	&VIZIER\\

33	&IRAS\,15556		&48778.44	&	&9.42	&8.34	&6.89	&4.55	&	&SAAO		&T.~Lloyd Evans\\
	&			&50642.32	&	&9.51	&8.38	&6.88	&4.49	&	&SAAO		&T.~Lloyd Evans\\
	&			&		&10.877	&9.440	&	&6.887	&	&	&DENIS		&VIZIER\\
	&			&		&	&9.423	&8.350	&6.940	&	&	&2\,MASS	&VIZIER\\

34	&IRAS\,16230		&49124.51	&	&9.10	&8.48	&7.79	&6.29	&	&SAAO		&T.~Lloyd Evans\\
	&			&48819.36	&	&9.37	&8.77	&8.12	&6.26	&	&SAAO		&T.~Lloyd Evans\\
	&			&		&	&9.002	&8.510	&7.812	&	&	&2\,MASS	&VIZIER\\

35	&IRAS\,17038		&50641.37	&	&7.74	&7.00	&6.58	&5.54	&	&SAAO		&T.~Lloyd Evans\\
	&			&49473.66	&	&8.36	&7.64	&7.16	&6.11	&	&SAAO		&T.~Lloyd Evans\\
	&			&		&9.434	&7.791	&	&6.369	&	&	&DENIS		&VIZIER\\
	&			&		&	&7.622	&7.050	&6.523	&	&	&2\,MASS	&VIZIER\\

36	&IRAS\,17233		&50204.56	&	&10.03	&8.89	&7.28	&4.86	&	&SAAO		&T.~Lloyd Evans\\
	&			&49474.60	&	&10.89	&9.68	&7.72	&5.09	&	&SAAO		&T.~Lloyd Evans\\
	&			&		&	&10.423	&9.592	&8.371	&	&	&2\,MASS	&VIZIER\\

37	&IRAS\,17243		&		&	&8.17	&7.46	&6.52	&4.75	&	&SAAO		&T.~Lloyd Evans (mean)\\
	&			&		&9.314	&8.177	&6.592	&	&	&	&DENIS		&VIZIER\\
	&			&		&	&8.035	&7.358	&6.462	&	&	&2\,MASS	&VIZIER\\

38	&89\,Her		&		&	&4.998	&4.239	&3.632	&	&	&2\,MASS	&VIZIER\\

39	&AI\,Sco		&42997.35	&	&6.74	&5.94	&5.14	&3.79	&	&SAAO		&\citet{85LloydEvans}\\
	&			&46245.48	&	&6.62	&5.98	&5.33	&3.81	&	&SAAO		&\citet{87Goldsmith}\\
	&			&		&	&6.864	&6.179	&5.485	&	&	&2\,MASS	&VIZIER\\

40	&IRAS\,18123		&		&	&8.04	&7.44	&6.86	&5.39	&	&SAAO		&T.~Lloyd Evans (mean)\\
	&			&		&	&7.974	&7.399	&6.729	&	&	&2\,MASS	&VIZIER\\

41	&IRAS\,18158		&48778.52	&	&10.52	&9.31	&8.01	&6.40	&	&SAAO		&T.~Lloyd Evans\\
	&			&50204.60	&	&11.15	&9.80	&8.27	&6.17	&	&SAAO		&T.~Lloyd Evans\\
	&			&		&12.341	&10.958	&	&8.291	&	&	&DENIS		&VIZIER\\
	&			&		&	&10.788	&9.489	&8.120	&	&	&2\,MASS	&VIZIER\\

42	&AC\,Her		&		&	&	&	&	&4.4	&4.0	&Kitt Peak NO	&\citet{70Gehrz}\\
	&			&		&	&	&	&	&	&	&Arizona system	&\\
	&			&		&	&	&	&5.2	&4.7	&3.6	&Mt Lemmon	&\citet{72bGehrz}\\
	&			&		&	&	&	&	&	&	&Arizona system	&\\
	&			&42456.87	&	&5.97	&5.52	&5.28	&4.73	&	&Tenerife	&\citet{85LloydEvans}\\
	&			&46243.52	&	&5.89	&5.39	&5.16	&4.66	&3.90	&SAAO		&\citet{87Goldsmith}\\
	&			&46248.47	&	&6.21	&5.66	&5.40	&4.87	&4.55	&SAAO		&\citet{87Goldsmith}\\
	&			&		&	&5.700	&5.338	&5.075	&	&	&2\,MASS	&VIZIER\\

43	&AD\,Aql		&46253.53	&	&9.62	&9.14	&8.95	&	&	&SAAO		&\citet{87Goldsmith}\\
	&			&		&	&9.543	&9.134	&9.002	&	&	&2\,MASS	&VIZIER\\

44	&IRAS\,19125		&		&	&7.48	&6.47	&5.38	&3.52	&	&SAAO		&T.~Lloyd Evans (mean)\\
	&			&		&	&8.356	&7.442	&7.067	&	&	&2\,MASS	&VIZIER\\

45	&EP\,Lyr		&		&	&	&	&	&6.6	&	&Mt Lemmon	&\citet{72bGehrz}\\
	&			&		&	&	&	&	&	&	&Arizona system	&\\
	&			&		&	&8.534	&8.153	&8.030	&	&	&2\,MASS	&VIZIER	\\

46	&IRAS\,19157		&48779.58	&	&8.89	&8.12	&6.95	&5.13	&	&SAAO		&T.~Lloyd Evans\\
	&			&48820.44	&	&8.94	&8.16	&6.92	&5.32	&	&SAAO		&T.~Lloyd Evans\\
	&			&		&9.713	&8.966	&	&7.109	&	&	&DENIS		&VIZIER\\
	&			&		&	&8.872	&8.205	&7.020	&	&	&2\,MASS	&VIZIER\\

47	&IRAS\,20056		&45897.49	&	&9.99	&8.89	&7.31	&4.77	&	&SAAO		&\citet{88Menzies}\\
	&			&46693.29	&	&10.55	&9.20	&7.43	&4.78	&	&SAAO		&\citet{88Menzies}\\
	&			&		&	&10.097	&8.964	&7.471	&	&	&2\,MASS	&VIZIER\\

48	&R\,Sge			&		&	&	&	&	&4.5	&3.3	&Kitt Peak NO	&\citet{70Gehrz}\\
	&			&		&	&	&	&	&	&	&Arizona system	&\\
	&			&		&	&	&	&	&4.6	&3.6	&Mt Lemmon	&\citet{72bGehrz}\\
	&			&		&	&	&	&	&	&	&Arizona system	&\\
	&			&46248.54	&	&7.10	&6.50	&5.80	&4.50	&	&SAAO		&\citet{87Goldsmith}\\
	&			&		&	&7.818	&7.234	&6.539	&	&	&2\,MASS	&VIZIER	\\

49	&V\,Vul			&		&	&	&	&	&4.2	&3.4	&Mt Lemmon	&\citet{72bGehrz}\\
	&			&		&	&	&	&	&	&	&Arizona system	&\\
	&			&		&	&6.557	&6.019	&5.396	&	&	&2\,MASS	&VIZIER\\

50	&HD\,213985		&		&	&8.18	&7.60	&6.68	&5.00	&4.45	&ESO		&C.~Waelkens\\
	&			&		&	&8.276	&7.609	&6.705	&	&	&2\,MASS	&VIZIER\\

51	&BD$+$39$^{\circ}$4926	&		&	&8.539	&8.394	&8.338	&	&	&2\,MASS	&VIZIER\\
\end{supertabular*}
\end{footnotesize}
\end{center}
\vspace{0.5cm}
\tablecaption{IRAS photometry points at $12$, $25$, $60$ and
$100\,\mu$m. Note however that in some cases the data points are upper
limits (L); these observations are probably contaminated by
interstellar cirrus clouds. Other data are lower limits (:). Remark
also that for BD$+$39$^{\circ}$4926 we don't have IRAS
datapoints.\label{DATA:IRAS}}
\begin{center}
\tablehead{
\hline
\hline
No	&Name			&$F_{12}$ (Jy)	&$F_{25}$ (Jy)	&$F_{60}$ (Jy)	&$F_{100}$ (Jy)\\
	&			&$12\,\mu$m	&$25\,\mu$m	&$60\,\mu$m	&$100\,\mu$m\\
\hline}
\tabletail{\hline}
\begin{supertabular}{llllll}
1	&TW\,Cam		&8.25		&5.60		&1.84		&1.79L\\

2	&RV\,Tau		&22.52		&18.05		&6.50		&2.44:\\

3	&IRAS\,05208		&9.53		&9.30		&2.78		&1.40\\

4	&DY\,Ori		&12.43		&14.89		&4.18		&11.46L\\

5	&CT\,Ori		&6.14		&5.55		&1.24		&1.59L\\

6	&SU\,Gem		&7.91		&5.68		&2.19		&11.99L\\

7	&UY\,CMa		&3.49		&2.46		&0.57		&1.00L\\

8	&HD\,44179		&421.60		&456.10		&173.10		&66.19\\

9	&HD\,46703		&0.46		&0.38:		&0.40L		&1.15L\\

10	&ST\,Pup		&3.51		&5.92		&1.24		&1.13L\\

11	&HD\,52961		&4.53		&2.22		&0.95		&0.87:\\

12	&SAO\,173329		&2.28		&1.83		&0.64		&10.45L\\

13	&U\,Mon			&124.30		&88.43		&26.59		&9.54\\

14	&AR\,Pup		&131.30		&94.32		&26.12		&11.96:\\

15	&IRAS\,08544		&180.30		&158.80		&56.25:		&28.43L\\

16	&IRAS\,09060		&3.78		&2.53		&0.65		&1.00L\\

17	&IRAS\,09144		&15.31		&12.14		&2.86		&24.68L\\

18	&IW\,Car		&101.00		&96.24		&34.52		&13.46\\

19	&IRAS\,09400		&7.85		&5.48		&1.58		&8.34L\\

20	&IRAS\,09538		&2.33		&1.60		&0.53		&1.34L\\

21	&HR\,4049		&48.25		&9.55		&1.77		&1.69L\\

22	&IRAS\,10174		&41.80		&60.39		&16.13:		&174.30L\\

23	&IRAS\,10456		&189.10		&115.50		&30.86		&10.89\\

24	&HD\,95767		&22.13		&15.65		&10.90		&58.88\\

25	&GK\,Car		&2.85		&2.45		&0.79		&12.41L\\

26	&IRAS\,11472		&11.35		&14.18		&1.78		&1.00L\\

27	&RU\,Cen		&5.32		&10.96		&5.65		&2.09\\

28	&SX\,Cen		&5.92		&3.59		&1.10		&1.55L\\

29	&HD\,108015		&32.46		&33.23		&7.99		&2.41\\

30	&IRAS\,13258		&1.90		&2.97		&1.46		&3.82:\\

31	&EN\,TrA		&13.20		&10.30		&4.11		&2.12\\

32	&IRAS\,15469		&48.81		&42.12		&15.50		&278.10L\\

33	&IRAS\,15556		&19.86		&17.80		&7.10		&189.80L\\

34	&IRAS\,16230		&4.29		&3.34		&0.82		&14.55L\\

35	&IRAS\,17038		&8.05		&9.67		&4.49		&43.34L\\

36	&IRAS\,17233		&17.02		&13.40		&3.67		&32.08L\\

37	&IRAS\,17243		&10.76		&8.77		&3.69		&5.41:\\

38	&89\,Her		&97.52		&54.49		&13.42		&6.04\\

39	&AI\,Sco		&17.59		&11.38		&2.95		&46.76L\\

40	&IRAS\,18123		&10.72		&11.02		&4.21		&1.85L\\

41	&IRAS\,18158		&3.00		&2.64		&1.03		&1.56L\\

42	&AC\,Her		&41.43		&65.33		&21.37		&8.04\\

43	&AD\,Aql		&2.49		&2.72		&0.65L		&24.46L\\

44	&IRAS\,19125		&28.89		&26.51		&7.81		&28.05L\\

45	&EP\,Lyr		&0.31L		&0.50		&0.40L		&2.10L\\

46	&IRAS\,19157		&8.88		&7.16		&2.45		&11.05L\\

47	&IRAS\,20056		&17.52		&18.01		&5.38		&2.84L\\

48	&R\,Sge			&10.60		&7.54		&2.12		&1.72L\\

49	&V\,Vul			&12.35		&5.69		&1.29		&7.25L\\

50	&HD\,213985		&5.57		&4.66		&2.11		&1.01L\\
\end{supertabular}
\end{center}
\vspace{0.5cm}
\tablecaption{Data of the Midcourse Space eXperiment (MSX). The
instrument on board MSX is the SPIRIT III (Spatial Infrared Imaging
Telescope III). The approximate effective wavelengths of the 6 MSX
filters are in $\mu$m.\label{DATA:MSX}}
\begin{center}
\tablehead{
\hline
\hline
No	&Name			&B1 (Jy)	&B2 (Jy)	&A (Jy)		&C (Jy)		&D (Jy)		&E (Jy)\\
	&			&$4.29\,\mu$m	&$4.35\,\mu$m	&$8.28\,\mu$m	&$12.13\,\mu$m	&$14.65\,\mu$m	&$21.34\,\mu$m\\
\hline}
\tabletail{\hline}
\begin{supertabular}{llllllll}
4	&DY\,Ori		&		&		&9.073		&11.58		&11.89		&13.97\\

5	&CT\,Ori		&		&		&2.618		&3.928		&1.421		&6.761\\

6	&SU\,Gem		&		&		&8.283		&7.666		&7.178		&7.505\\

10	&ST\,Pup		&		&		&4.419		&5.743		&5.464		&7.199\\

13	&U\,Mon			&		&42.93		&56.45		&88.71		&72.38		&73.29\\

14	&AR\,Pup		&72.73		&81.04		&114.3		&118.0		&100.2		&91.48\\

15	&IRAS\,08544		&76.62		&51.92		&139.0		&150.7		&143.6		&134.2\\

17	&IRAS\,09144		&		&7.285		&10.69		&11.61		&10.31		&9.341\\

19	&IRAS\,09400		&		&		&4.814		&3.847		&3.072		&3.034\\

22	&IRAS\,10174		&16.89		&19.14		&25.84		&33.91		&28.70		&45.77\\

23	&IRAS\,10456		&139.5		&147.8		&165.4		&148.1		&126.7		&106.3\\

24	&HD\,95767		&		&12.33		&18.96		&18.70		&16.33		&13.09\\

25	&GK\,Car		&		&		&2.774		&2.630		&2.172		&2.086\\

32	&IRAS\,15469		&34.65		&35.82		&42.91		&45.97		&40.16		&37.48\\

33	&IRAS\,15556		&		&		&16.63		&17.50		&17.20		&14.98\\

35	&IRAS\,17038		&		&		&10.45		&12.66		&10.93		&11.09\\

36	&IRAS\,17233		&		&		&13.94		&15.28		&12.87		&10.74\\

37	&IRAS\,17243		&		&		&9.254		&9.280		&8.408		&7.775\\

39	&AI\,Sco		&		&		&15.26		&14.89		&12.72		&8.562\\

44	&IRAS\,19125		&		&23.43		&24.55		&26.48		&23.90		&22.54\\
\end{supertabular}
\end{center}
\vspace{0.5cm}
\tablecaption{$850\,\mu$m fluxes from observations with the
Submillimetre Common-User Bolometer Array (SCUBA) at the James Clerk
Maxwell Telescope (JCMT), Mauna Kea, Hawaii.\label{DATA:SCUBA}}
\begin{center}
\tablehead{
\hline
\hline
No	&Name			&\multicolumn{2}{c}{$F_{850}$ (mJy)}	&Reference\\
\hline}
\tabletail{\hline}
\begin{supertabular}{llr@{\,$\pm$\,}ll}
1	&TW\,Cam		&11.0	&2.3				&\citet{05DeRuyter}\\

2	&RV\,Tau		&50.3	&3.6				&\citet{05DeRuyter}\\

6	&SU\,Gem		&7.5	&2.5				&\citet{05DeRuyter}\\

7	&UY\,CMa		&2.4	&2.1				&\citet{05DeRuyter}\\

11	&HD\,52961		&2.8	&1.9				&new data\\

13	&U\,Mon			&181.6	&2.6				&\citet{05DeRuyter}\\

21	&HR\,4049		&8.7	&2.8				&new data\\
	&			&10.4	&2.2				&\citet{03Dominik}\\

38	&89\,Her		&40.9	&2.4				&new data\\

42	&AC\,Her		&99.4	&3.8				&\citet{05DeRuyter}\\

47	&IRAS\,20056		&21.8	&1.8				&\citet{02Gledhill}\\
\end{supertabular}
\end{center}

\end{document}